\documentclass[useAMS,usenatbib]{mnras}

\usepackage{hyperref}

\usepackage{epsfig}
\usepackage{myaasmacros}
\usepackage{amsmath}
\usepackage{amsfonts}
\usepackage{amssymb}
\usepackage{subfig}
\usepackage{color}
\usepackage{changes}

\newcommand{\Msun}{\hbox{M$_\sun$}}
\newcommand{\hii}{\hbox{H\,{\sc ii}}}
\newcommand{\ha}{\hbox{H$\alpha$}}
\newcommand{\hb}{\hbox{H$\beta$}}
\newcommand{\oiii}{\hbox{[O\,{\sc iii}]$\lambda5007$}}
\newcommand{\oiiifour}{\hbox{[O\,{\sc iii}]$\lambda4959$}}
\newcommand{\oiiiboth}{\hbox{[O\,{\sc iii}]$\lambda\lambda4959,5007$}}
\newcommand{\oiiiauroral}{\hbox{[O\,{\sc iii}]$\lambda4363$}}
\newcommand{\oi}{\hbox{[O\,{\sc i}]$\lambda6300$}} 
\newcommand{\niiauroral}{\hbox{[N\,{\sc ii}]$\lambda5755$}}
\newcommand{\nii}{\hbox{[N\,{\sc ii}]$\lambda6584$}}

\newcommand{\sii}{\hbox{[S\,{\sc ii}]$\lambda6724$}}

\newcommand{\lcii}{\hbox{[C\,{\sc ii}]}}
\newcommand{\loii}{\hbox{[O\,{\sc ii}]}}

\newcommand{\oiiihb}{\hbox{[O\,{\sc iii}]/H$\beta$}}
\newcommand{\oiihb}{\hbox{[O\,{\sc ii}]/H$\beta$}}
\newcommand{\niiha}{\hbox{[N\,{\sc ii}]/H$\alpha$}}
\newcommand{\siiha}{\hbox{[S\,{\sc ii}]/H$\alpha$}}

\newcommand{\niioii}{\hbox{[N\,{\sc ii}]/[O\,{\sc ii}]}}
\newcommand{\niisii}{\hbox{[N\,{\sc ii}]/[S\,{\sc ii}]}}
\newcommand{\oiiioii}{\hbox{[O\,{\sc iii}]/[O\,{\sc ii}]}}
\newcommand{\oiiioiihb}{\hbox{([O\,{\sc iii}]+[O\,{\sc ii}])/H$\beta$}}

\newcommand{\civd}{\hbox{C\,\textsc{iv}\,$\lambda\lambda1548,1551$}}
\newcommand{\civ}{\hbox{C\,\textsc{iv}\,$\lambda1550$}}
\newcommand{\heii}{\hbox{He\,\textsc{ii}\,$\lambda1640$}}
\newcommand{\oiiiuvd}{\hbox{O\,\textsc{iii}]$\lambda\lambda1661,1666$}}
\newcommand{\oiiiuv}{\hbox{O\,\textsc{iii}]\,$\lambda1663$}}
\newcommand{\niii}{\hbox{N\,\textsc{iii}]\,$\lambda1750$}}
\newcommand{\silii}{\hbox{Si\,\textsc{ii}]\,$\lambda1814$}}
\newcommand{\siliiid}{\hbox{[Si\,\textsc{iii}]$\lambda1883$+Si\,\textsc{iii}]$\lambda1892$}}
\newcommand{\siliii}{\hbox{Si\,\textsc{iii}]\,$\lambda1888$}}
\newcommand{\ciiid}{\hbox{[C\,\textsc{iii}]$\lambda1907$+C\,\textsc{iii}]$\lambda1909$}}
\newcommand{\ciii}{\hbox{C\,\textsc{iii}]\,$\lambda1908$}}
\newcommand{\cii}{\hbox{C\,\textsc{ii}]\,$\lambda2326$}}
\newcommand{\oiid}{\hbox{[O\,{\sc ii}]$\lambda\lambda3726,3729$}}
\newcommand{\oii}{\hbox{[O\,\textsc{ii}]\,$\lambda3727$}}

\newcommand{\logoh}{\hbox{$\mathrm{12+\log(O/H)}$}}
\newcommand{\logU}{\hbox{$\log U$}}

\newcommand{\Te}{\hbox{$T_{\mathrm e}$}}
\newcommand{\nel}{\hbox{$n_{\mathrm e}$}}

\newcommand{\cloudy}{\textsc{cloudy}}

\newcommand{\mappings}{\textsc{mappings}}

\newcommand{\JWST}{\textit{JWST}}

\title[Evolution of metallicity calibrations] 
{High-redshift metallicity calibrations for JWST spectra: insights
  from line emission in cosmological simulations}    
\author[Hirschmann, Charlot \& Somerville]{Michaela Hirschmann$^{1,2}$\thanks{E-mail:
    michaela.hirschmann@epfl.ch}, Stephane Charlot$^{3}$ and Rachel
  S. Somerville$^4$\\
$^{1}$Institute for Physics, Laboratory for Galaxy Evolution and
Spectral Modelling, EPFL,
Observatoire de Sauverny,\\
\indent \hspace{0.08cm} Chemin Pegasi 51, 1290 Versoix, Switzerland\\
$^{2}$INAF, Osservatorio Astronomico di Trieste, Via Tiepolo 11,
34131 Trieste, Italy\\
$^{3}$Sorbonne Universit\'e, CNRS, UMR 7095, Institut d'Astrophysique 
de Paris, 98 bis bd Arago, 75014 Paris, France\\
$^{4}$Center for Computational Astrophysics, Flatiron Institute, 162 5th Ave, New York, NY 10010, USA \\
}

\begin{document}

\date{Accepted ???. Received ??? in original form ???}

\pagerange{\pageref{firstpage}--\pageref{lastpage}} \pubyear{2002}

\maketitle

\label{firstpage}

\begin{abstract}

Optical emission-line ratios are traditionally used to
estimate gas metallicities from observed galaxy spectra. While
such estimators have been calibrated primarily at low
redshift, they are commonly used to study high-redshift galaxies,
where their applicability may be questioned.
We use comprehensive emission-line catalogues of galaxies from
the IllustrisTNG simulation including ionization by stars, active nuclei 
and shocks to reassess the calibrations of both optical and
ultraviolet metallicity estimators at redshifts $0\la z \la8$.
For present-day galaxies, the predicted optical-line calibrations
are consistent with previously published ones, while we find
different ultraviolet-line ratios, such as \heii/\ciii, to provide powerful 
metallicity diagnostics. 
At fixed metallicity, most emission-line ratios are predicted to strongly 
increase or decrease with redshift (with the notable exception of 
N2O2=\nii/\oii), primarily because of a change in ionization parameter.
The predicted dependence of R3=\oiii/\hb\ and R23=(\oii+\oiiiboth)/\hb, and to
a slightly lesser extent R2=\oii/\hb\ and O32=\oiii/\oii, on O abundance
for galaxies at $z=4-8$ agrees remarkably well with \Te-based
measurements in 14 galaxies observed with \JWST. This success leads us
to provide new calibrations of optical and ultraviolet metallicity estimators 
specifically designed for galaxies at $z>4$, to guide interpretations of future, 
high-redshift spectroscopic surveys. 
We further demonstrate that applying classical $z=0$ calibrations to
high-redshift galaxies can bias O-abundance estimates downward 
by up to 1~dex, leading to the conclusion of a stronger evolution of the
mass-metallicity relation than the actual one.
\end{abstract}


\section{Introduction}\label{intro}

Shedding light on the metal enrichment history of galaxies is an
essential step toward
understanding galaxy formation and evolution. Specifically,
gas-phase metallicity has been observed to strongly correlate with
galaxy stellar mass in the nearby universe \citep{Lequeux79, Kinman81,
  VilaCostas92, Baldry02, Tremonti04}. This ``mass-metallicity''
relation is thought to be shaped by a balance between gas accretion
from the circumgalactic and intergalactic media and gas outflows
driven by different physical processes, such as feedback from massive
stars and supernovae, and active galactic nuclei (AGN; 
see, e.g., the model predictions of \citealp{Hirschmann13, Hirschmann16,
  Dave17, Torrey19, DeLucia20}, and the review by
\citealp{Maiolino19}). In fact, state-of-the-art
  cosmological simulations adopting different prescriptions for these
  processes have been shown to diverge greatly in their
  predictions of the mass-metallicity relation,
  especially at high redshift \citep{Somerville15, Naab16}.
  Reliable measurements of the mass-metallicity
  relation at different cosmic epochs are essential to better constrain
  these uncertain physical processes in galaxy-formation simulations.

Observationally, the gas-phase metallicity of a star-forming galaxy can be
estimated from a wide variety of nebular emission lines at ultraviolet (UV)
to infrared (IR) wavelengths. There are four main approaches to this, which
rely on: (i) metal-recombination lines; (ii) 
collisionally-excited lines combined with estimates of the electronic
temperature and density (the `direct-\Te' approach); (iii) empirical 
calibrations of line-intensity ratios as a function of metallicity; and (iv) 
theoretical calibrations of line-intensity ratios as a function of metallicity. 
The recombination-line approach, which provides the most accurate 
metallicity estimates, can be used in 
practice only in nearby \hii\ regions, where weak metal-recombination lines,
such as \lcii$\lambda$4267 and \loii$\lambda$4650, are 
measurable \citep[e.g.,][]{Esteban14}. The direct-\Te\ method relies
on measurements of \Te-sensitive auroral-line intensities, such as
\oiiiauroral\ and \niiauroral, which also tend to be weak, except in
metal-poor 
galaxies \citep[][and references therein]{Kewley19}. For these
reasons, various authors have 
proposed `empirical' relations to estimate gas-phase 
metallicity (oxygen abundance) from intensity ratios of strong, 
easily-measurable emission lines, such as \oii, \hb\ and \oiii, which are 
calibrated  using samples of \hii\ regions and galaxies for which 
auroral-line measurements are also available 
\citep[e.g.][]{Pagel80, Pettini04, Pilyugin16, Marino16}. More recently, 
\citet{Curti20} proposed an updated set of empirical relations including 
a comprehensive set of optical emission lines,  
calibrated using a large sample of galaxies from 
the Sloan Digital Sky Survey \citep[SDSS,][]{Abazajian09}. Yet, the direct-\Te\
metallicity measurements on which these calibrations rely have known 
caveats, as they depend for example indirectly on models to correct for 
unobserved states of ionization and exist primarily for low-redshift galaxies
\citep[e.g.,][]{Gutkin16, Kewley19, Cameron22}.    

An alternative to the above empirical approach is to appeal to 
photoionization models to relate observable intensity ratios of strong
emission lines to gas metallicity. This has been explored by different 
groups using the public photoionization codes \cloudy\ \citep{Ferland17} 
and \mappings\  \citep{Sutherland17}, for various assumptions about
the nature of the ionizing sources \citep[stars, AGN, radiative 
shocks; e.g.,][]{Charlot01, Kewley02, Kobulnicky04, Nagao11,
Gutkin16, Feltre16, Byler18}. While this approach allows 
one to fully investigate the influence of the many adjustable 
model parameters on predicted emission-line intensities, the wide 
ranges of theoretically allowed parameters and degeneracy of solutions 
generally imply large uncertainties on the estimated quantities
\citep[e.g.,][]{Chevallard16, Vidal22}. Hence, the four approaches described
above to estimate gas-phase metallicities in galaxies each have their
own strengths and weaknesses. Unfortunately, estimates based on different 
empirical and model metallicity calibrations can be significantly discrepant 
\citep[e.g.,][]{Kewley08, Stasinska05,  Peimbert17, Kewley19},
implying large uncertainties in the derived mass-metallicity relation of nearby galaxies.   

Over the past fifteen years, major efforts have been devoted to the
exploration of the gas-phase metallicities of galaxies at higher redshift, 
mainly by means of near-infrared (IR) spectroscopic surveys
\citep[e.g.][]{Maiolino08, Troncoso14, Zahid14, Erb16, 
Shapley17, Maiolino19}. Such studies often rely on metallicities estimated from 
observed line ratios using calibrations derived at $z=0$. These studies 
generally find that a mass-metallicity relation was already in place 
at $z \ga 3$ and evolved down to $z=0$ such that the average gas-phase
metallicity at a fixed stellar mass decreased with time 
\citep{Zahid13, Maiolino08, Kobulnicky04}, in a way qualitatively consistent
with expectations from various models and simulations
\citep[e.g.,][]{Hirschmann16, Dave17, Torrey19}. 

Despite this achievement, the robustness of metallicity estimates of
high-redshift galaxies obtained using calibrations derived at $z=0$ 
remains debated, mainly because of the potentially very different
conditions in the interstellar medium (ISM) of such galaxies relative
to their low-redshift counterparts \citep{Kewley19, Maiolino19}. 
In fact, \citet{Bian18} find that metallicities derived using the direct-\Te\
method in local SDSS analogues of $z\sim2$ galaxies are discrepant from
those derived using standard $z=0$ metallicity calibrations. Another 
complication is that metallicity and ionization parameter can have 
similar effects on strong-line luminosity ratios, which some
methods have been developed to overcome
\citep[e.g.,][]{Poetrodjojo18, Izotov19, Kewley19}.   Nevertheless, to
date, no robust metallicity calibrations could be derived
for large samples of high-redshift galaxies, for the following main two
reasons: (i) auroral lines are typically too faint to be detected in
high-redshift  spectra, preventing direct-\Te\ metallicity estimates, 
although small samples are starting to be gathered out to $z\sim9$ 
\citep[e.g.][]{Patricio18, Sanders23, Curti23}; and (ii) predictions from
photoionization models span wide ranges of adjustable parameters, 
which are unconstrained outside any cosmological context. 

The newly commissioned {\it James Webb Space Telescope} 
(\JWST) and its near-IR spectrograph NIRSpec
\citep{Jakobsen22, Ferruit22} provide a revolutionary means
of exploring the rest-frame UV- and optical-line properties
of galaxies out to very high redshift. For example, the faint \oiiiauroral\ 
auroral line has already been detected out to $z\sim9$ by, e.g., \citet{Curti23} and
\citet{Sanders23}. Both studies find that the metallicities derived 
using the direct-\Te\ method in high-redshift galaxies differ from
those that would be derived from strong emission-line luminosities
using typical $z\sim0$ calibrations. {\it This implies that a  
reassessment of metallicity calibrations  at high redshift is required.} 
Together with the results from other studies based on the direct-\Te\ 
method and/or empirical calibrations
\citep[e.g.][]{Langeroodi22, Heintz22, Schaerer22, Trump23}, these first high-redshift
data from \JWST\ suggest that galaxies at $z\sim6$--9 have metallicities
either consistent with the extrapolation of the
mass-metallicity relation at z$\sim$2--3, or below this relation.

In the case of very distant galaxies, the strong rest-frame optical lines are 
shifted outside the window of \JWST/NIRSpec. This occurs at $z\ga7$ 
for \ha\ and \nii\ and at $z\ga9$ for \hb\ and \oiii. In such cases, rest-frame 
UV emission lines can be observed \citep[e.g.][]{Bunker23} and provide 
valuable diagnostics of the physical conditions in the ISM \citep[e.g.][]{Stark14}.
Still, the capability of UV-line ratios to trace metallicities is less well 
explored and understood than that of optical-line ratios, despite important 
efforts in photoinization modelling \citep[e.g.,][]{Gutkin16, Byler18, Kewley19}. 

A promising new alternative to classical calibrations of metallicity
estimators at high redshift is offered by simulations of nebular-line
emission of galaxy populations  in a full cosmological context
\citep[e.g.,][]{Orsi14, Hirschmann17,   Hirschmann19,
  Hirschmann22}. This approach overcomes the simplifying
assumptions inherent in the direct-\Te\ method and reduces the 
wide range of adjustable parameters (such as SFR, ionization 
parameter, gas density, gas metallicity and C/O abundance ratio) 
spanned by grids of 
photoionization models to physically motivated combinations. 
Yet, despite the increasing number of theoretical studies in this
field over the last few years \citep[e.g.][]{Orsi14, Shimizu16,
  Hirschmann17,   Hirschmann19, Shen20, Wilkins20, Baugh22, Garg22},
so far, none has focused on the calibration of strong-line 
emission to derive metallicities from high-redshift galaxy spectra. 

In this paper, we fill this gap by appealing to the same methodology
as in our previous work  \citep[][]{Hirschmann17, Hirschmann19,
  Hirschmann22} to model in a self-consistent 
way the emission-line properties of galaxy populations from the
IllustrisTNG simulations; we incorporate the contribution to nebular
emission from young star clusters, AGN, post-asymptotic-giant-branch
(PAGB) stellar populations and fast radiative shocks. 
We achieve this by coupling photoionization models for young stars 
\citep{Gutkin16}, AGN \citep{Feltre16}, PAGB stars \citep{Hirschmann17} 
and fast, radiative shocks \citep{Alarie19} with the IllustrisTNG cosmological 
hydrodynamic simulations. Our methodology offers a unique way to
address important questions about the derivation of interstellar
metallicities from optical- and UV-line ratios of galaxies at
different cosmic epochs:
\begin{itemize}
\item Are the relations between optical-line ratios 
  and metallicity predicted by IllustrisTNG simulations for 
  present-day galaxy populations consistent with published 
  calibrations?
\item How do the relations between strong-line ratios 
  and metallicity predicted by IllustrisTNG simulations 
  evolve from $z=0$ to $z=8$? Is the application 
  of metallicity calibrations derived at $z=0$ at high-redshift appropriate?
 \item Can UV-line ratios enable a robust derivation of gas-phase 
   metallicities for galaxy populations observable at redshifts
   $z>4$ in ongoing and planned \JWST/NIRSpec surveys? 
\end{itemize}
Answers to these questions will provide valuable insights into
the chemical enrichment of galaxy populations observed out to cosmic
dawn with \JWST, and hence, into the cosmic evolution of the mass-metallicity
relation.

The paper is structured as follows. We start by briefly describing the
theoretical framework of our study in Section~\ref{theory}, including
the IllustrisTNG simulation set, the photoionization models and the
coupling methodology between simulations and emission-line
models. Section~\ref{opticalz0} presents our main findings
about the relations between strong optical-line ratios and interstellar
metallicity in present-day simulated galaxies, and how these
relations compare with observationally derived relations/calibrations. In
Section~\ref{opticalhighz}, we explore the cosmic evolution of the
relations between optical-line ratios and metallicity for IllustrisTNG
galaxies out to $z=7$ and provide fitting functions for calibrations 
applicable to $z>3$ galaxies. Finally, in
Section~\ref{UVlineratios}, we explore different UV-line ratios as tracers
of interstellar metallicity in high-redshift galaxies, as
guidance for future \JWST\ observations. We address possible caveats of
the usage of line-ratio calibrations to derive metallicities and
discuss advantages and caveats of our approach in Section~\ref{discussion}. 
Section~\ref{summary} summarizes our main results. 

\section{Theoretical framework}\label{theory} 

In this paper, we take advantage of the modelled optical and UV
emission lines of galaxy populations in a wide redshift range, as
described in \citet{Hirschmann22}, using the IllustrisTNG100 and
IllustrisTNG50 simulation suite \citep[TNG100 and TNG50
hereafter,][]{Pillepich18a, Springel18,   Nelson18, Naiman18,
  Marinacci18, Nelson19, Pillepich19}. In the 
following paragraphs, we briefly summarize the simulation details
(Section~\ref{TNG}),   the emission-line models and coupling
methodology (Section~\ref{ELmodels}, \citealp{Gutkin16,   Feltre16,
  Hirschmann17, Hirschmann19,  Alarie19, Hirschmann22}),
referring the reader to the original studies for more details.     

\subsection{IllustrisTNG}\label{TNG}

IllustrisTNG is a suite of publicly available, large volume, cosmological,
gravo-magnetohydrodynamical simulations, run with the moving-mesh code 
Arepo \citep{Springel10}, and composed of three simulations with
different volumes and resolutions: TNG300,
TNG100 and TNG50, assuming the currently favoured Planck cosmology. In
this work, we consider only the TNG100 and TNG50 simulations
due to their higher resolution. The IllustrisTNG simulations include a
comprehensive model for galaxy formation physics \citep{Weinberger17,
  Pillepich18a}, which has been tuned to match observational
constraints on the galaxy stellar-mass function and stellar-to-halo
mass relation, the total gas-mass content within the virial radius of
massive galaxy groups, the stellar-mass/stellar-size relation and the
relation between black-hole (BH) mass and galaxy mass at $z=0$.
In an IllustrisTNG simulation, the properties of galaxies, galaxy groups,
subhaloes and haloes (identified using the FoF and Subfind 
substructure-identification algorithms, see \citealp{Davis85} and
\citealp{Springel01}, respectively), are computed `on the fly' and saved 
for each snapshot. In addition, an on-the-fly cosmic shock finder coupled
to the code \citep{Schaal15} uses a ray-tracing method to identify
shock surfaces and measure their properties.
The IllustrisTNG simulations have been shown in various studies to  
provide a fairly realistic representation of the properties of galaxies 
evolving across cosmic time \citep[e.g.,][]{Torrey19}. For more
details on the simulations, we refer the reader to the original studies above.

\subsection{Emission-line models and their coupling with simulated galaxies}\label{ELmodels}

We use the same methodology as in \citet{Hirschmann17, Hirschmann19,
  Hirschmann22} to compute nebular-line emission of
galaxies from the post-processing of the TNG50 and TNG100
simulations.
For the entire analysis, we consider only well-resolved galaxies containing
at least $\sim1000$ star particles, corresponding to stellar masses 
greater than $10^8 \Msun$ and $3 \times 10^9 \Msun$ in the TNG50 and 
TNG100 simulations, respectively.
For all these galaxies and their progenitors
at $z>0$, nebular emission from young star clusters \citep{Gutkin16}, narrow-line 
regions (NLR) of AGN \citep{Feltre16} and PAGB stars \citep{Hirschmann17} is
computed using version c13.03 of the photoionization  code \textsc{Cloudy} 
(\citealp{Ferland13}), while the emission from fast radiative shocks \citep{Alarie19} 
is computed using \textsc{Mappings V} \citep{Sutherland17}. All photoionization 
calculations considered in this work were performed adopting a common set of 
element abundances down to metallicities of a few per cent of Solar
\citep[from][]{Gutkin16}. For more details on the employed libraries
of nebular-emission models, parameterised in terms of different
stellar and ISM parameters, we refer the reader to \citet{Hirschmann22}.

We combine the IllustrisTNG simulations of galaxy populations described in 
Section~\ref{TNG} with these photoionization models by associating, 
at each time step, each simulated
galaxy with the appropriate \hii-region, AGN-NLR, PAGB and radiative-shock models,
which, taken together, constitute the total nebular emission of the
galaxy. We achieve this using the procedure described in
\citet{Hirschmann17, Hirschmann19, Hirschmann22}, by
self-consistently matching the model parameters available from the  
simulations with those of the emission-line models. The ISM
and stellar parameters of simulated galaxies are evaluated by considering
all `bound' gas cells and star particles (as identified by the Subfind algorithm;
Section~\ref{TNG}) for the coupling with the \hii-region and PAGB models, 
and within 1~kpc around the BH for the coupling with the AGN-NLR
models. Since the simulations do not allow us to distinguish between metals
in the dust and gas phases, we gather all metals under the term ``interstellar 
metallicity" in the remainder of this paper.

A few photoionization model parameters cannot be defined from the 
simulation, such as the slope of the AGN ionizing spectrum ($\alpha$),
the hydrogen gas density in individual ionized regions ($n_{\mathrm{H}}$),
the dust-to-metal mass ratio ($\xi_{\rm d}$) and the pre-shock density
($n_{\mathrm{H,shock}}$). For these parameters we adopt the same
values as in \citet{Hirschmann22}, which were shown in that work to reproduce many
observational diagnostics of emission lines as summarized below.

\subsection{Total emission-line luminosities, line ratios and
  equivalent widths of IllustrisTNG galaxies}\label{ELlums} 

The procedure described above allows us to compute the contributions of young stars,
an AGN, PAGB stars and fast radiative shocks to the luminosities
of various emission lines (such as $L_{\mathrm{H}\alpha}$,
$L_{\mathrm{H}\beta}$,  $L_{\mathrm{[OIII]}}$, etc.) in a simulated
galaxy. The {\it total} emission-line luminosities of the galaxy can
then be calculated by summing  these four contributions. For line
luminosity ratios, we adopt for simplicity the notation
$L_{\mathrm{[OIII]}}/L_{\mathrm{H}\beta}\equiv\oiiihb$. In this study, we
focus on exploring line ratios built from five optical lines, 
\hb, \oiii, \oiiifour,\footnote{Note that with \hbox{[O\,{\sc iii}]},
  we will always refer to  \oiii\ unless specified differently.} \ha, \nii\ and \sii\,
and nine UV lines, \civd\ (hereafter simply \civ),
\heii, \oiiiuvd\ (hereafter simply \oiiiuv), \niii\ (multiplet), \silii,
\siliiid\ (hereafter simply \siliii), \ciiid\ (hereafter simply
\ciii), \cii\ and \oiid\ (hereafter simply \oii).

\citet{Hirschmann22} have shown that the predicted basic properties of 
the emission lines of TNG galaxies are consistent with those observed, 
such as local optical line-ratio diagnostic diagrams, the evolution of the 
\oiiihb\ ratio and the evolution of the optical line-luminosity functions.  
We note that in the present paper, unless otherwise stated, we do not 
consider attenuation by dust outside \hii\ regions and compare our 
predictions with observed emission-line ratios corrected for this effect.

\begin{figure*}
  \begin{flushleft}
\epsfig{file=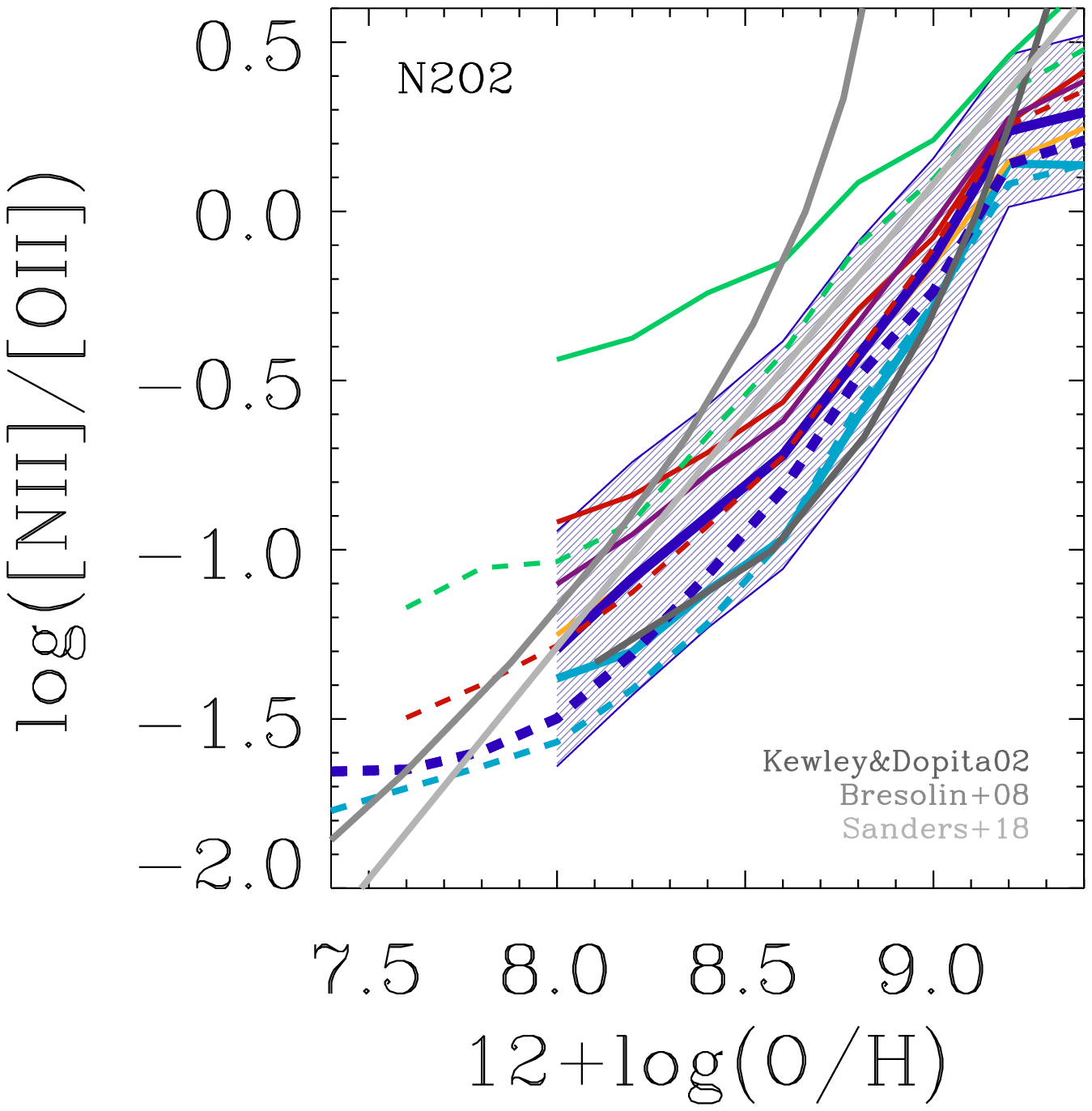,
  width=0.26\textwidth}\hspace{-0.4cm}
\epsfig{file=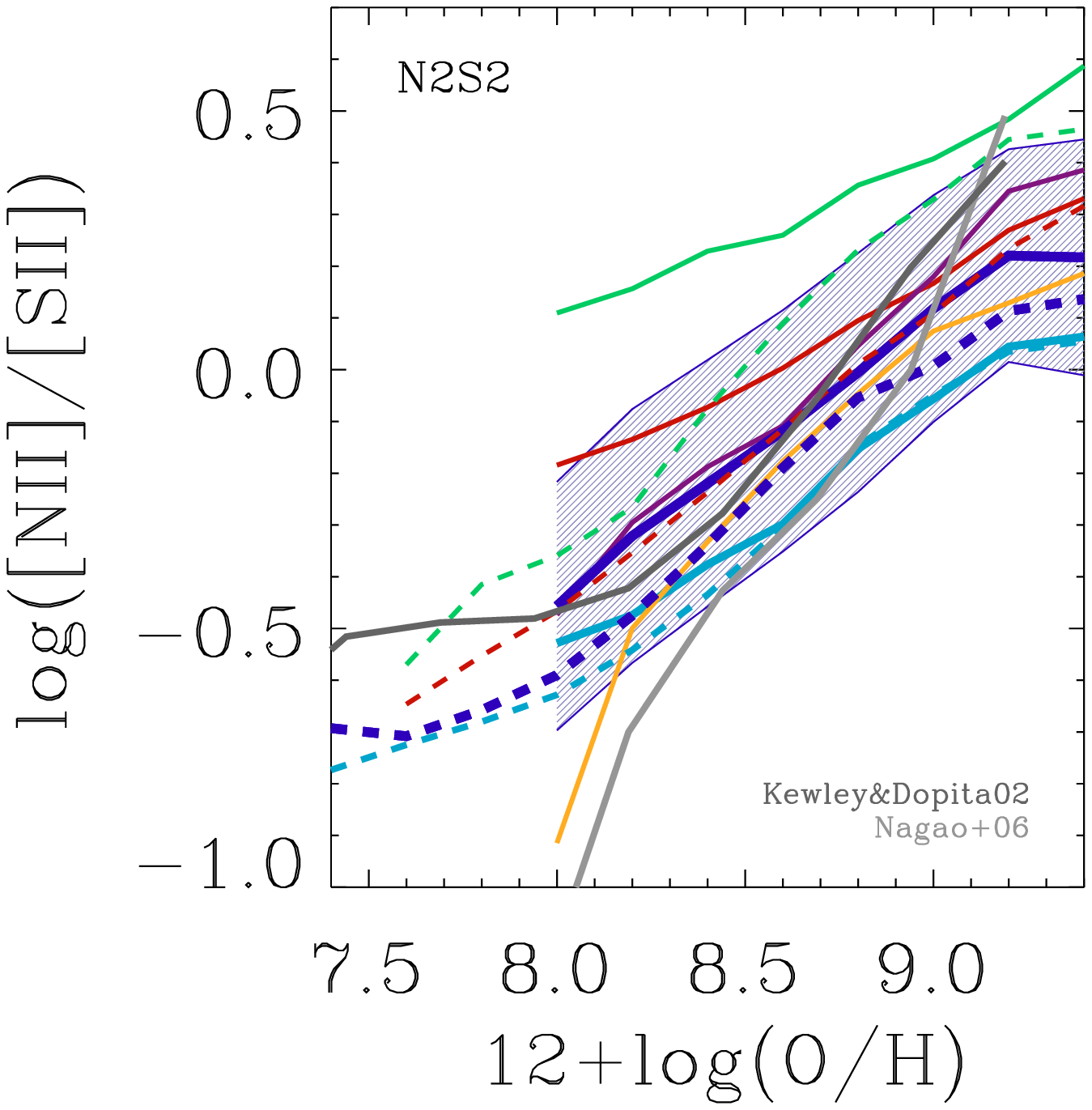,
  width=0.26\textwidth}\hspace{-0.4cm}
\epsfig{file=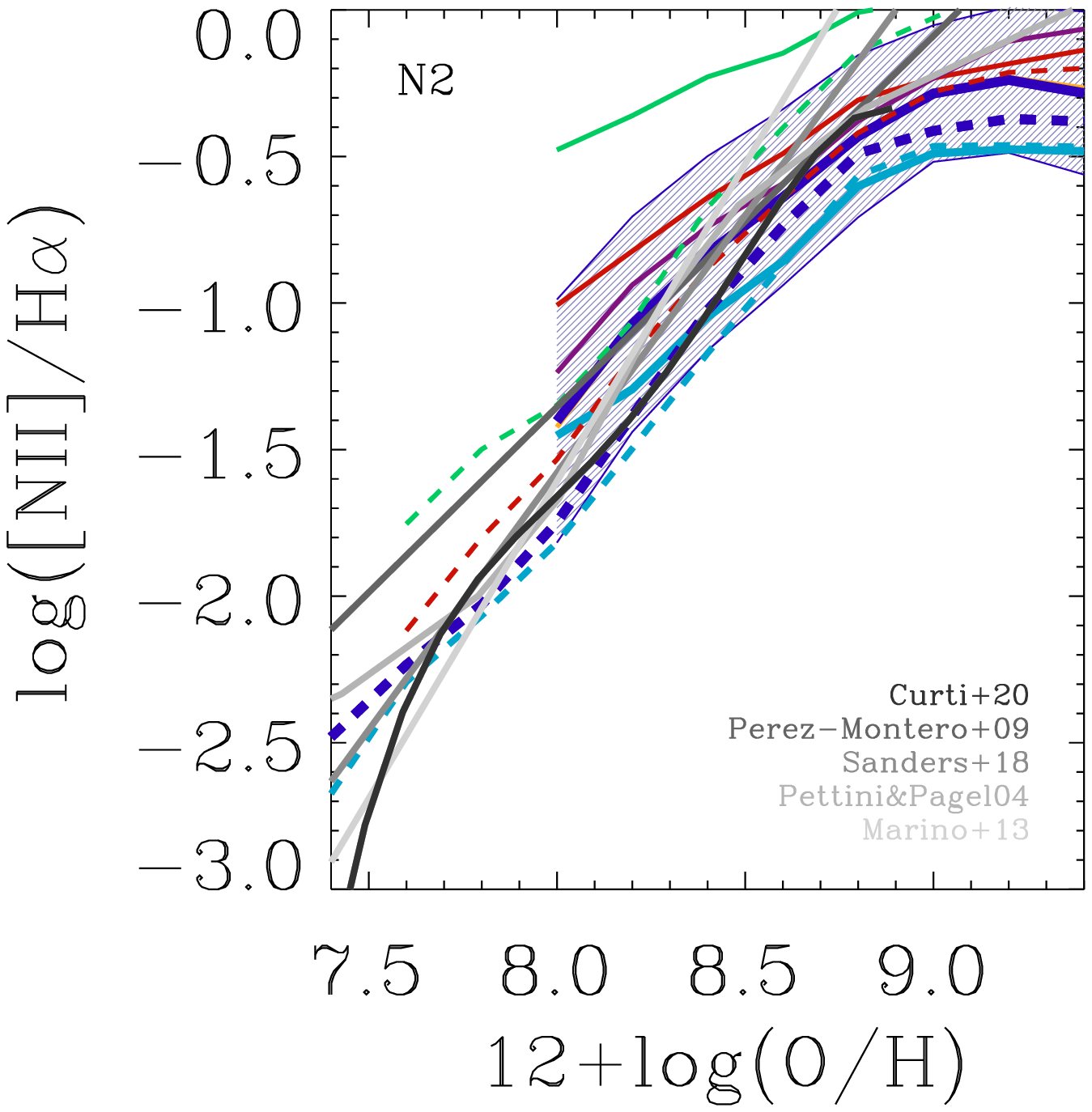,
  width=0.26\textwidth}\hspace{-0.4cm}
\epsfig{file=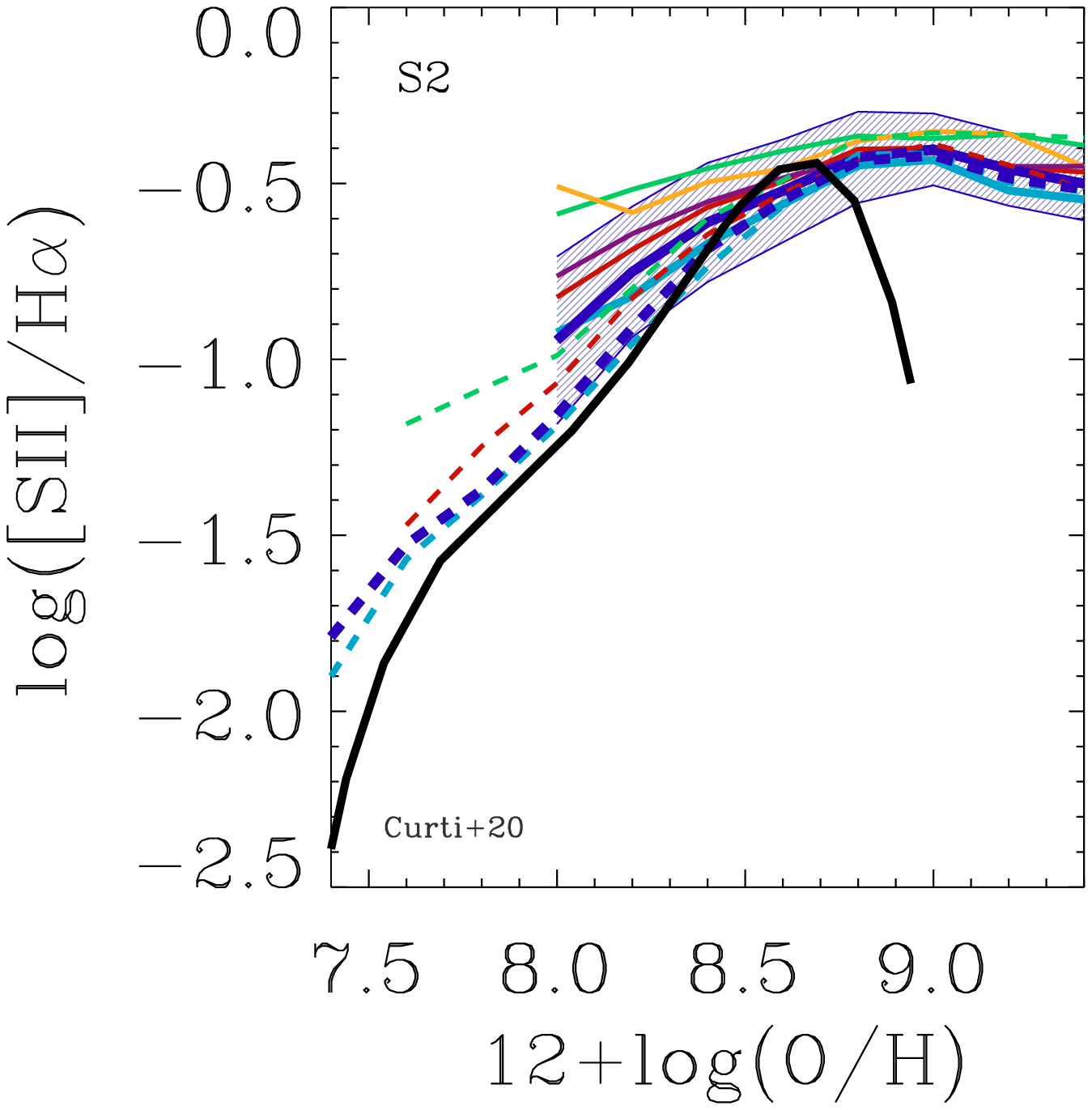,
  width=0.26\textwidth}\hspace{-0.4cm}
\epsfig{file=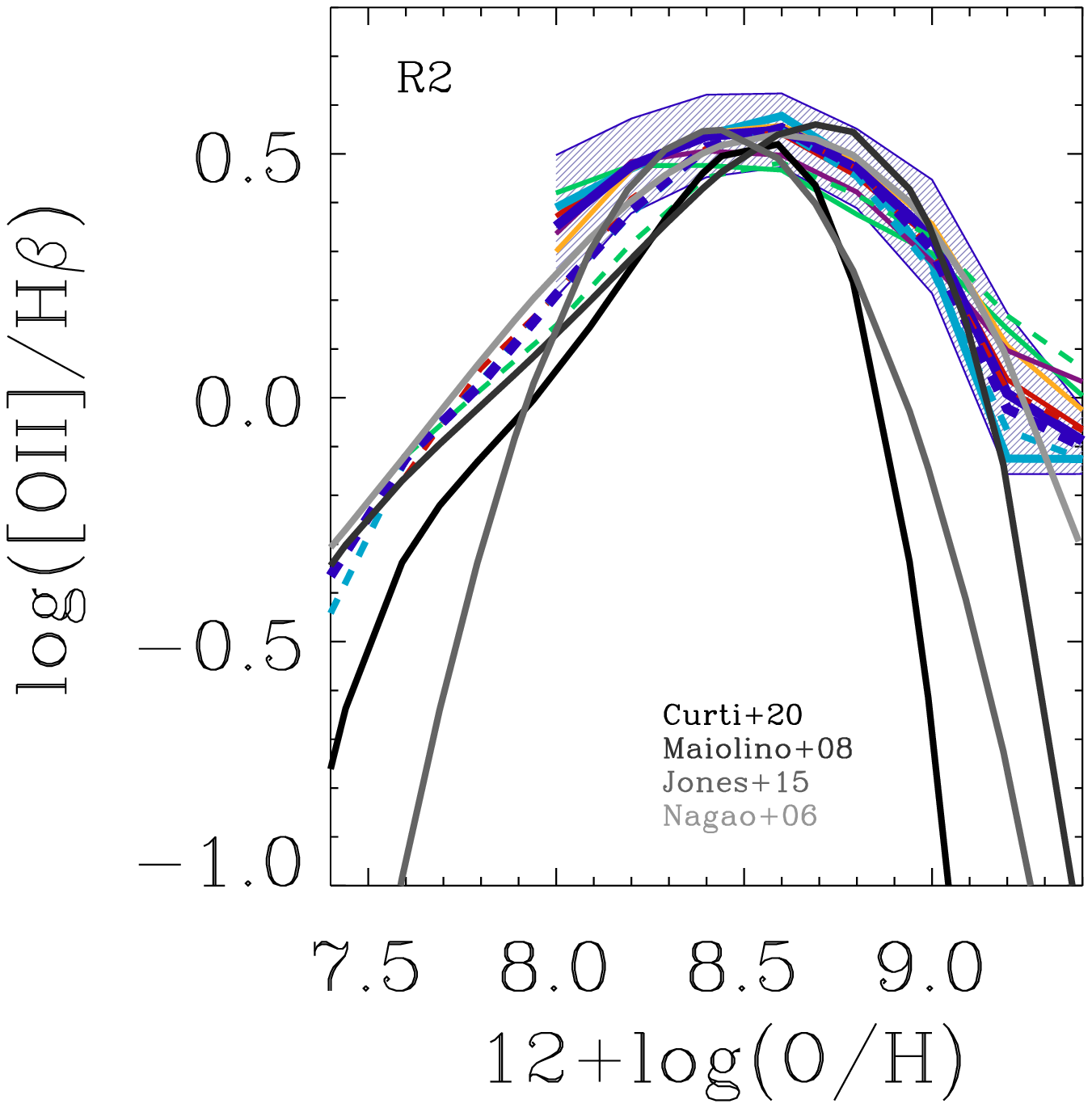,
  width=0.26\textwidth}\hspace{-0.4cm}
\epsfig{file=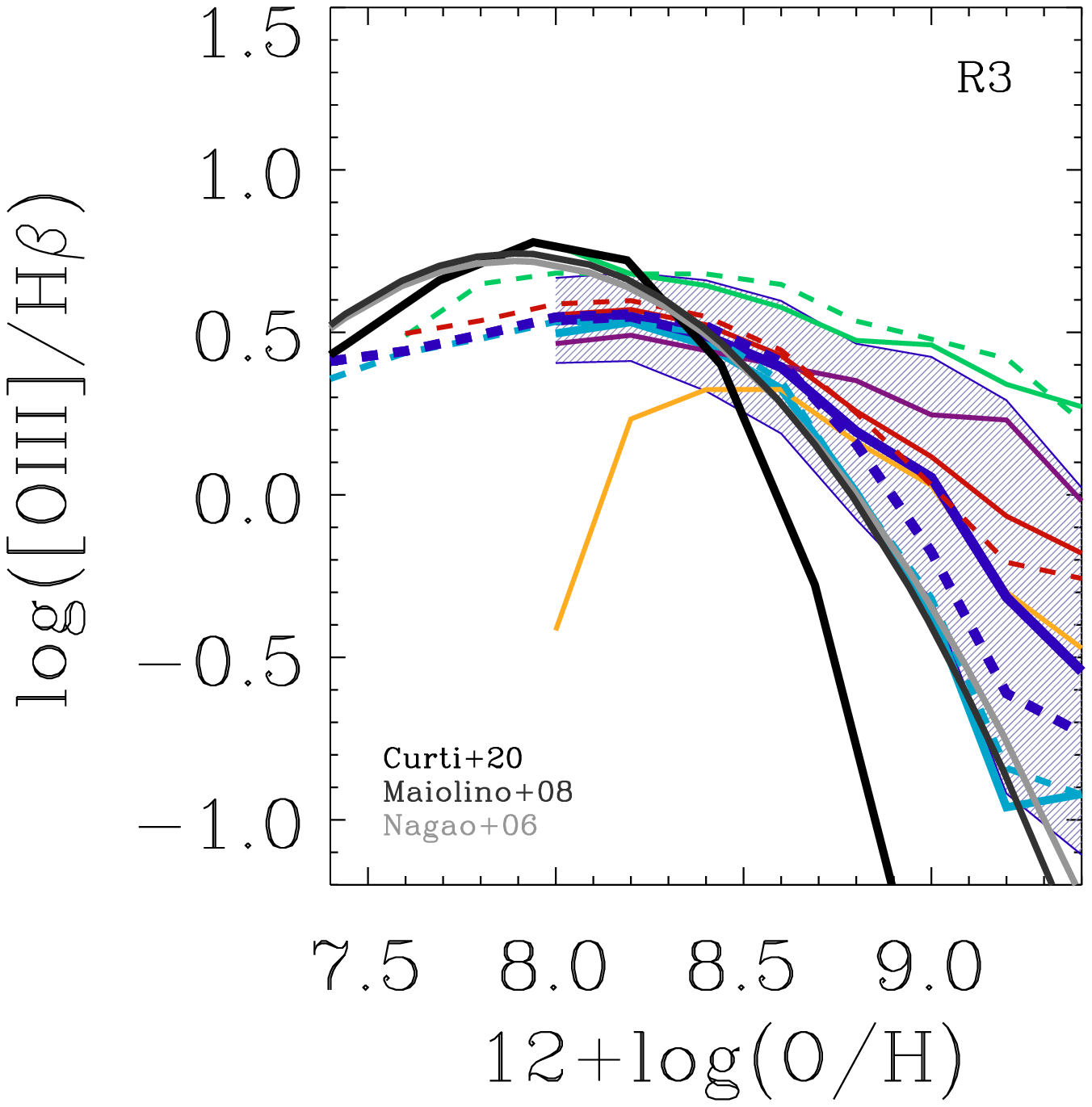,
  width=0.26\textwidth}\hspace{-0.4cm}
\epsfig{file=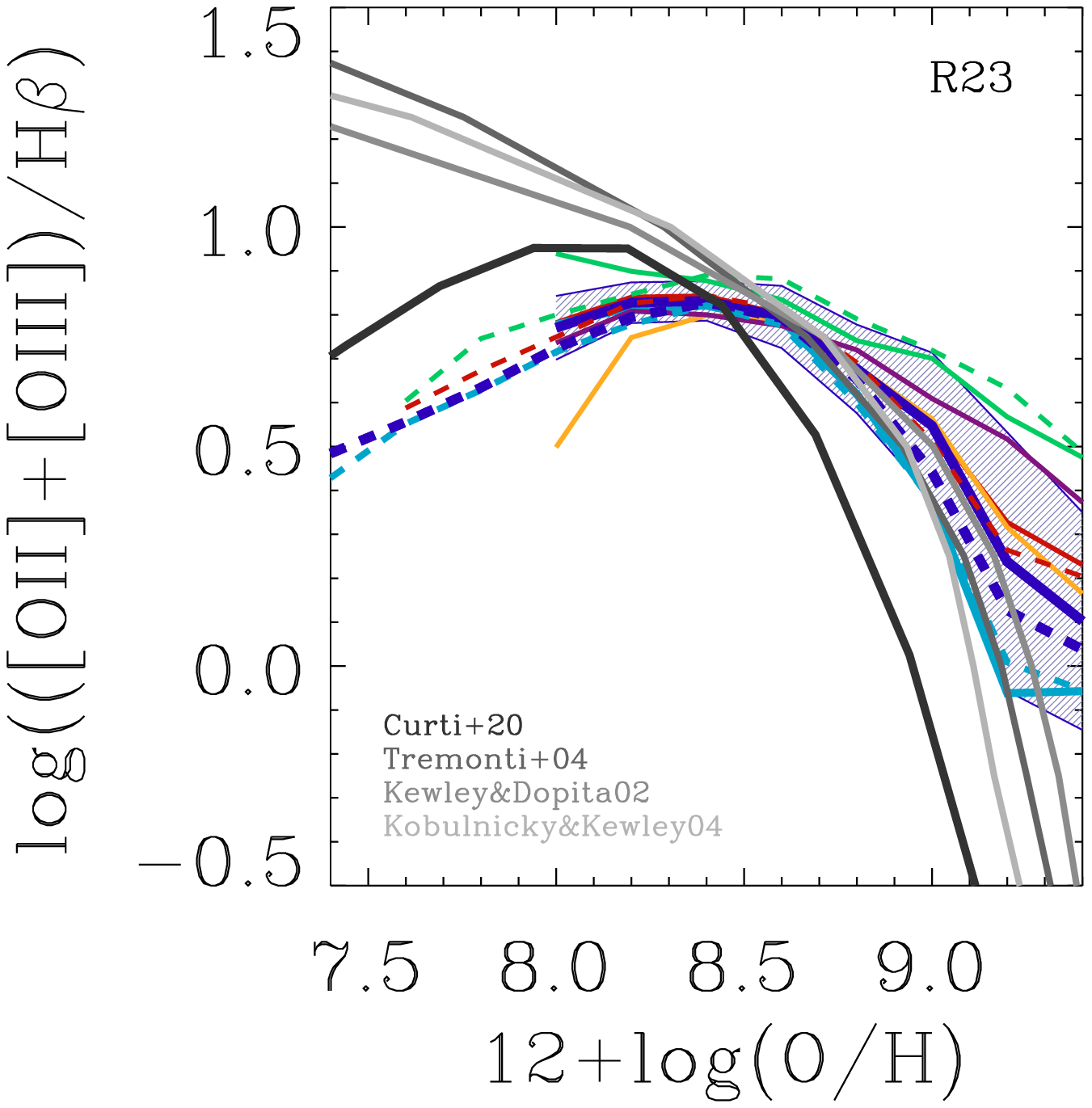,
  width=0.26\textwidth}\hspace{-0.4cm}
\epsfig{file=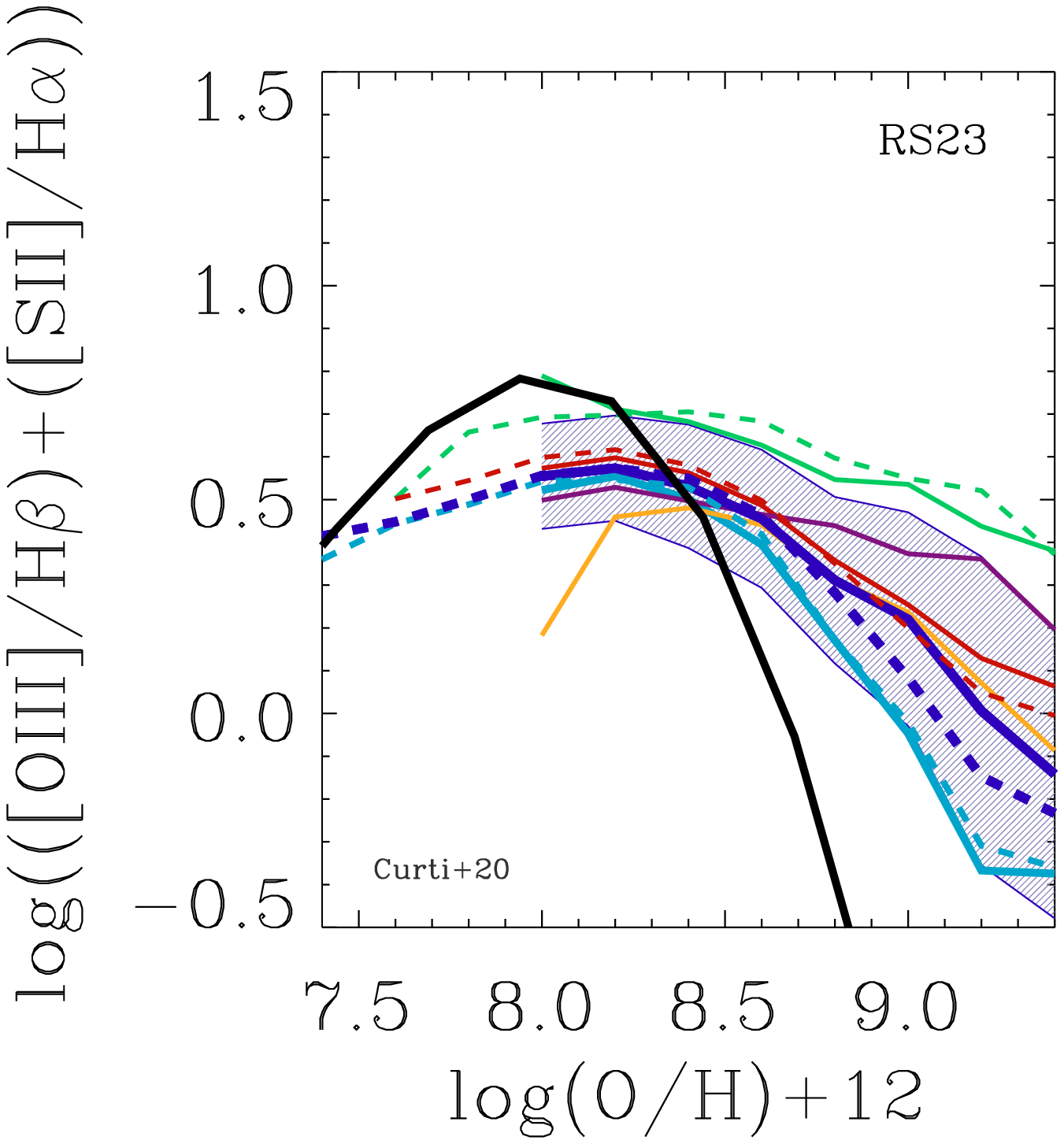,
  width=0.26\textwidth}\hspace{-0.4cm}
\epsfig{file=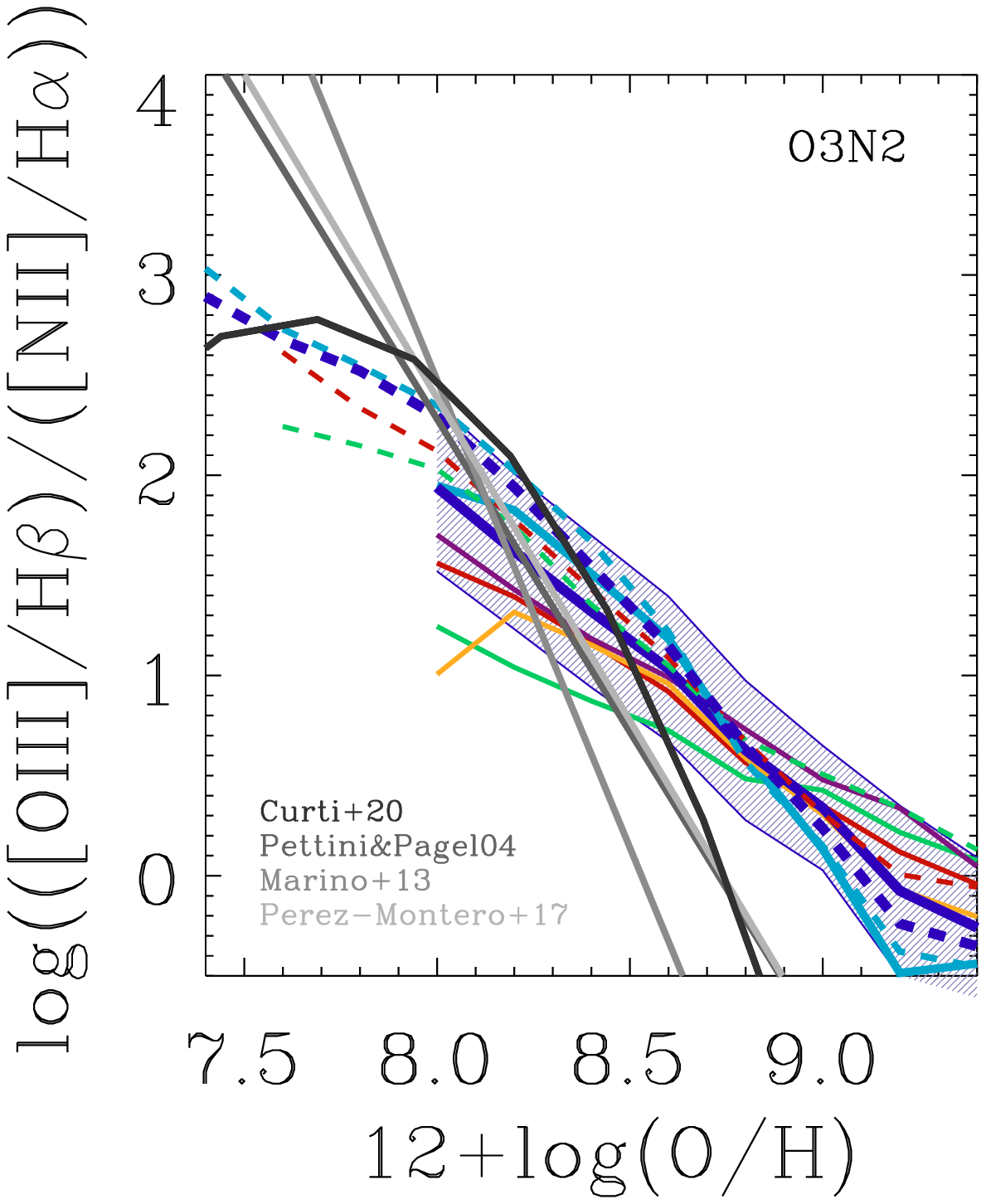,
  width=0.26\textwidth}\hspace{-0.4cm}
\epsfig{file=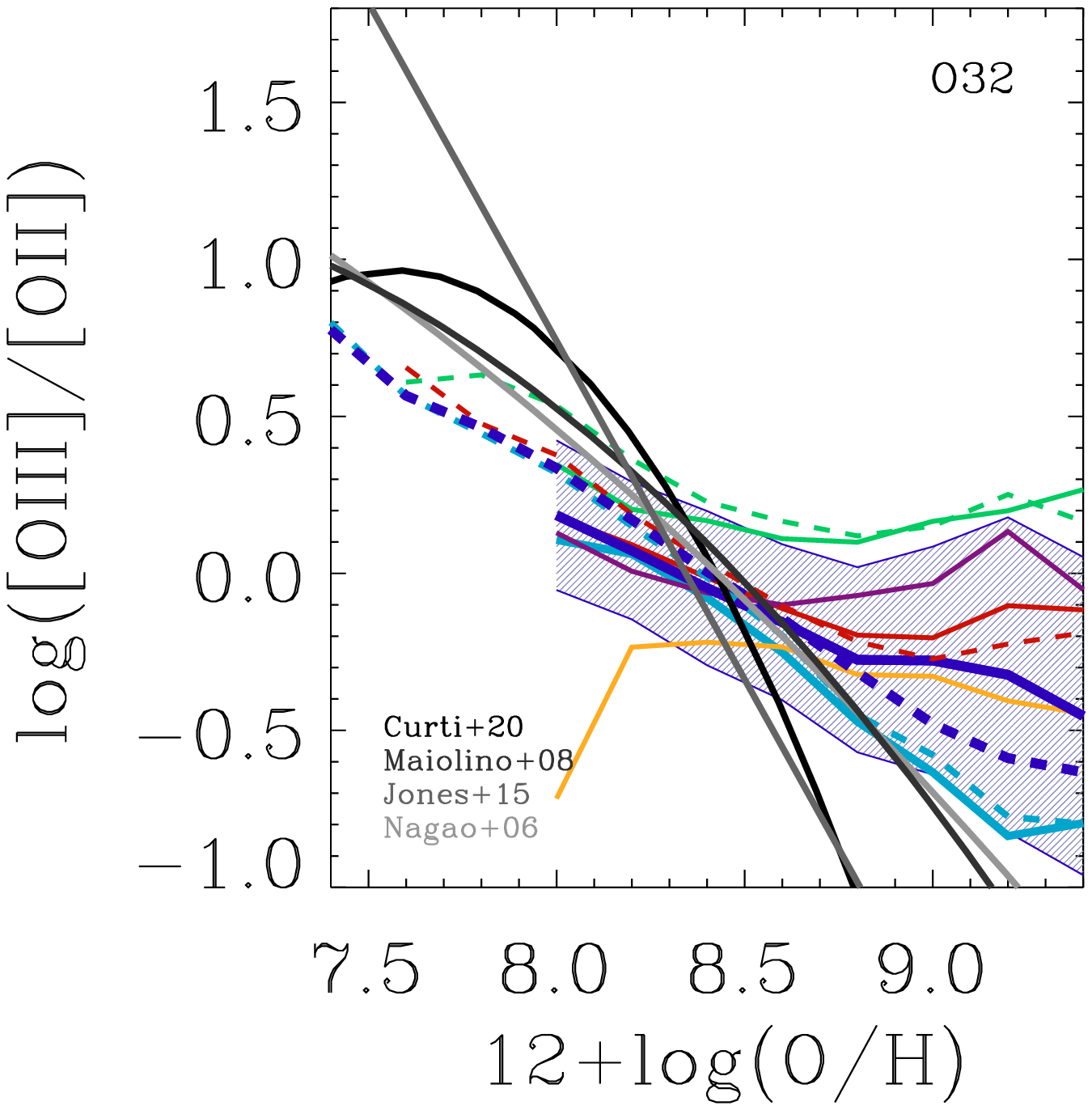,
  width=0.26\textwidth}\hspace{-0.4cm}
\epsfig{file=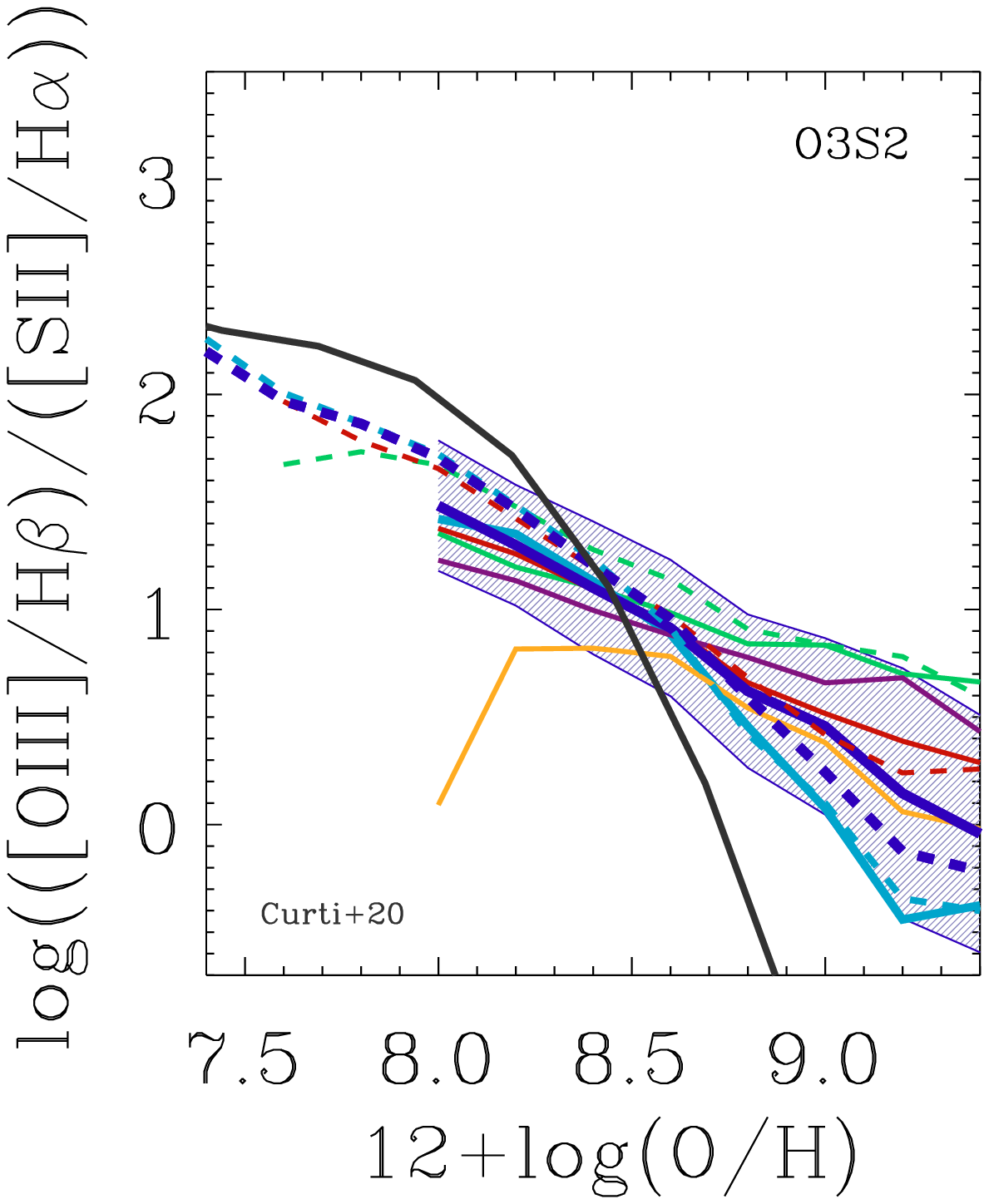,
  width=0.26\textwidth}\hspace{-0.4cm}
\epsfig{file=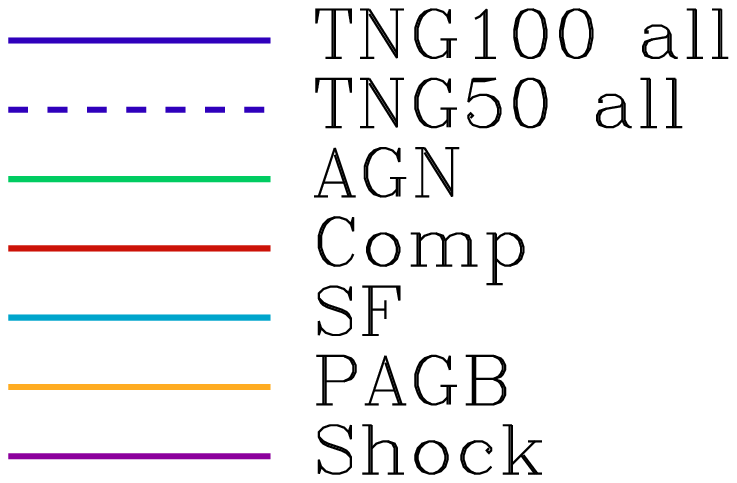,
  width=0.26\textwidth}\hspace{-0.4cm}
\end{flushleft}
\caption{Median N2O2, N2S2, N2, S2, R2,
  R3, R23, RS23, O3N2, O32 and O3S2 ratios (panels from top left
  to bottom right) versus the interstellar oxygen abundance \logoh\
  for the TNG50 (dashed lilac line in each panel) and TNG100 (solid
  lilac line) galaxy populations, together with the 1$\sigma$ scatter
  about the TNG100 relations (lilac shaded area). Also shown in each
  panel are the separate relations for SF-dominated (blue), composite
  (red), AGN- (green), PAGB- (orange) and shock-dominated (purple)
  galaxies. The predicted relations are futher compared with different 
calibrations derived from photoionization models \citep{Kewley02, 
Kobulnicky04} and the direct-\Te\ method \citep[different grey and
black solid lines from][]{Pettini04, Maiolino08, Perez-Montero09,
Marino13, Sanders18, Curti20}.}\label{met_lineratio_z0}       
\end{figure*}

\section{Results}\label{results} 

\subsection{Optical emission-line ratios as tracers of interstellar
  metallicity in present-day galaxies}\label{opticalz0} 

In this section, we start by exploring the relations between interstellar gas phase metallicity and
strong  optical-line ratios, often used as metallicity indicators, predicted
for present-day galaxies in the IllustrisTNG simulations. We consider in particular:
 \niioii\ (hereafter N2O2), \niisii\ (N2S2),
\niiha\ (N2), \oiihb\ (R2), \oiiihb/(\niiha) (O3N2), \oiiioii\ (O32),
\oiiihb/(\siiha) (O3S2), 
\oiiihb\ (R3), \oiiioiihb\ (here \hbox{[O\,{\sc iii}]} is sum of \oiii\ and
\oiiifour\ as in literature; R23), \oiiihb + \siiha\ (RS23) and 
\siiha\ (S2; note that the abbrevations of these line-ratios
specified in the brackets will be used for the remainder of the
paper). 
We  also investigate how these relations differ for galaxies dominated by
different ionizing sources. Following \citet{Hirschmann22}, we distinguish
between five galaxy types, based on the predicted ratio of
BH accretion rate (BHAR) to star formation rate (SFR) and the
H$\beta$-line luminosity. Specifically, SF-dominated, composite,
AGN-dominated, PAGB-dominated and shock-dominated galaxies are defined     
as follows:  
\begin{itemize}
 \item SF-dominated galaxies: BHAR/SFR $< 10^{-4}$ and H$\beta_{\rm
     SF+AGN} > $ H$\beta_{\rm PAGB}$ + H$\beta_{\rm shock}$;
\item Composite galaxies: $10^{-4}<$ BHAR/SFR  $< 10^{-2}$ and
  H$\beta_{\rm SF+AGN} > $ H$\beta_{\rm PAGB} $ + H$\beta_{\rm shock}$;
\item AGN-dominated galaxies: BHAR/SFR $> 10^{-2}$ and H$\beta_{\rm
    SF+AGN} > $ H$\beta_{\rm PAGB}$ + H$\beta_{\rm shock}$;
\item PAGB-dominated galaxies: H$\beta_{\rm SF+AGN}$ + H$\beta_{\rm
    shock} < $ H$\beta_{\rm PAGB}$;
  \item Shock-dominated galaxies: H$\beta_{\rm SF+AGN}$ + H$\beta_{\rm
    PAGB} < $ H$\beta_{\rm shock}$.
\end{itemize}
We also investigate how the predicted relations between interstellar metallicity
and observable line ratios compare with various empirical \citep{Pettini04, 
Nagao06, Maiolino08, Bresolin08, Perez-Montero09, Marino13, Jones15, 
Sanders18, Curti20} and theoretical \citep{Kewley02, Kobulnicky04, Tremonti04,
  Perez-Montero17} calibrations.

Fig.~\ref{met_lineratio_z0} shows the median N2O2, N2S2,
N2, S2, R2, R3, R23, RS23, O3N2, O32 and O3S2 ratios (panels from top left to
bottom right) against interstellar oxygen abundance, \logoh,
for the TNG50 (dashed lavender line in each panel) and TNG100
(solid lavender line) galaxy populations, together with the 1$\sigma$ scatter about
the TNG100 relations  (shaded area). Also shown in each panel 
are the separate relations for SF-dominated (blue),
composite (red), AGN- (green), PAGB- (orange) and shock-dominated
(purple) galaxies.

Irrespective of galaxy type, the N2O2, N2S2, N2 and S2 ratios
(first row in Fig.~\ref{met_lineratio_z0}) strongly increase with
increasing oxygen abundance, mainly as a consequence of
secondary nitrogen production and increasing sulfur abundance
toward higher metallicity \citep[the simulation follows the differential
  enrichment in several chemical elements; see][]{Hirschmann22}.
While the relations for SF-dominated
galaxies are close to those for all galaxies, composite and
AGN-dominated galaxies exhibit progressively higher line
ratios at fixed oxygen abundance, reflecting the sensitivity of 
\nii\ and \sii\ to the hard ionizing radiation from the central accreting 
BH. Similarly, the hot ionizing radiation of PAGB- and shock-dominated 
galaxies produces slightly higher N2O2, N2 and S2 ratios. 

The R2, R3, R23 and RS23 ratios (middle row of
Fig.~\ref{met_lineratio_z0}) exhibit a more complex dependence on
oxygen abundance: for metallicities lower than $\logoh\approx 8$, 
these line ratios increase with increasing O/H because of the rising
abundance of oxygen; instead, for higher metallicities, the line ratios 
drop because oxygen, which is a major gas coolant through
infrared fine-structure transitions, causes \Te\ to drop, resulting in 
fewer collisional excitations of optical transitions. R2 hardly changes 
among different galaxy types, since O and H have similar ionization
energies ($\sim13.6$\,eV). In contrast, since the production of O$^{2+}$ 
requires energies in excess of 35.1\,eV, R3, R23 and RS23 increase 
from SF- to PAGB- and shock-dominated, composite and AGN-dominated 
galaxies (the drop in these ratios for PAGB-dominated galaxies at low 
metallicities is caused by a shortening of the lifetime of hot PAGB stars).

The behaviour of the O3N2, O32 and O3S2 ratios (bottom row of
Fig.~\ref{met_lineratio_z0}) follows directly from the arguments 
above: as \logoh\ increases, the rise in \nii\ and \sii\ 
conspires with the drop in \oiii\ to make O3N2, O32 and O3S2 
decrease steeply. SF-dominated galaxies have properties similar to
the median of the population (so do PAGB-dominated galaxies at
high metallicities), while composite, AGN- and shock-dominated
galaxies show comparatively elevated ratios (via enhanced
O$^{2+}$/O$^+$) at fixed oxygen abundance. 

Overall, a noteworthy conclusion from Fig.~\ref{met_lineratio_z0} 
is that the relations between oxygen abundance and optical-line
ratios often used as metallicity indicators are predicted to
depend significantly on galaxy type at $z=0$. 
This implies that calibrations of metallicity estimators
using restricted samples of \hii\ regions and star-forming 
galaxies do not apply to every galaxy. In fact, specific calibrations
have been proposed for AGN narrow-line regions \citep{Dors17, Dors21,
  Carvalho20}. Yet, since only a minor fraction ($\sim10$ per cent) of
all galaxies host AGN, in practice, calibrations of metallicity estimators 
including all galaxy types are close to those for SF-dominated galaxies
(as illustrated by Fig.~\ref{met_lineratio_z0}).

\begin{figure*}
  \begin{flushleft}
\epsfig{file=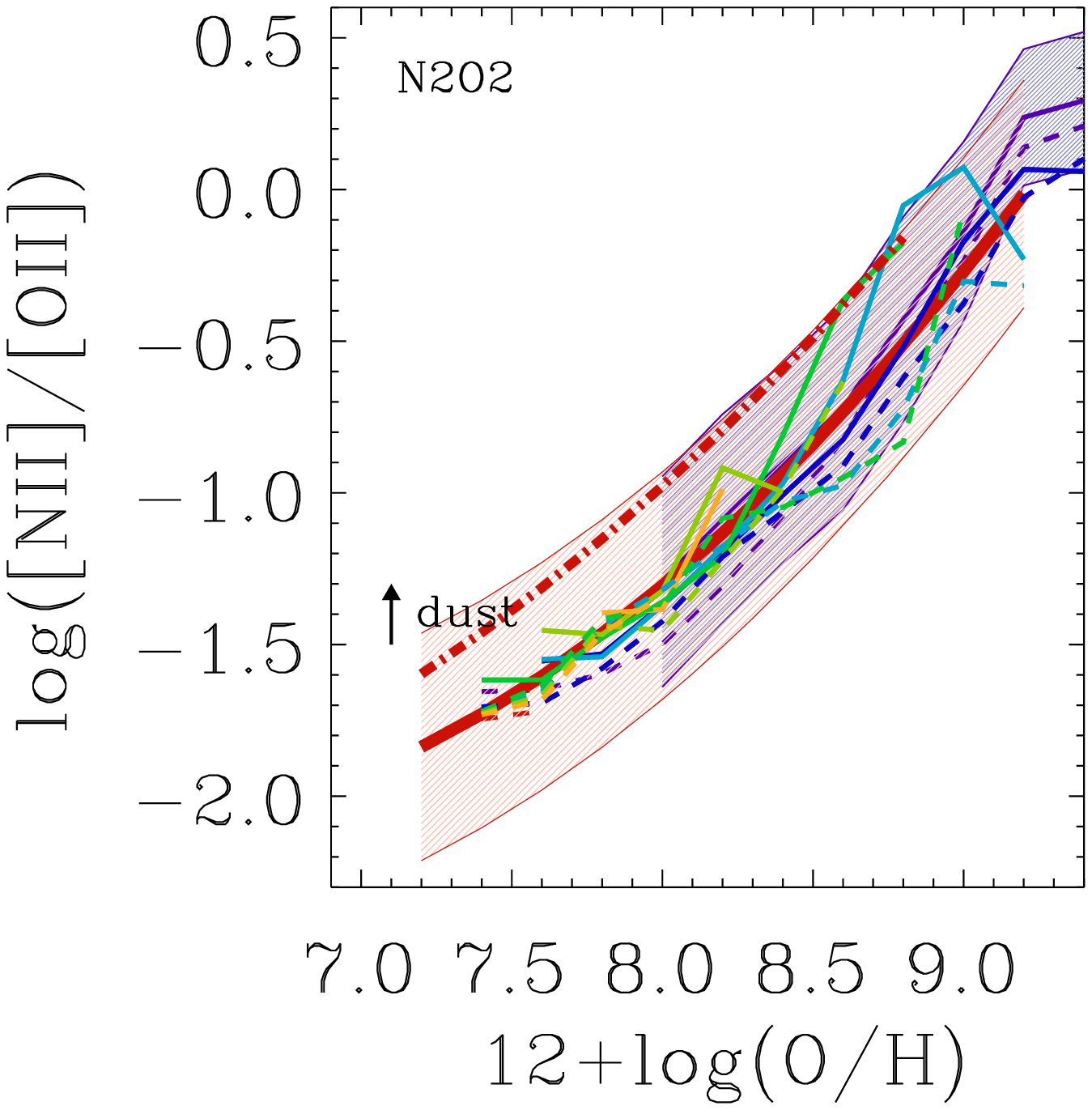,
  width=0.26\textwidth}\hspace{-0.4cm}
\epsfig{file=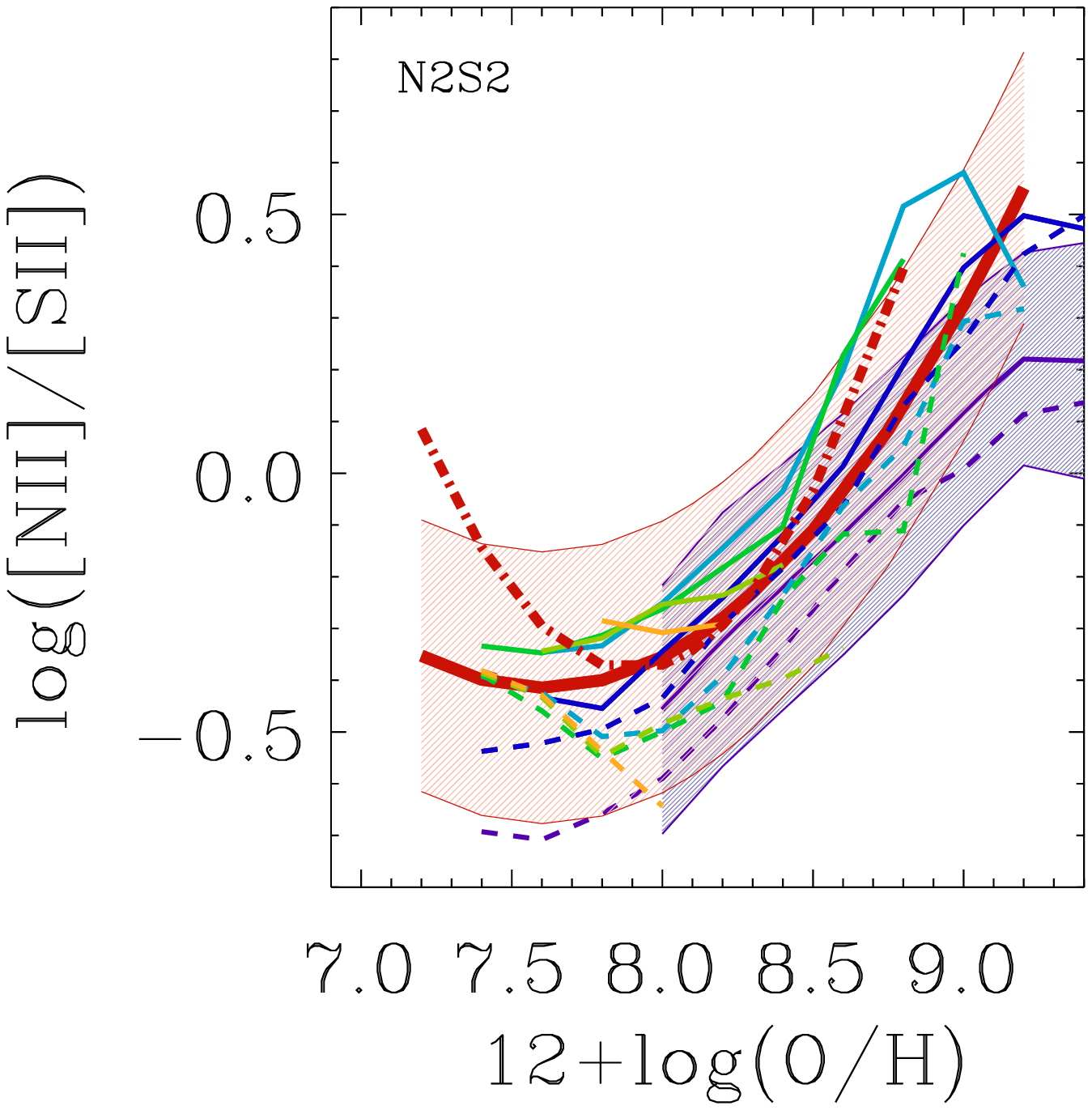,
  width=0.26\textwidth}\hspace{-0.4cm}
\epsfig{file=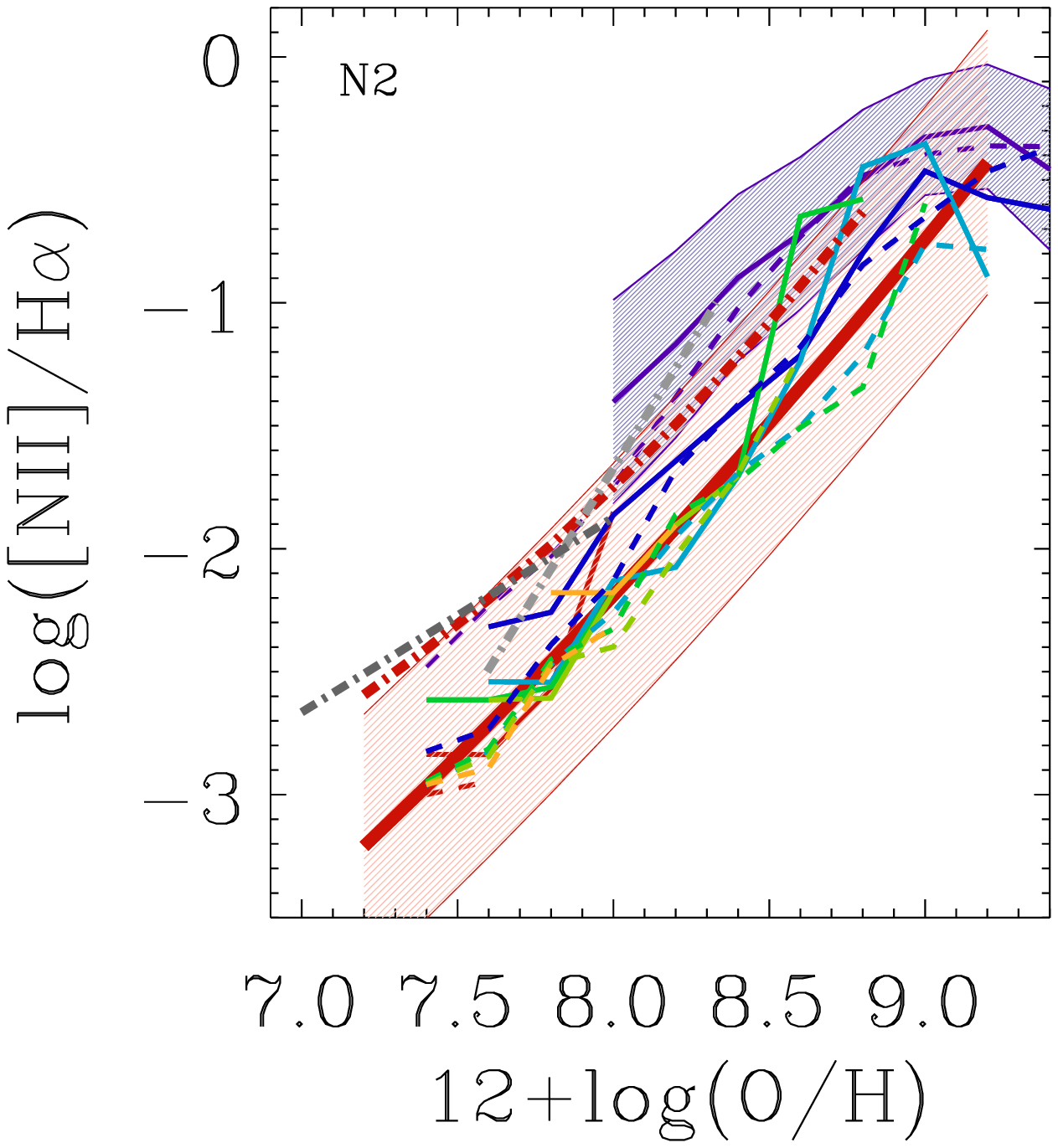,
  width=0.26\textwidth}\hspace{-0.4cm}
\epsfig{file=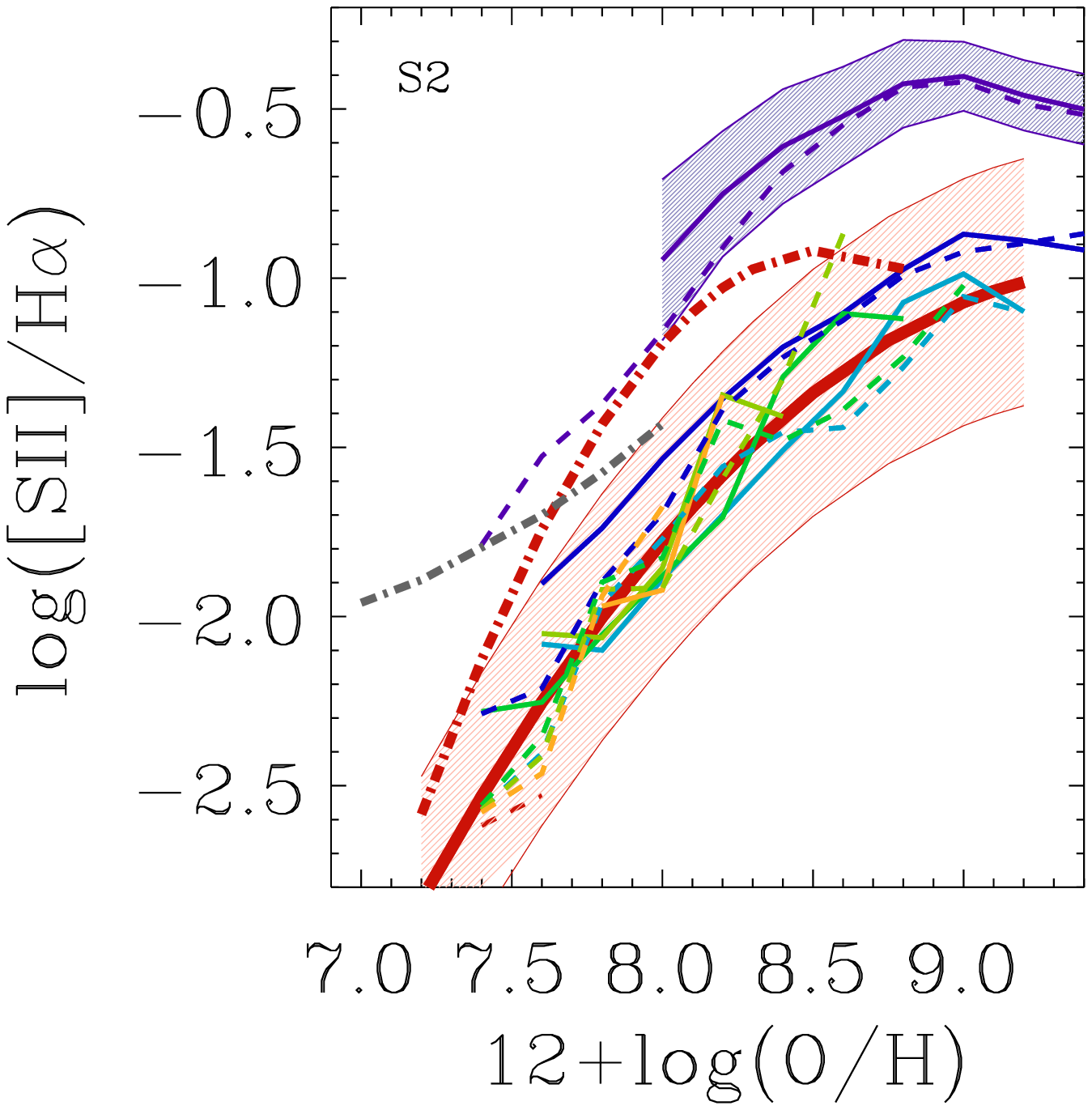,
  width=0.26\textwidth}\hspace{-0.4cm}
\epsfig{file=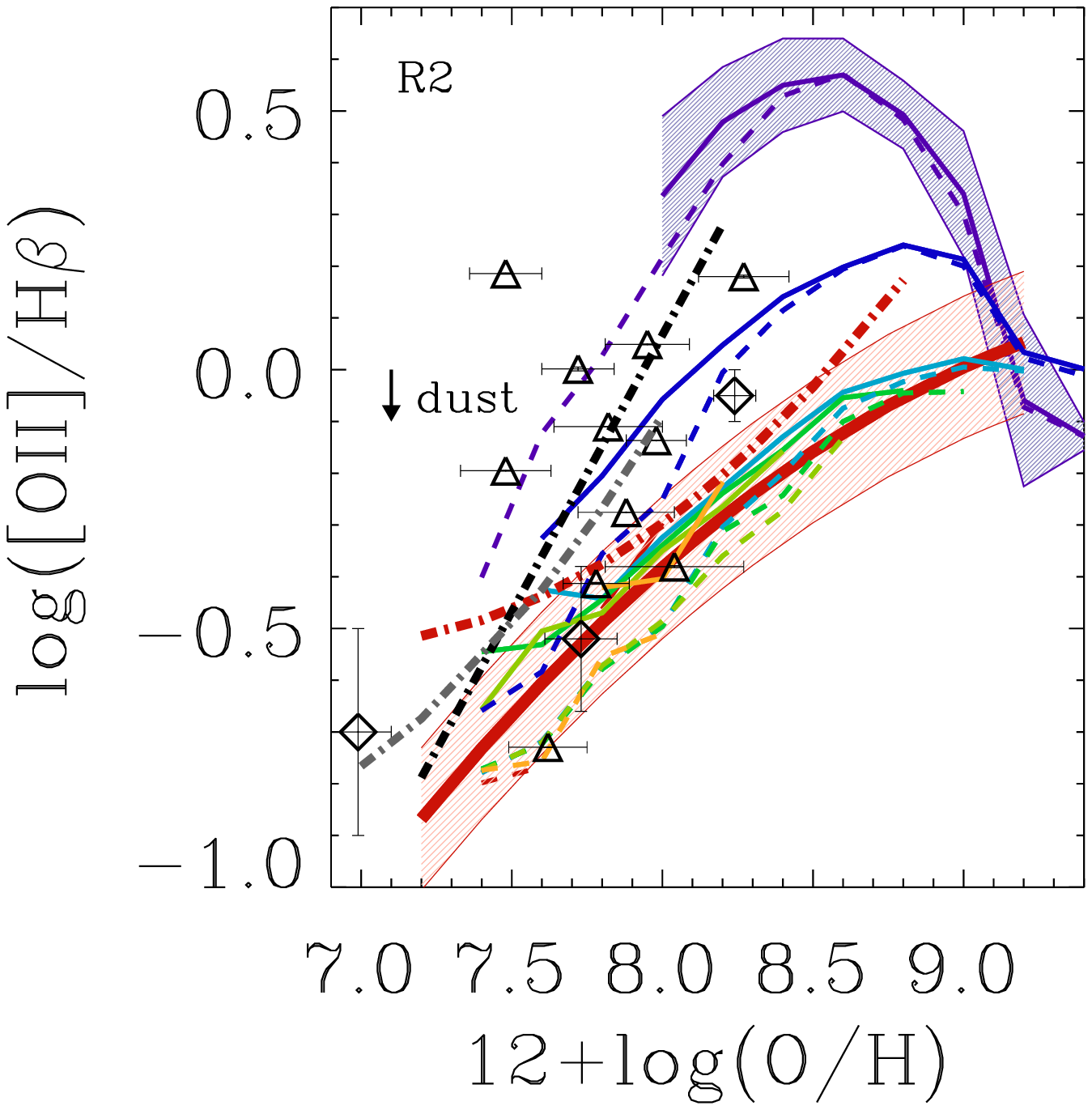,
  width=0.26\textwidth}\hspace{-0.4cm}
\epsfig{file=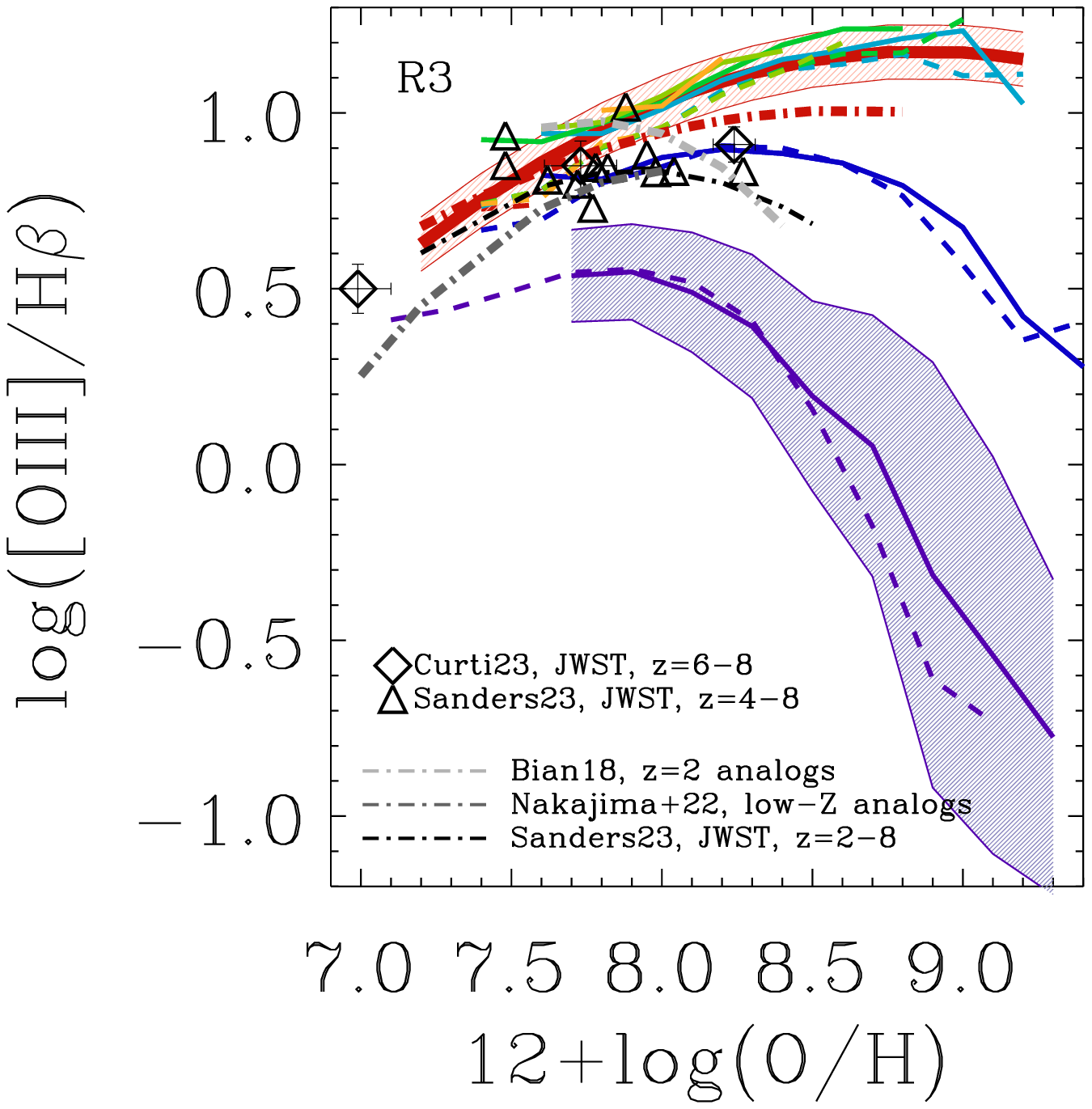,
  width=0.26\textwidth}\hspace{-0.4cm}
\epsfig{file=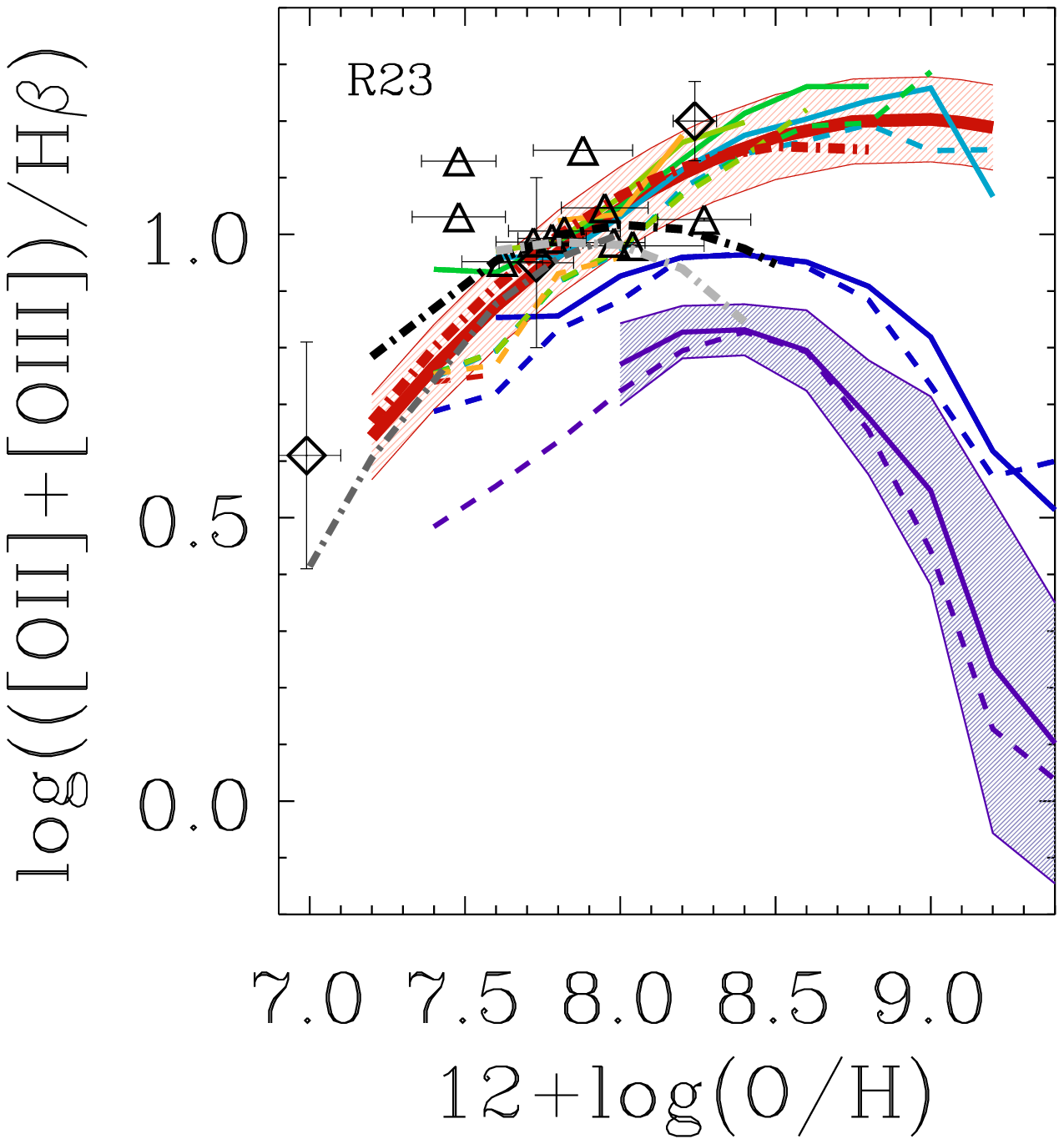,
  width=0.26\textwidth}\hspace{-0.4cm}
\epsfig{file=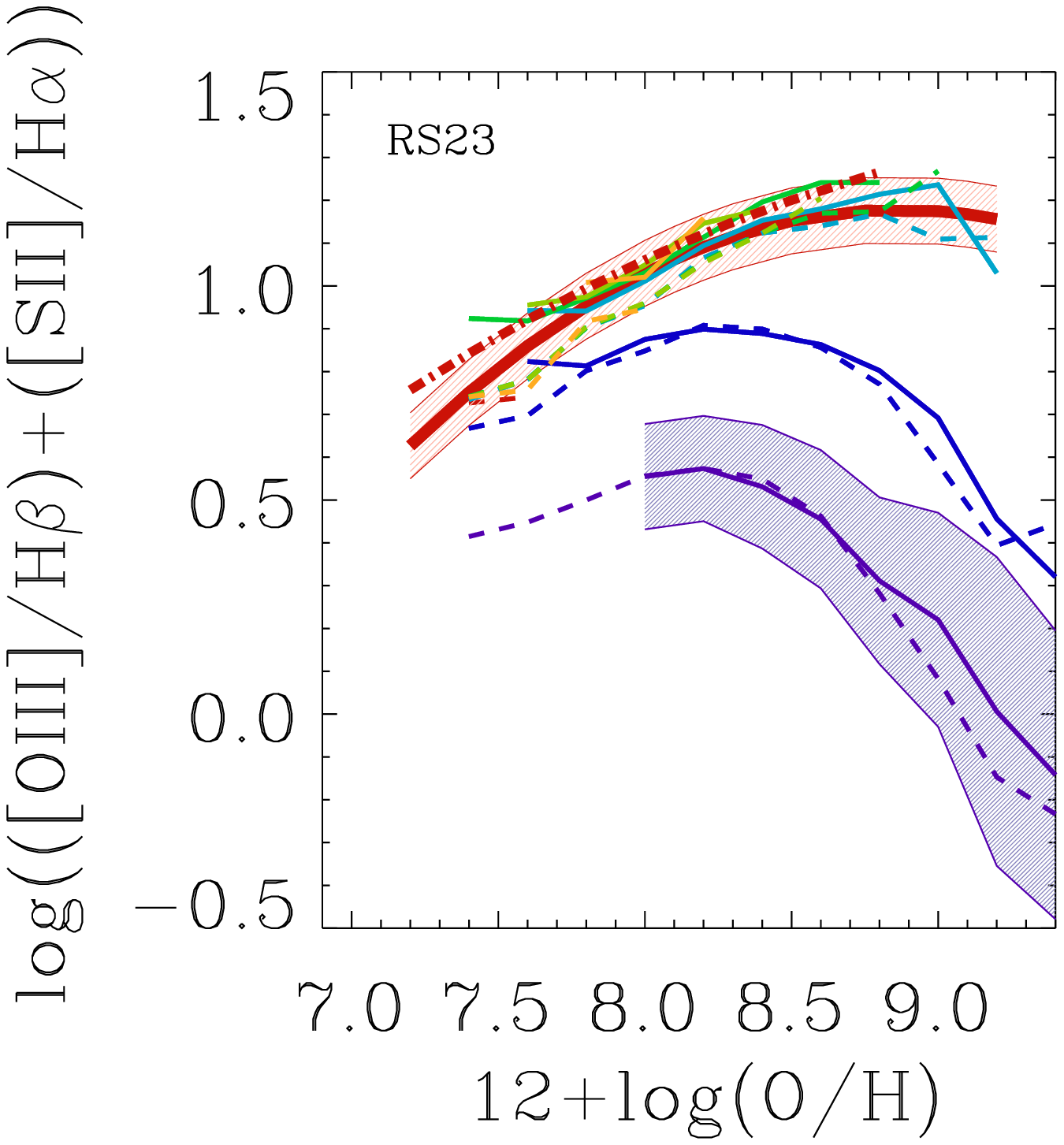,
  width=0.26\textwidth}\hspace{-0.4cm}
\epsfig{file=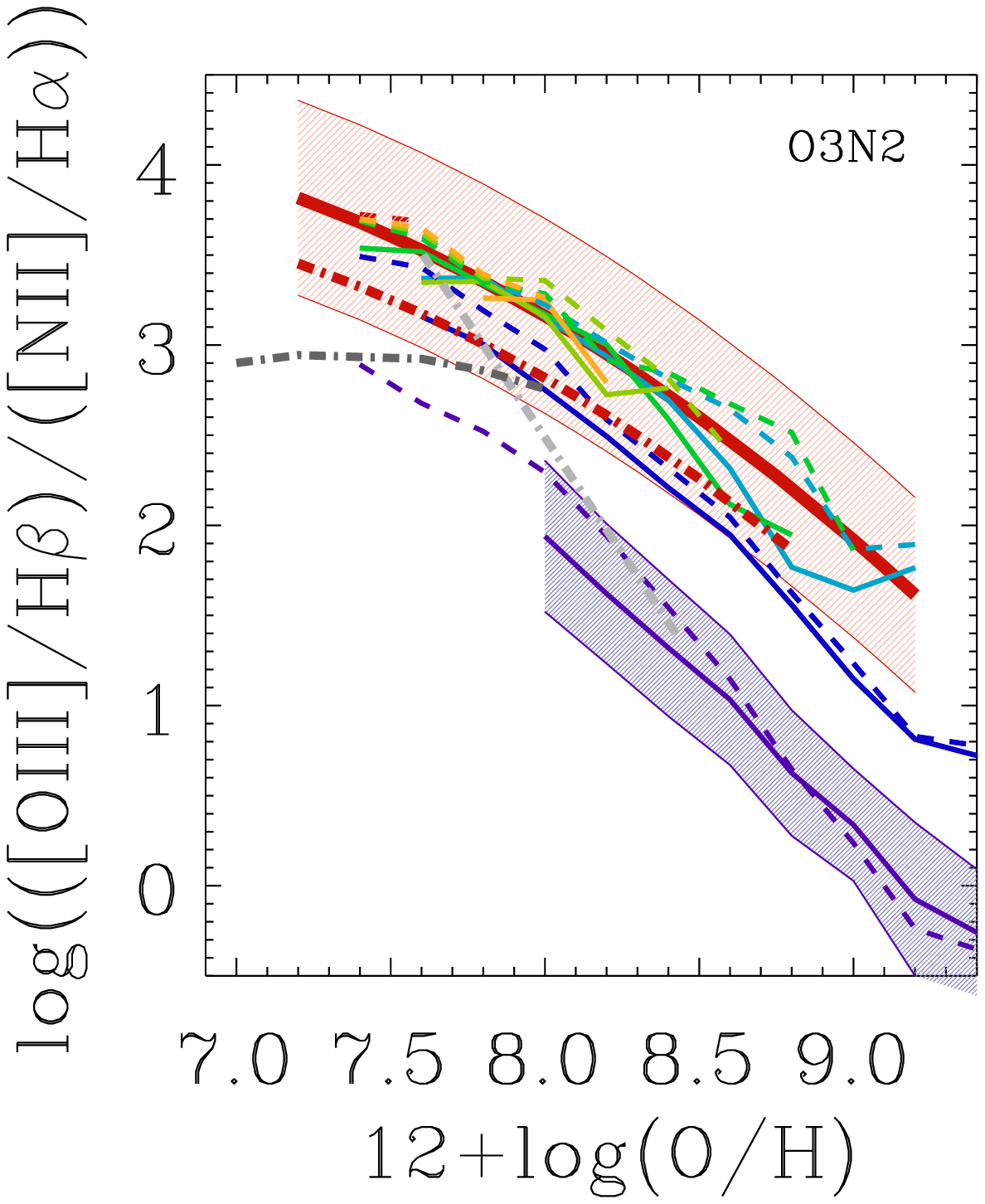,
  width=0.26\textwidth}\hspace{-0.4cm}
\epsfig{file=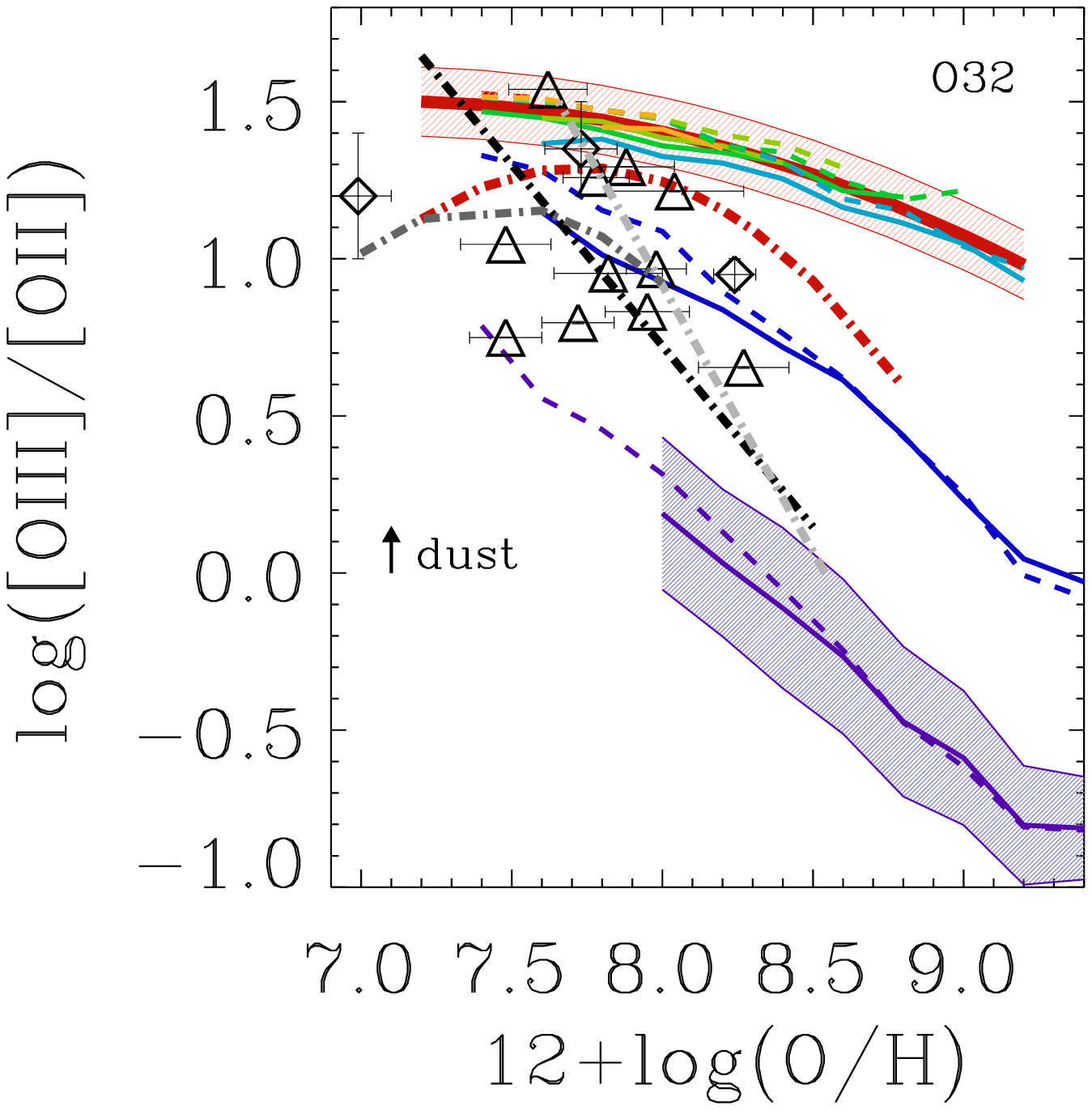,
  width=0.26\textwidth}\hspace{-0.4cm}
\epsfig{file=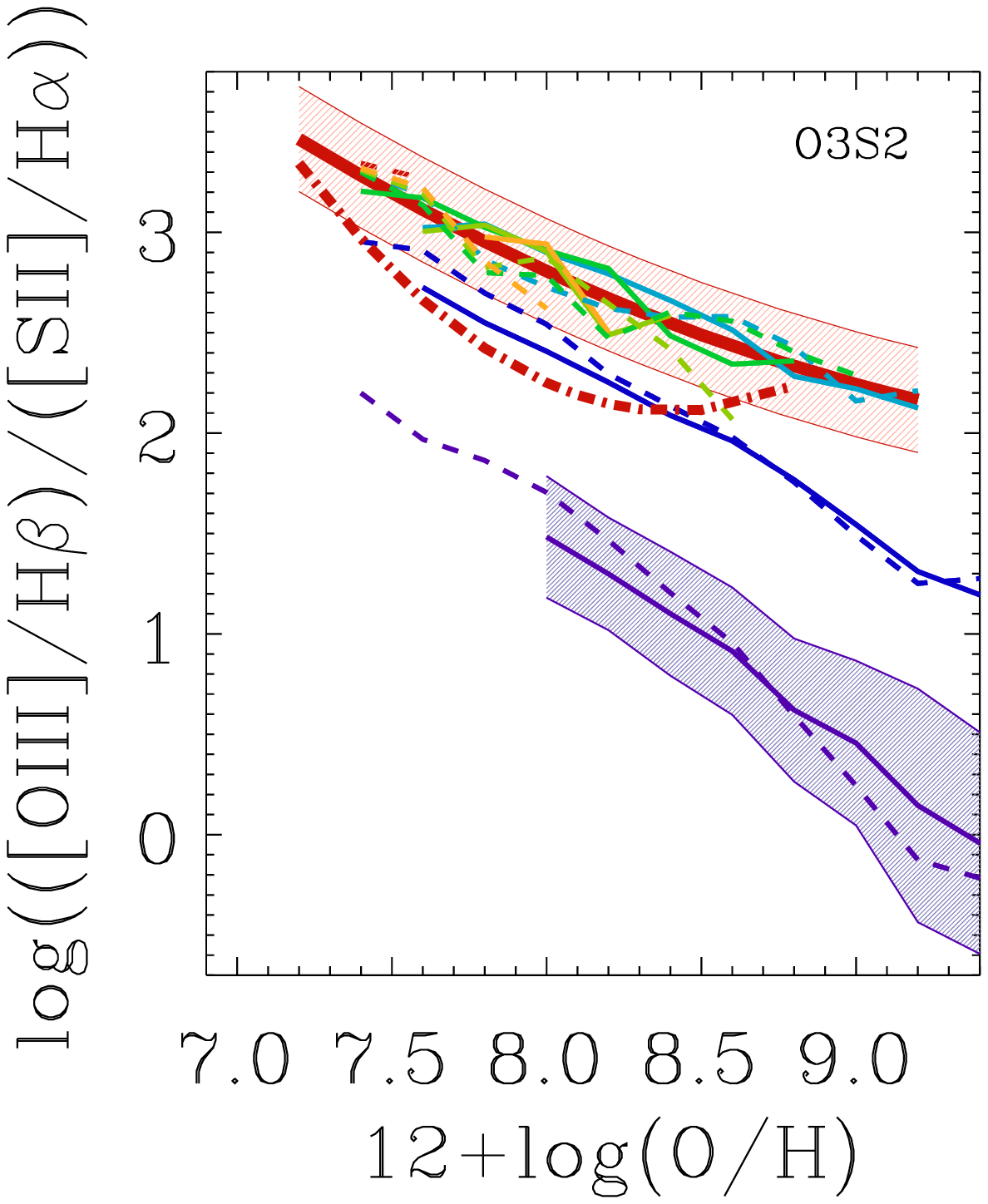,
   width=0.26\textwidth}\hspace{-0.4cm}
\epsfig{file=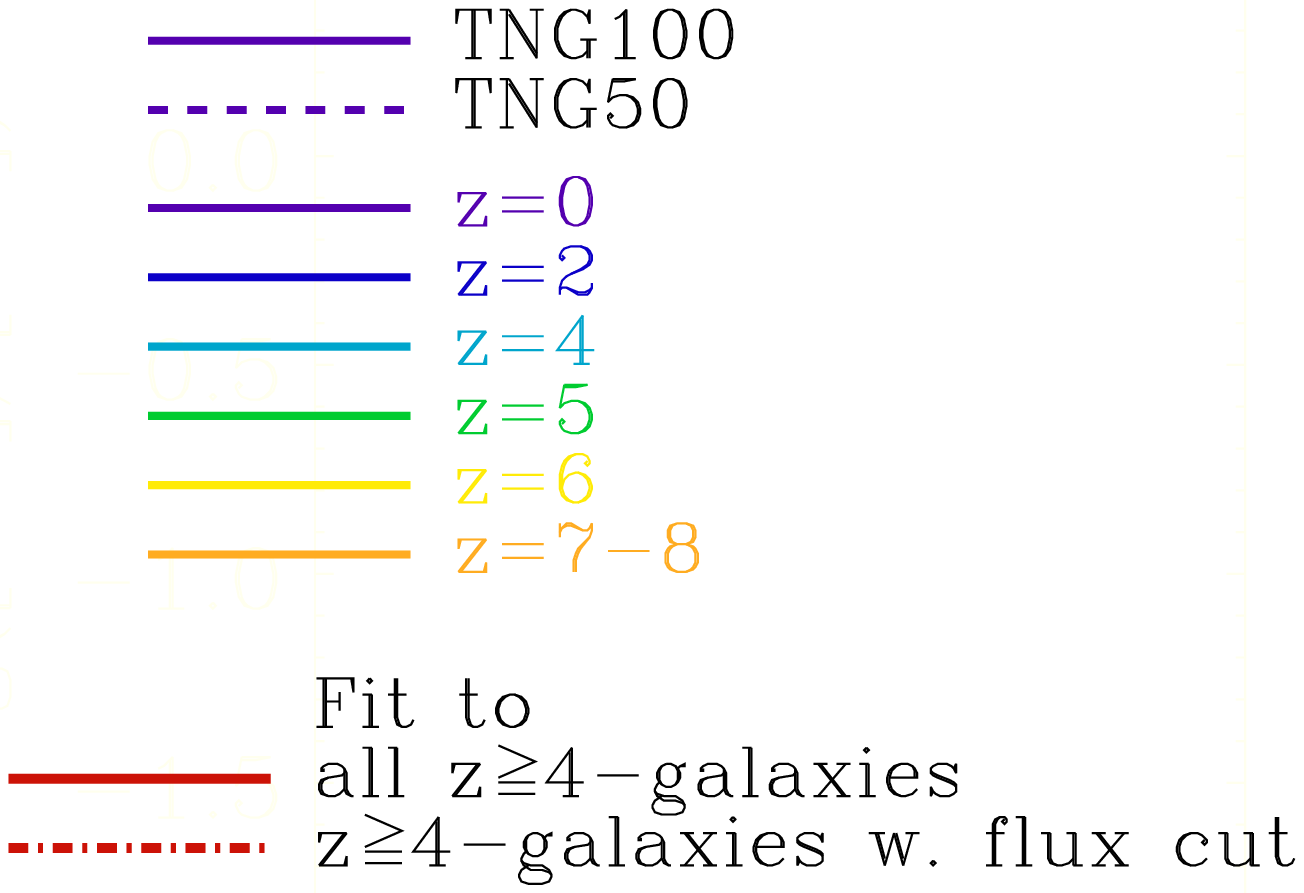,
  width=0.26\textwidth}\hspace{-0.4cm}
\end{flushleft}
\caption{Analogue of Fig.~\ref{met_lineratio_z0}, but now only for the
  global TNG50 (dashed lines) and TNG100 galaxy populations (solid
  lines) and at different redshifts (lilac: z = 0, dark blue: z = 2,
  light blue: z = 4, turquoise: z = 5, green: z = 6, orange: z = 7--8).
  Overplotted are fits to the predicted $z \geq 4$ relations of all
  galaxies (thick red line in each panel, with the fitted error shown
  by the red shaded area) and of galaxies above a flux limit of
  $3\times 10^{-17}$~erg\ s$^{-1}$\ cm$^{-2}$ (thick, red dashed
  line). Model predictions are compared to T$_e$-based measurements of
  \logoh\   for local SDSS analogues of z $\sim$ 2 galaxies from
  \citet[][light-grey, dashed-dotted lines in the O3N2, O32, R3 and
  R23 panels]{Bian18},   and for local extremely metal-poor galaxies
  in the Subaru EMPRESS   survey from \citet[][dark-grey,
  dashed-dotted lines in the R2, R3,   R23, O3N2 and O32
  panels]{Nakajima22}. Also shown are the results   inferred from
  JWST/NIRSpec spectroscopy for three galaxies at z 
  $\sim$ 6--8 from \citet[][black diamonds with error bars in the R2,
  O32, R3 and R23 panels]{Curti23}, 11 galaxies at z $\sim$ 4--9 from
  \citet[][black triangles with error bars in the R2, O32, R3 and R23
  panels]{Sanders23} and their associated proposed new calibrations of
  the R2, R3, R23 and O32 metallicity estimators at redshifts z $\sim$
  2--9 (black dashed-dotted lines).}\label{met_lineratio_evol}  
\end{figure*}

In the various panels of Fig.~\ref{met_lineratio_z0}, we compare
the relations predicted by the IllustrisTNG simulations with different 
calibrations derived from photoionization models \citep{Kewley02, 
Kobulnicky04} and the direct-\Te\ method \citep[grey and
black solid lines from][]{Pettini04, Maiolino08, Perez-Montero09,
Marino13, Sanders18, Curti20}. We note that, in our models, the
quantity \logoh\  includes oxygen in both the gas and dust phases 
(Section~\ref{ELmodels}), while observational determinations
pertain in general to purely gas-phase oxygen abundances. Since
the gas-phase oxygen abundance is about 0.05--0.25~dex lower 
than the total interstellar O abundance \citep[depending on 
metallicity and dust-to-metal ratio; see table~2 of][]{Gutkin16}, we 
indicate by a horizontal arrow in the top-left panel of 
Fig.~\ref{met_lineratio_z0} the typical small amount, 
$\Delta\mathrm{(O/H)}\approx-0.15$, by which the predicted relations 
should be shifted for a more accurate comparison with observational
relations.

On the whole, there is good
general agreement between published calibrations and those
predicted for SF-dominated IllustrisTNG galaxies (and population 
medians), confirming that our modelling approach provides a 
valuable means of investigating the optical emission-line properties of  
present-day galaxies \citep[see also][]{Hirschmann22}. However, 
some tensions exist, particularly with the \Te-based calibrations of
\citet[][black solid line]{Curti20} for metal-rich galaxies with 
$\logoh\ga9$. In this metallicity regime, the simulated R2, O3N2, 
O32, O3S2, R3, R23 and RS23 ratios are higher than expected
from the calibration. A possible explanation for this discrepancy may 
be the well-known tendency for the direct-\Te\ method to underestimate 
O/H in metal-rich galaxies \citep[e.g.,][]{Kewley19, Cameron22}.
Another discrepancy in Fig.~\ref{met_lineratio_z0} is that our models predict 
R3, R23 and RS23 ratios 0.2--0.3~dex lower than literature calibrations
(based on both the direct-\Te\ method and photoionization-models) in 
metal-poor galaxies with $\logoh\la8$. This may arise from: (i) different
assumptions in the photoionization modelling; (ii) caveats in the
derivation of low metallicities with the direct-\Te\ method; and (iii) caveats 
in our modelling approach (see Section~\ref{discussion} for a more detailed 
discussion).

\subsection{Cosmic evolution of the relations between optical-line ratios and metallicity}\label{opticalhighz} 

In the previous Section, we have seen that the relations between 
interstellar metallicity and strong optical-line ratios used as 
metallicity indicators predicted for IllustrisTNG galaxies at $z=0$ 
are qualitatively consistent with published calibrations. In this Section,
we investigate how these relations are predicted to evolve out to redshifts
$z \sim 8$. We compare our predictions with first measurements of
\logoh\ based on the direct-\Te\ method using \JWST/NIRSpec 
observations of galaxies at $4\la z\la9$ \citep{Curti23, Sanders23}.

\begin{table}
\centering
\begin{tabular}{ l | r | r | r | r}
Line ratio  &    $P_0$ 	&	$P_1$	&	$P_2$  	 &     $\chi^2$ \\
  \hline
N2O2  &      5.246 &  $-2.467$ & 0.206 & 1.230 \\
N2S2   &	21.482 &   $-5.759$ & 0.379 & 1.049 \\
N2	&	$-6.848$ & $-0.1875$ & 0.096 & 2.175 \\
R2 	&	$-14.945$ & 3.120  & $-0.162$ & 0.275 \\
O3N2 &	$-3.844$ & 2.759 & $-0.235$ & 1.082 \\
O32	&	0.521 & 0.4008 & $-0.036$ &  0.110 \\
O3S2 &	17.638  & $-3.001$  & 0.143 &  0.523 \\
R3	&	$-14.424$ & 3.521  & $-0.199$ &  0.154 \\
R23	&	$-14.004$ & 3.412  & $-0.191$ &  0.150 \\
RS23 &	$-14.455$ & 3.527  & $-0.199$ &  0.154 \\
S2	&	$-32.063$ & 6.521  & $-0.342$ &  0.728\\
\end{tabular}
\caption{Parameters of the fits to optical-line ratios of llustrisTNG 
galaxies at $z=4$--8 used as metallicity indicators 
(thick red lines in Fig.~\ref{met_lineratio_evol}) with 
quadratic functions of the form $y = P_0 + P_1\,x +
P_2\,x^2$, where $x =  \logoh$ and $y$ is the line ratio. The
rightmost column quantifies the goodness of fit.}\label{table1} 
\end{table}

\begin{figure}
  \begin{flushleft}
\epsfig{file=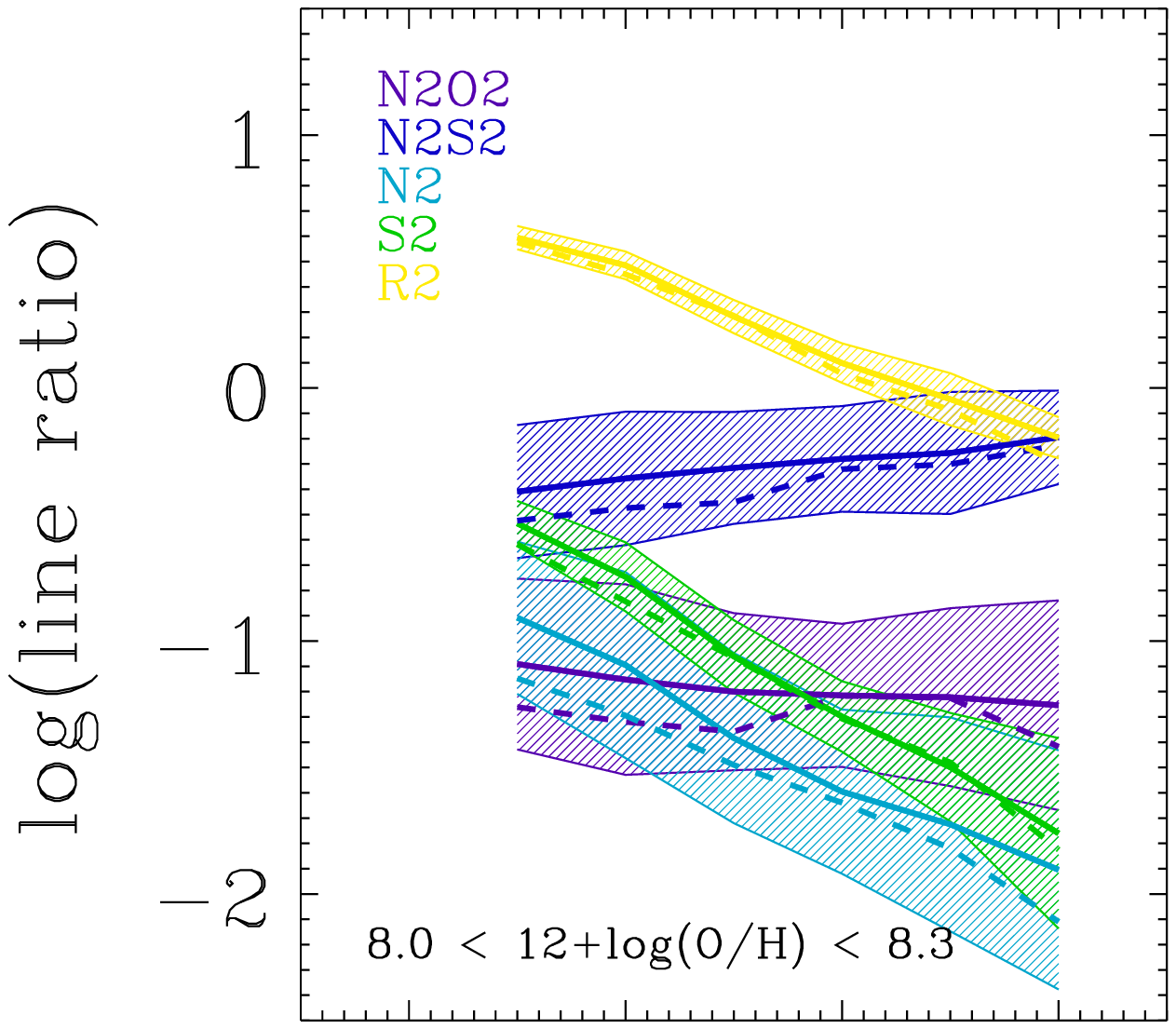,
  width=0.29\textwidth}\hspace{-2.cm}
 \epsfig{file=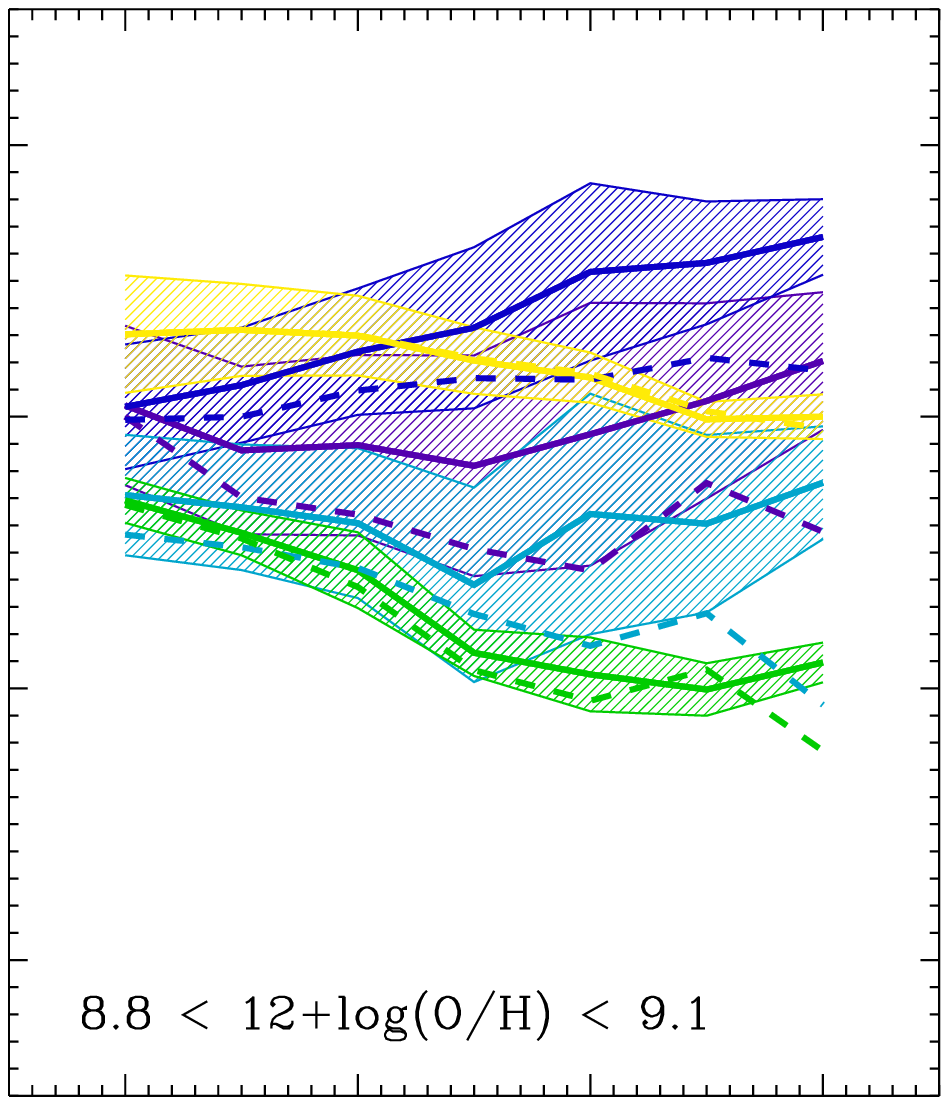,
  width=0.29\textwidth}\vspace{-1.38cm}
  \epsfig{file=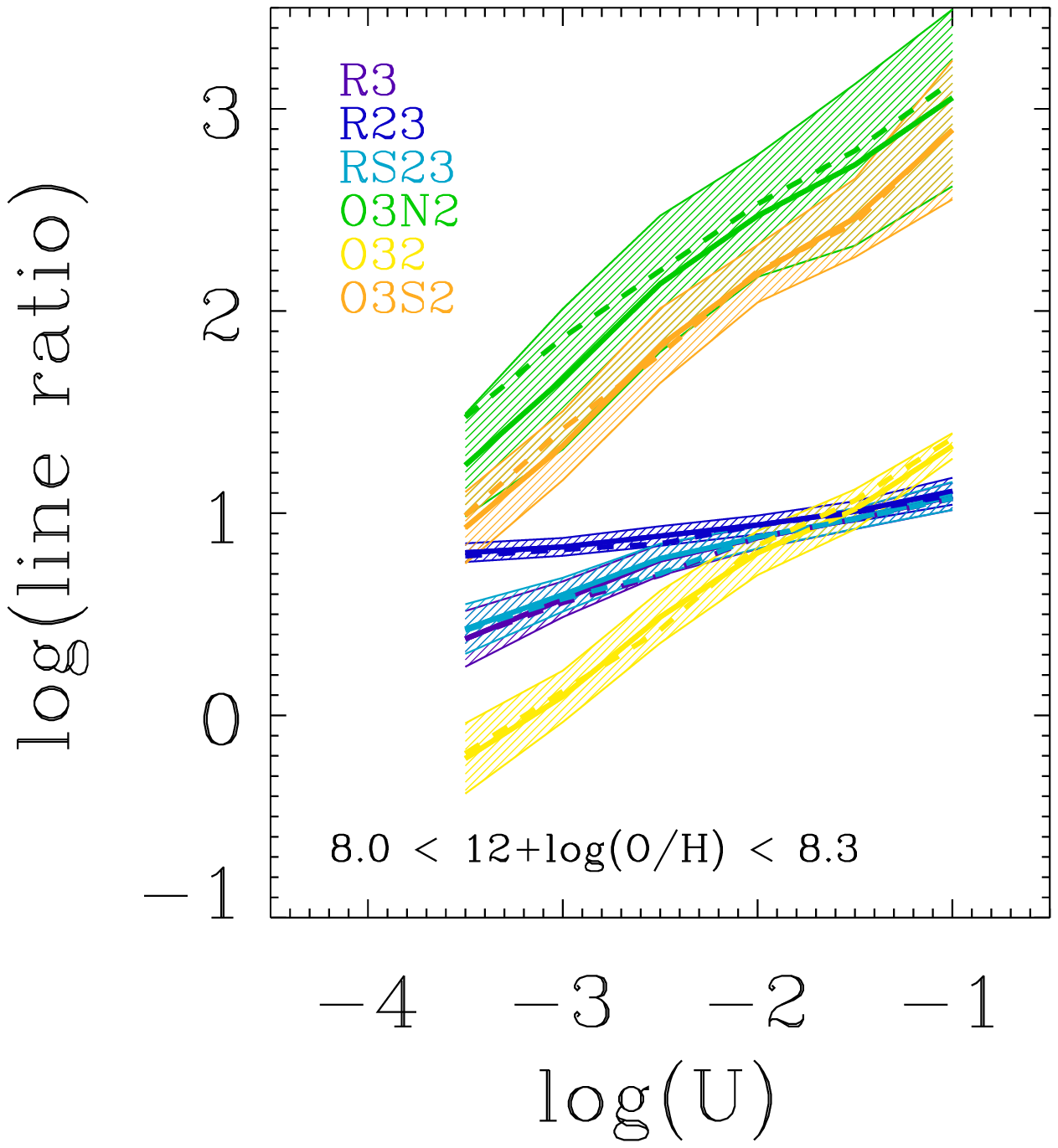,
    width=0.29\textwidth}\hspace{-2.cm}
\epsfig{file=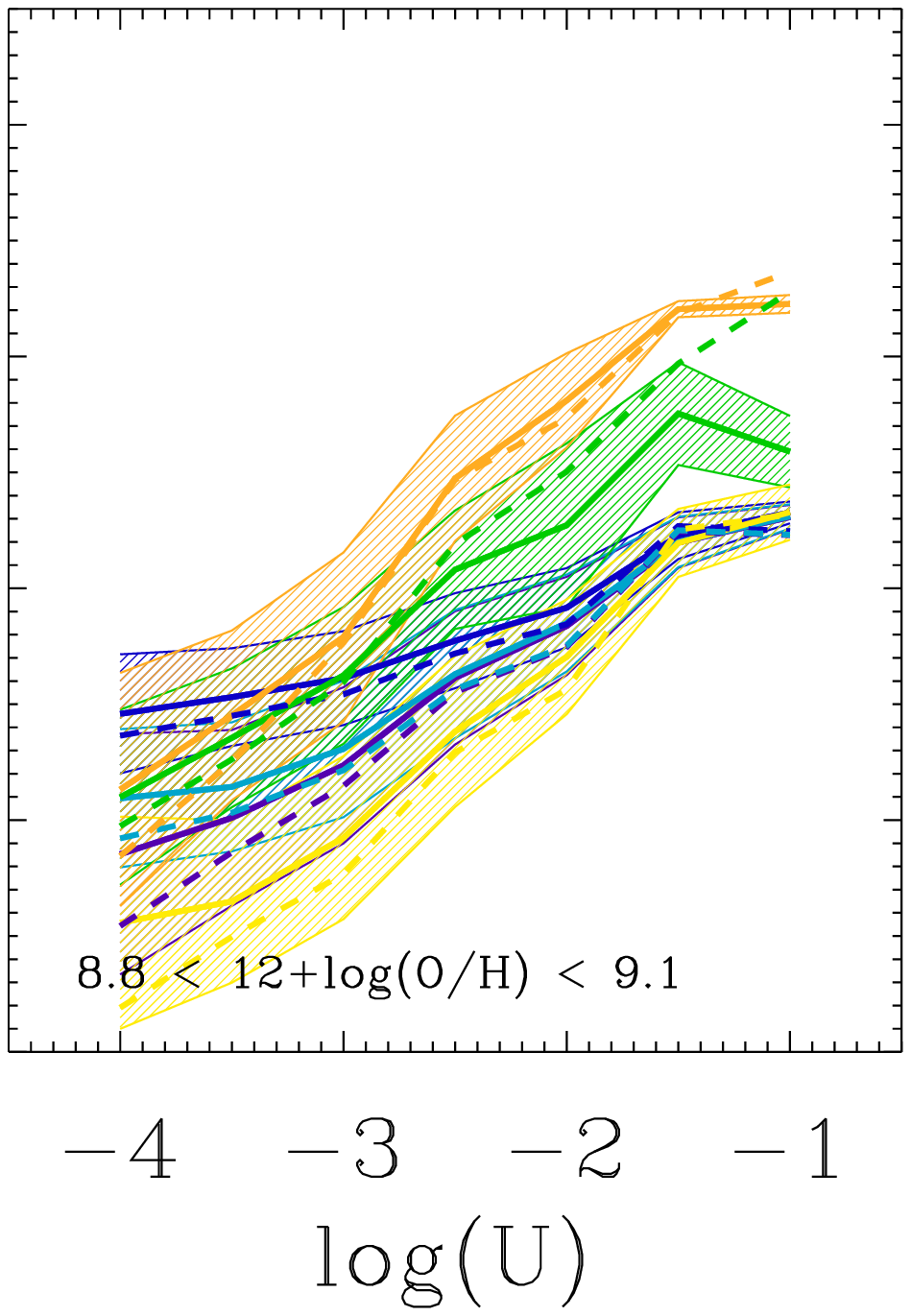,
  width=0.29\textwidth}\hspace{-1.cm}
  \end{flushleft}
\caption{Mean N2O2, N2S2, N2, S2 and R2 (top row) and R3, R23, RS23,
  O3N2 and O3S2 ratios (bottom row) and their 1$\sigma$ scatter
  (different coloured lines and shaded areas as indicated by the
  legend), versus the ionization parameter log(U) of TNG50 (dashed
  lines) and TNG100 (solid lines) galaxies at $z=0$--8 in two
  different metallicity bins (left column: $8.0<$ \logoh\ $<8.3$;
  right column: $8.8<$ \logoh\ $<9.1$).}\label{logU_lineratio}     
\end{figure}

Fig.~\ref{met_lineratio_evol} shows the analogue of Fig.~\ref{met_lineratio_z0}, 
but now only for the global TNG50 (dashed lines) and TNG100 galaxy 
populations (solid lines) and at different redshifts (lilac: $z=0$, dark
blue: $z=2$, light blue: $z=4$, turquoise: $z=5$, green: $z=6$,
orange: $z=7$--8). Where available, we report \Te-based measurements
of \logoh\ for local SDSS analogues of $z\sim2$ galaxies from
\citet[][light-grey, dashed-dotted lines in the O3N2, O32, R3 and R23
panels]{Bian18} and for local extremely metal-poor galaxies in the
Subaru EMPRESS survey from \citet[][dark-grey, dashed-dotted lines in 
the R2, R3, R23, O3N2 and O32 panels]{Nakajima22}.
Also shown are the results inferred from  \JWST/NIRSpec spectroscopy
for three galaxies at $z\sim$6--8 from \citet[][black diamonds with error
bars in the R2, O32, R3 and R23 panels]{Curti23} and 11 galaxies 
at $z\sim4$--9 from \citet[][black triangles with error bars
in the R2, O32, R3 and R23 panels]{Sanders23}, who also proposed 
new calibrations of the R2, R3, R23 and O32 metallicity estimators
at redshifts $z\sim2$--9 (black dashed-dotted lines).

Fig.~\ref{met_lineratio_evol} shows that, with the exception of N2O2 and N2S2, 
the dependence of optical metallicity estimators on \logoh\ is predicted
to evolve strongly between $z=0$ to $z=4$. At fixed oxygen abundance, N2, R2 
and S2 are up to 1~dex smaller at $z\sim 4$ than at the present day, while in 
contrast, R3, R23, RS23, O3N2, O32 and O3S2 increase, also by up to 
1~dex, over the same redshift interval. The reason for this predicted redshift 
evolution can be traced back to the larger ionization parameter of simulated 
high-redshift  galaxies relative to the present-day population, driven by higher
SFRs and higher gas densities \citep[see][]{Hirschmann22}.

Fig.~\ref{logU_lineratio} illustrates the dependence of optical-line ratios 
on ionization parameter, for N2O2, N2S2, N2, S2 and R2 (different coloured 
lines) in the top row and R3, R23, RS23, O3N2, O32 and O3S2 in the bottom
row, for IllustrisTNG galaxies in two narrow bins of oxygen abundances, 
$8.0<\logoh< 8.3$ (left column) and $8.8<\logoh< 9.1$ (right column). At fixed
\logoh, N2O2 and N2S2 depend only weakly on \logU\ because of the roughly similar
ionization energies of N versus N$^+$ (14.5 and 29.6\,eV), O versus O$^+$ (13.6 
and 35.1\,eV) and S versus S$^+$ (10.4 and 23.3\,eV), explaining the lack of 
evolution in the corresponding calibrations in Fig.~\ref{met_lineratio_evol}. 
In contrast, N2, S2 and R2 drop and O3N2, O32, O3S2, R3, R23 and RS32
rise as \logU\ increases because of the increasing abundance of N$^{2+}$,
S$^{2+}$ and O$^{2+}$ to the detriment of N$^+$, S$^+$ and O$^+$. 

Fig.~\ref{met_lineratio_evol} further shows that, beyond $z\sim4$, 
the dependence of optical metallicity estimators on \logoh\ does not 
 evolve much toward earlier cosmic epochs (light blue, green,
yellow and orange lines). 
We provide fits to the 
predicted $z \geq 4$ relations (thick red line in each panel of 
Fig.~\ref{met_lineratio_evol} with the fitted error shown by the red
shaded area) of the form of  $y = P_0 + P_1\,x + P_2\,x^2$, where 
$x = \logoh$ and $y$ is the line ratio under consideration. The fit 
parameters $P_0$, $P_1$ and $P_2$ as well as the goodness of 
the fit $\chi^2$ are listed in Table~\ref{table1}. These may be used
as more physically motivated and potentially more robust metallicity
calibrations for high redshift galaxies.  

So far, we have not considered attenuation of emission lines by dust,
while galaxies may contain significant amounts of dust, even at early
cosmic epochs \citep[e.g.,][]{Schneider04, Zavala22, Witstok23}. 
We estimate the impact of attenuation by dust on line ratios in
Fig.~\ref{met_lineratio_evol}, by adopting for simplicity the
dust attenuation curve of \citet{Calzetti00} with a V-band attenuation
of A$_V$ = 0.5~mag. Under these assumptions, only line ratios
including the \oii\ line, i.e., N2O2, R2 and O32, slightly increase or
decrease (by $\sim0.1$--0.2~dex) because of dust attenuation, as
quantitatively indicated by the black arrow in each of the three
panels. Dust attenuation may, instead, have a negligible impact ($<
0.01$~dex) on line ratios and their fitting curves not including
\oii. We note that the adoption of a
  fixed attenuation curve might not be a very realistic approximation,
and also that A$_V$ may depend strongly on both galaxy mass
and redshift. To account for these effects, a more sophisticated
modelling of dust attenuation would be required, which is beyond 
the scope of this study.

The predicted dependence of R23 on \logoh\ for galaxies at $z=4$--8 
in Fig.~\ref{met_lineratio_evol} agrees remarkably well with \Te-based 
measurements in three galaxies at $z=7$--8.5 by \citet{Curti23} and 
11 galaxies at $z=4$--9 by \citet{Sanders23}. It is also consistent 
with the relation found by \citet{Nakajima22} for local extremely 
metal-poor galaxies. As expected, the relation found by \citealp{Bian18} 
for local analogues of $z=2$ galaxies is more consistent with the 
predicted relation at $z=2$. The agreement between observed and 
predicted relations is also very good for the R3 estimator, at least
at metallicities below $\logoh\sim8$, while even the most distant 
metal-rich galaxies appear to fall closer to the predicted $z=2$ 
relation (but still in strong disagreement with the $z=0$ relation). 
The effect is more pronounced for the R2 and O32 estimators, for
which, from low to high metallicities, the observed relations deviate 
progressively from the predicted ones, by up to $\sim0.5$~dex above 
and $\sim1.0$~dex below, respectively.

To investigate whether this bias may be caused by observational 
selection effects, we compute the relations obtained when requiring 
that all line fluxes associated with a simulated metallicity estimator
exceed a given detection limit. For the purpose of illustration, we choose 
a limit of  $3\times10^{-17}$~erg\ s$^{-1}$\ cm$^{-2}$.\footnote{We note
that this limit is a few times higher than the typical line fluxes of galaxies in 
the \citet{Curti23} and \citet{Sanders23} samples. We checked that similar 
results would be obtained by adopting a much lower detection limit of 
$5\times10^{-18}$~erg\ s$^{-1}$\ cm$^{-2}$ when considering only SF-dominated 
galaxies -- more representative of these two observational samples -- to compute 
the IllustrisTNG calibrations in Fig.~\ref{met_lineratio_evol}. This is because 
AGN-dominated and composite galaxies are  generally brighter than 
SF-dominated ones at fixed stellar mass.}
Interestingly, the fit to the resulting flux-limited galaxy sample at $z\sim 4$--8
(thick dashed-dotted red line in Fig.~\ref{met_lineratio_evol}) is roughly the 
same as for the full sample in the case of R23 (retaining good agreement 
with JWST data), slightly improved at high metallicities for R3, and significantly 
improved for R2 and O32, thus reducing (while not completely alleviating) the 
tension between predicted and observed calibrations. This behaviour arises from 
the faintness of the \oii\ line relative to \hb\ and \oiii, which preferentially selects
out galaxies with low R2 and high O32, together with the fact that fainter 
galaxies tend to have higher \logU, and hence higher R3.

Given the small number statistics of the \citet{Curti23} and 
\citet{Sanders23} samples and the purely theoretical nature of
the model predictions (which have not calibrated against any high-redshift galaxy 
properties), we interpret the level of agreement between model and 
observations in Fig.~\ref{met_lineratio_evol} as an encouraging illustration 
of the success of our approach. This result emphasizes the importance of 
accounting for the redshift evolution of the calibration of optical-line
ratios as metallicity estimators, together with the potential bias of empirical 
metallicity calibrations caused by flux-detection limits.

\begin{figure*}
\epsfig{file=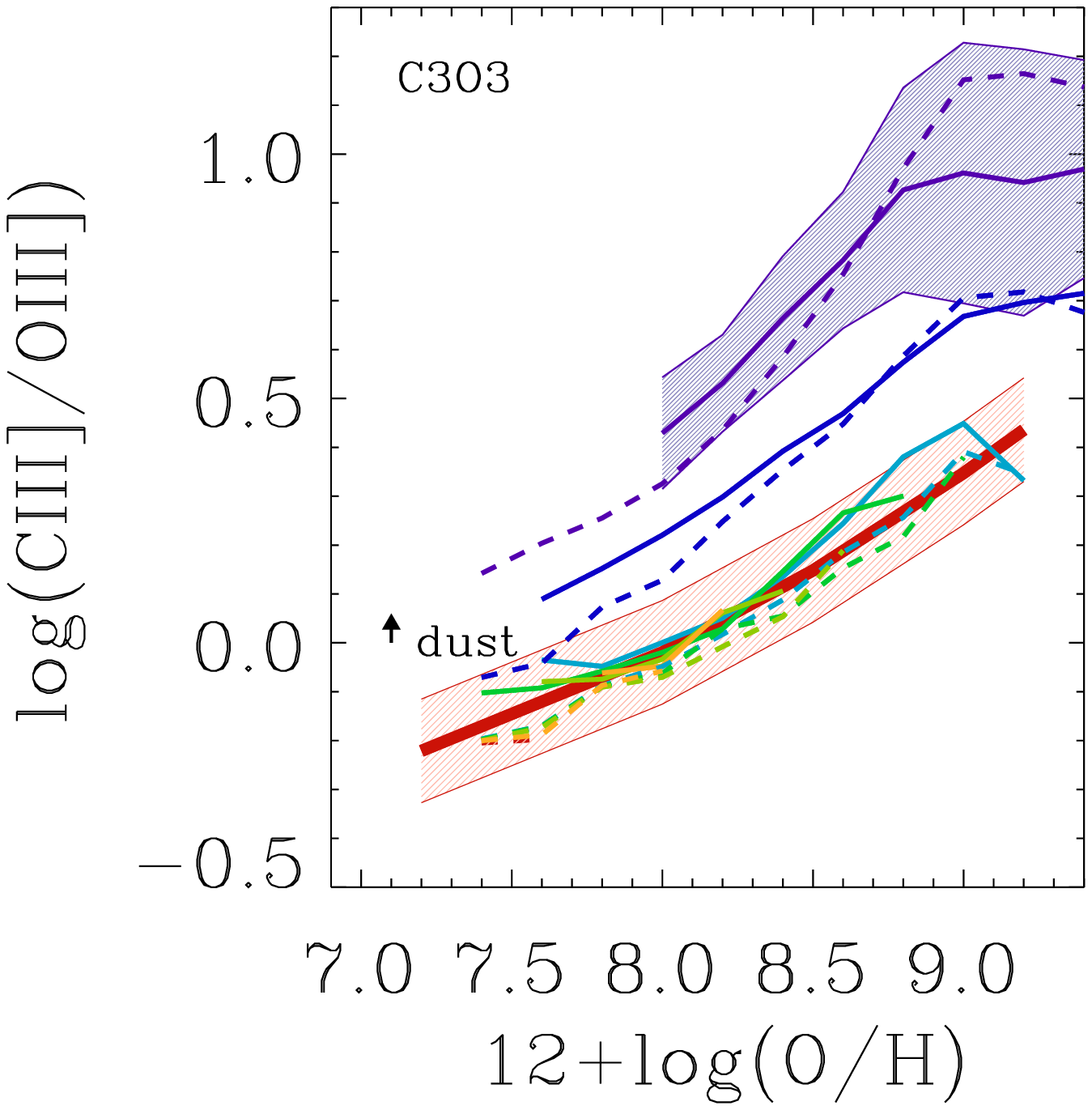,
  width=0.26\textwidth}\hspace{-.4cm}
\epsfig{file=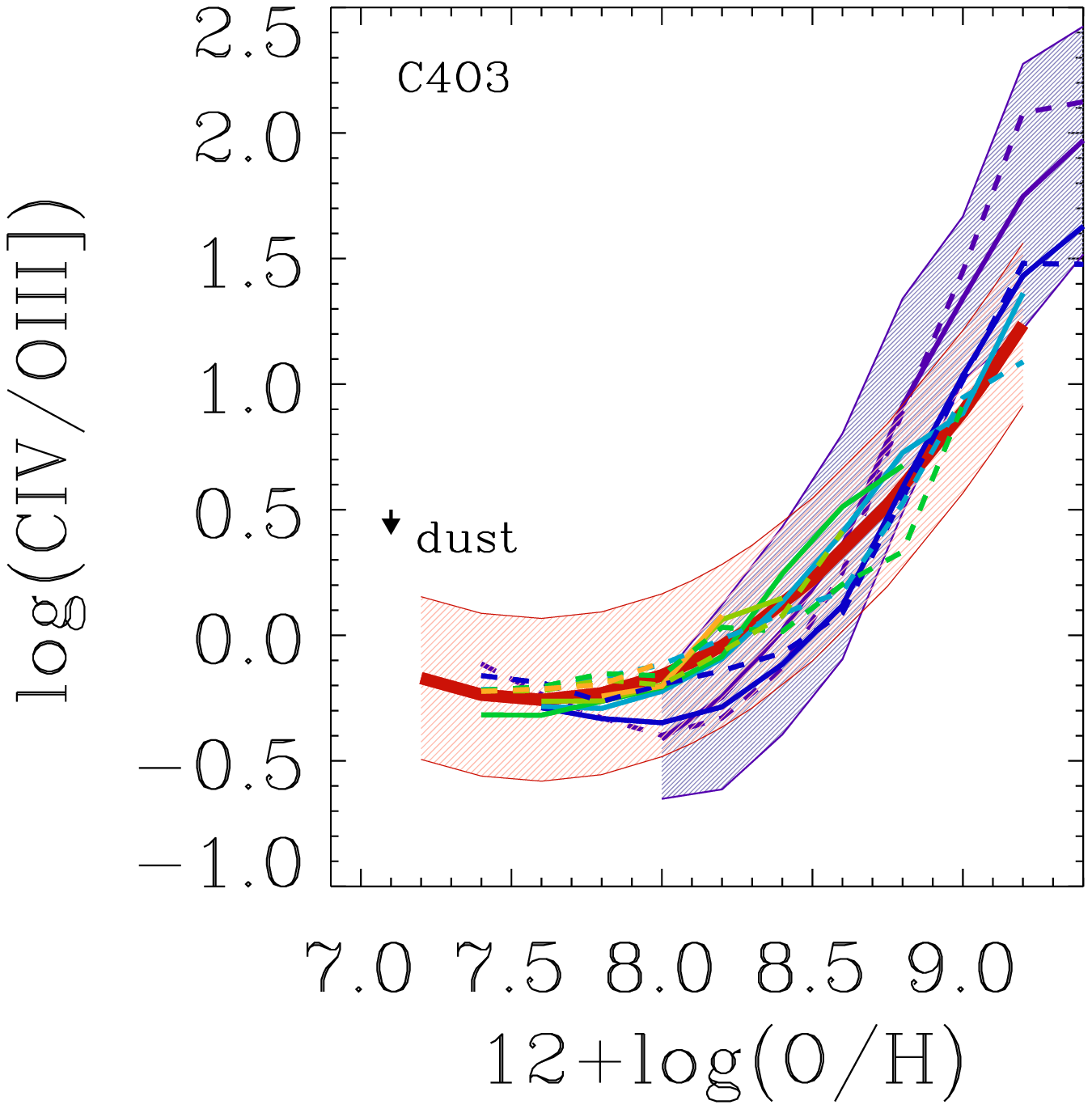,
  width=0.26\textwidth}\hspace{-.4cm}
\epsfig{file=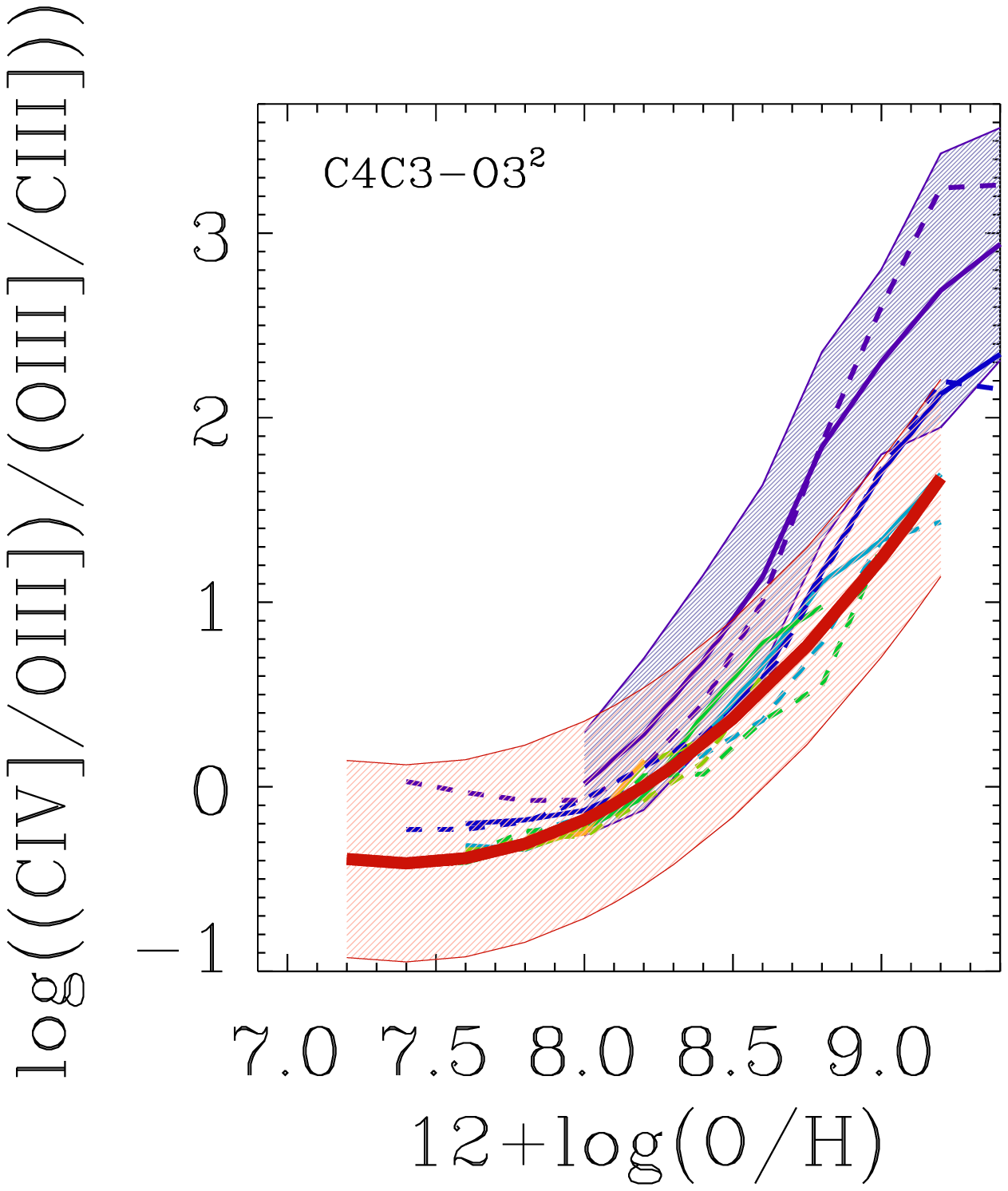,
  width=0.26\textwidth}\hspace{-.4cm}
\epsfig{file=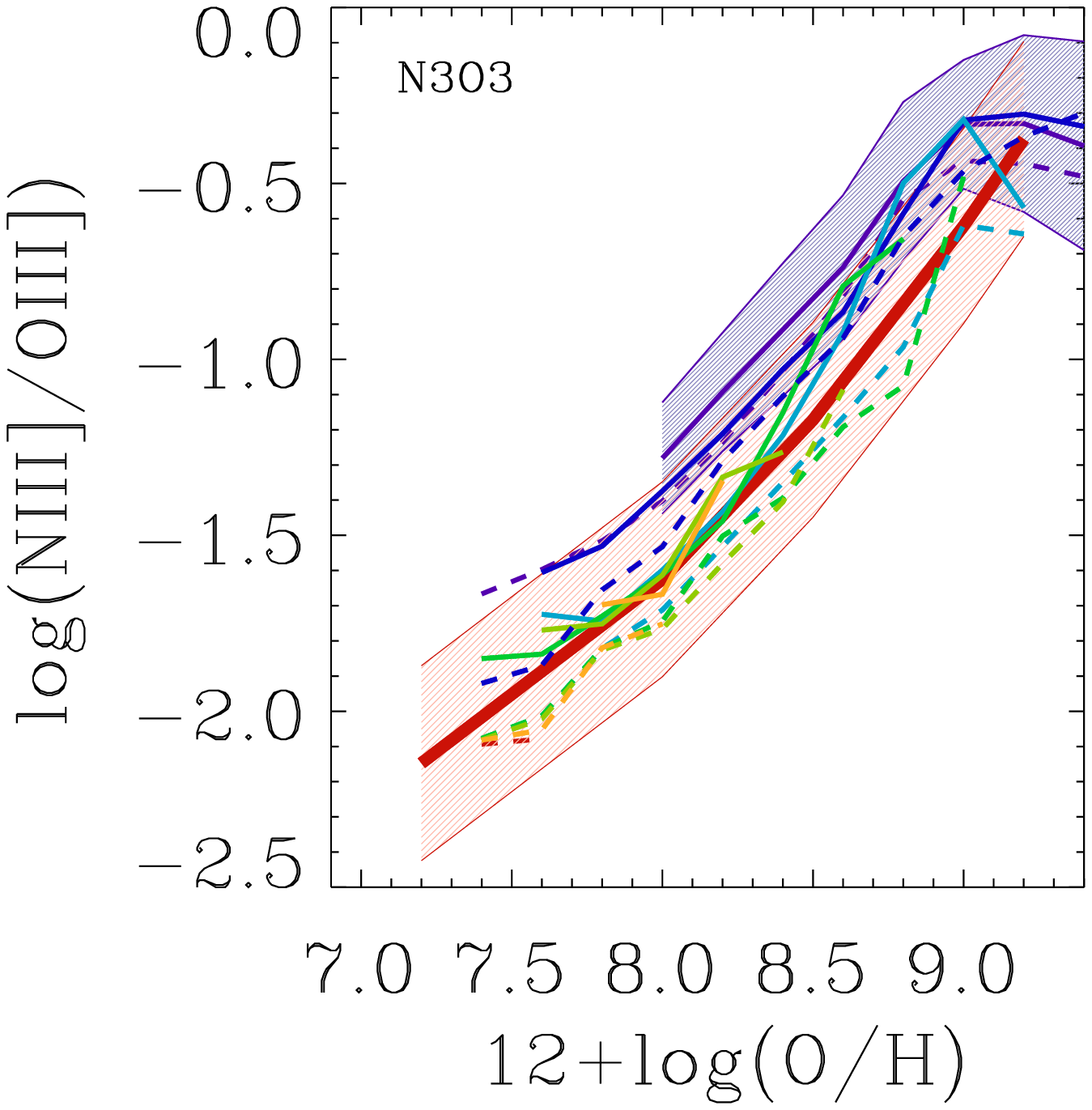,
  width=0.26\textwidth}\hspace{-.4cm}
\epsfig{file=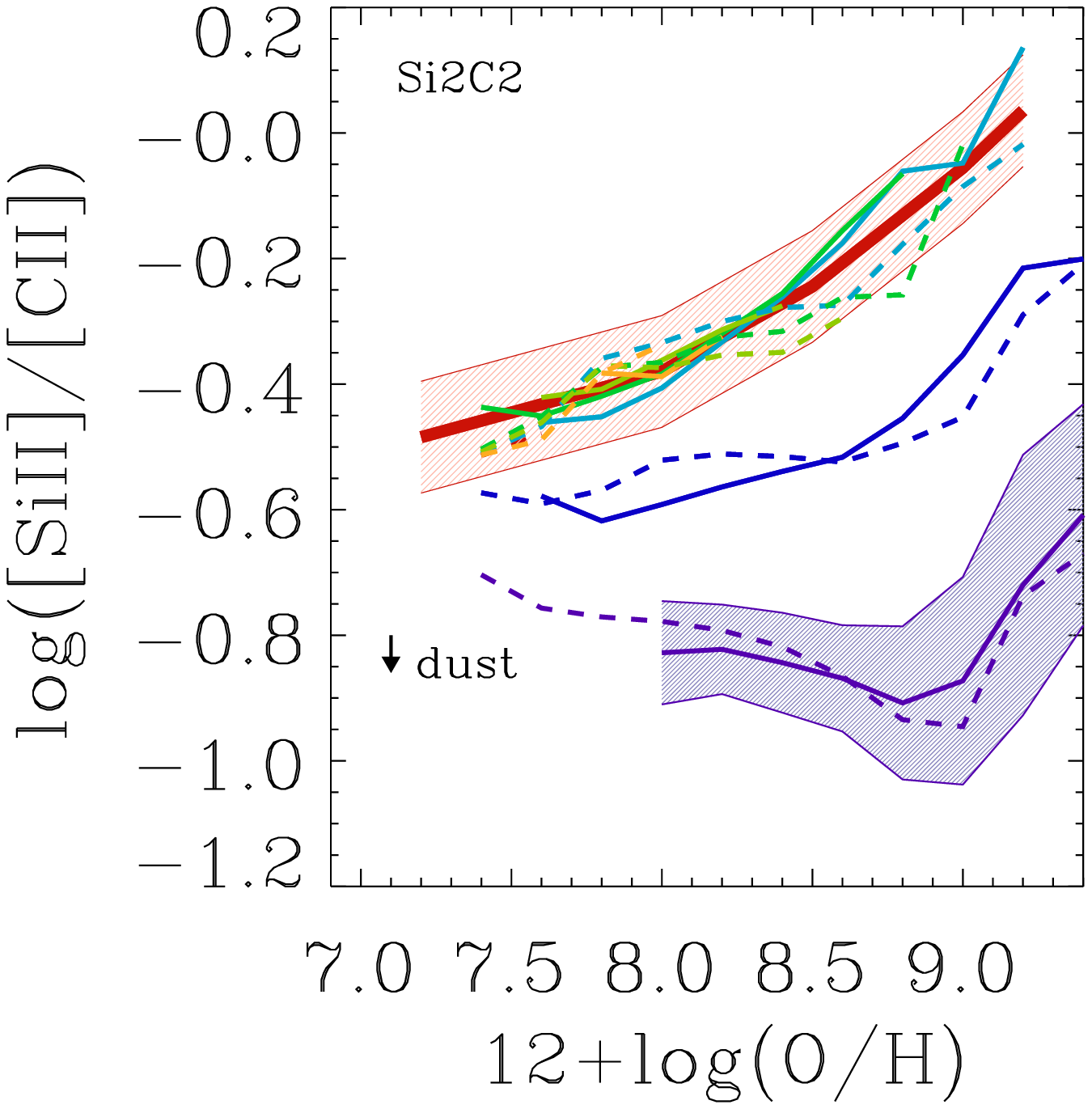,
  width=0.26\textwidth}\hspace{-.4cm}
\epsfig{file=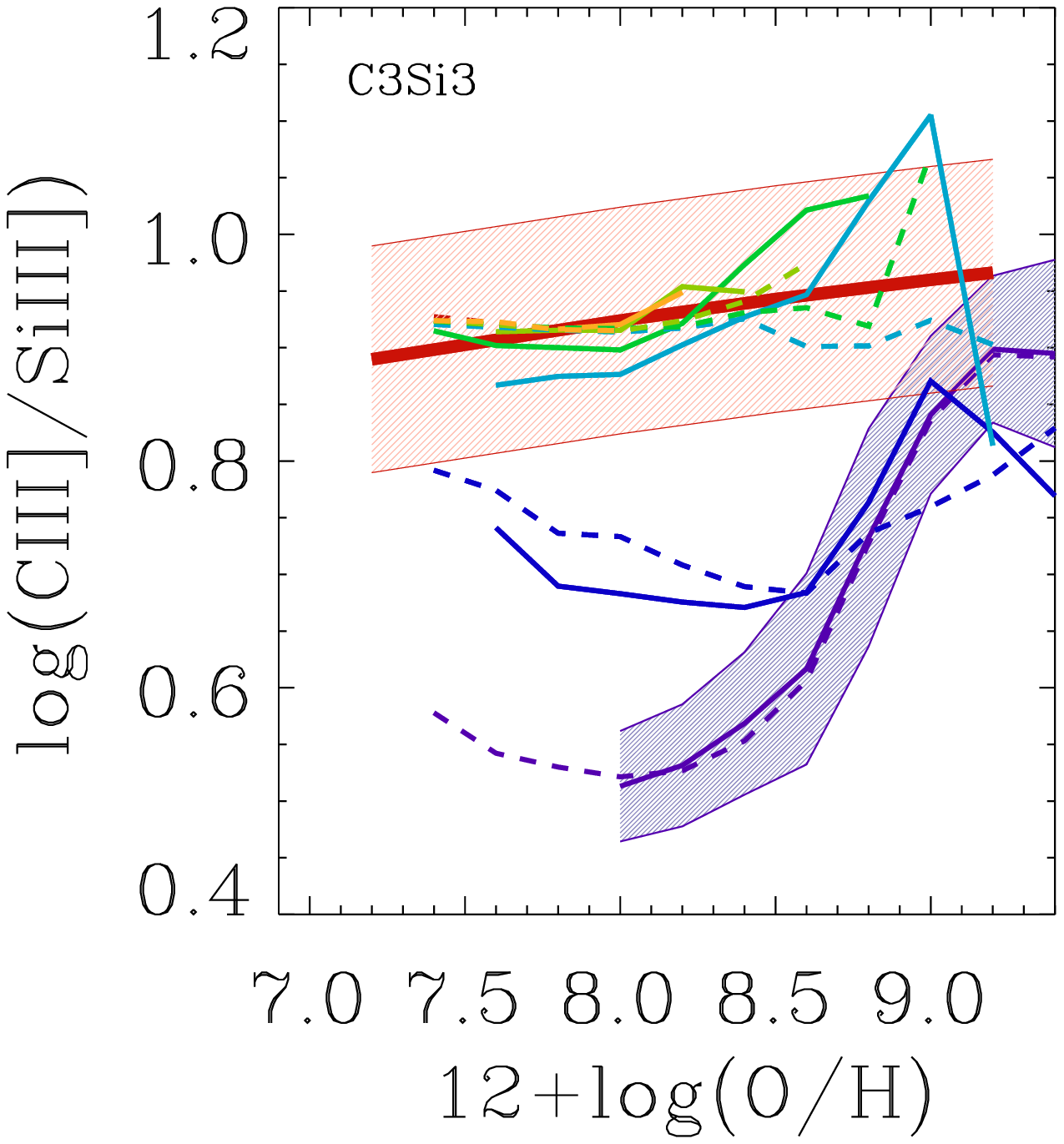,
  width=0.26\textwidth}\hspace{-.4cm}
\epsfig{file=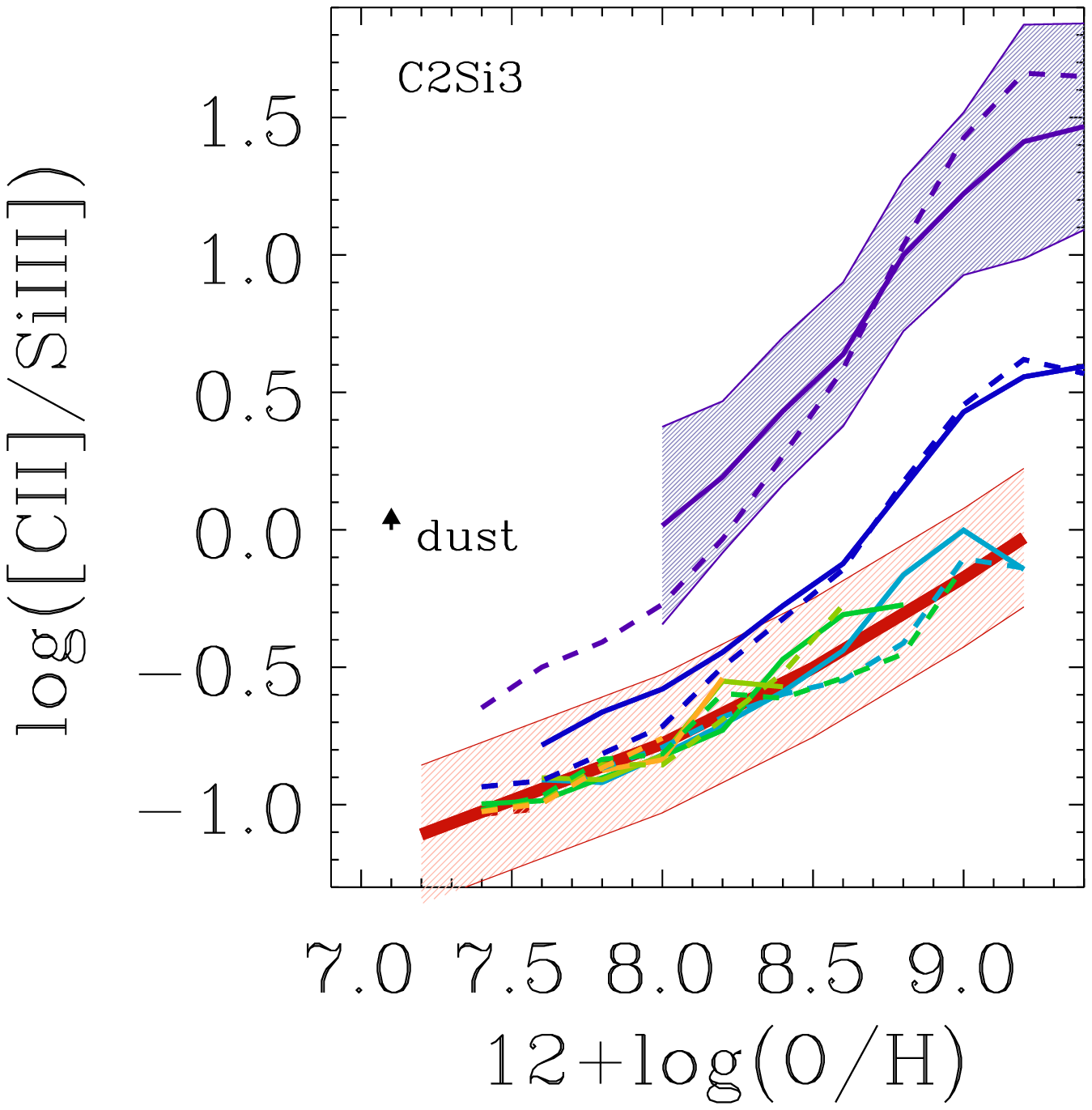,
  width=0.26\textwidth}\hspace{-.4cm}
\epsfig{file=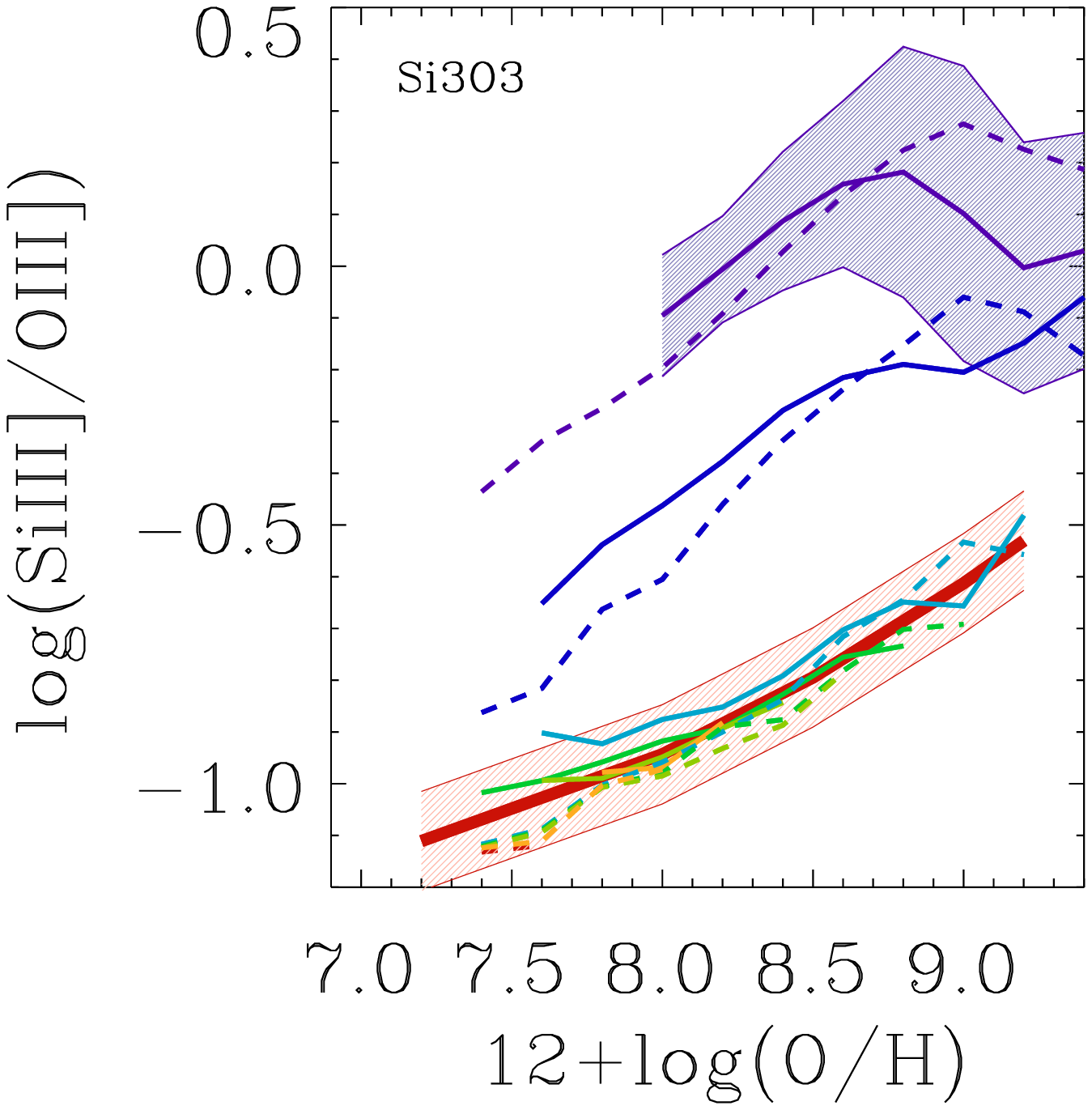,
  width=0.26\textwidth}\hspace{-.4cm}
\epsfig{file=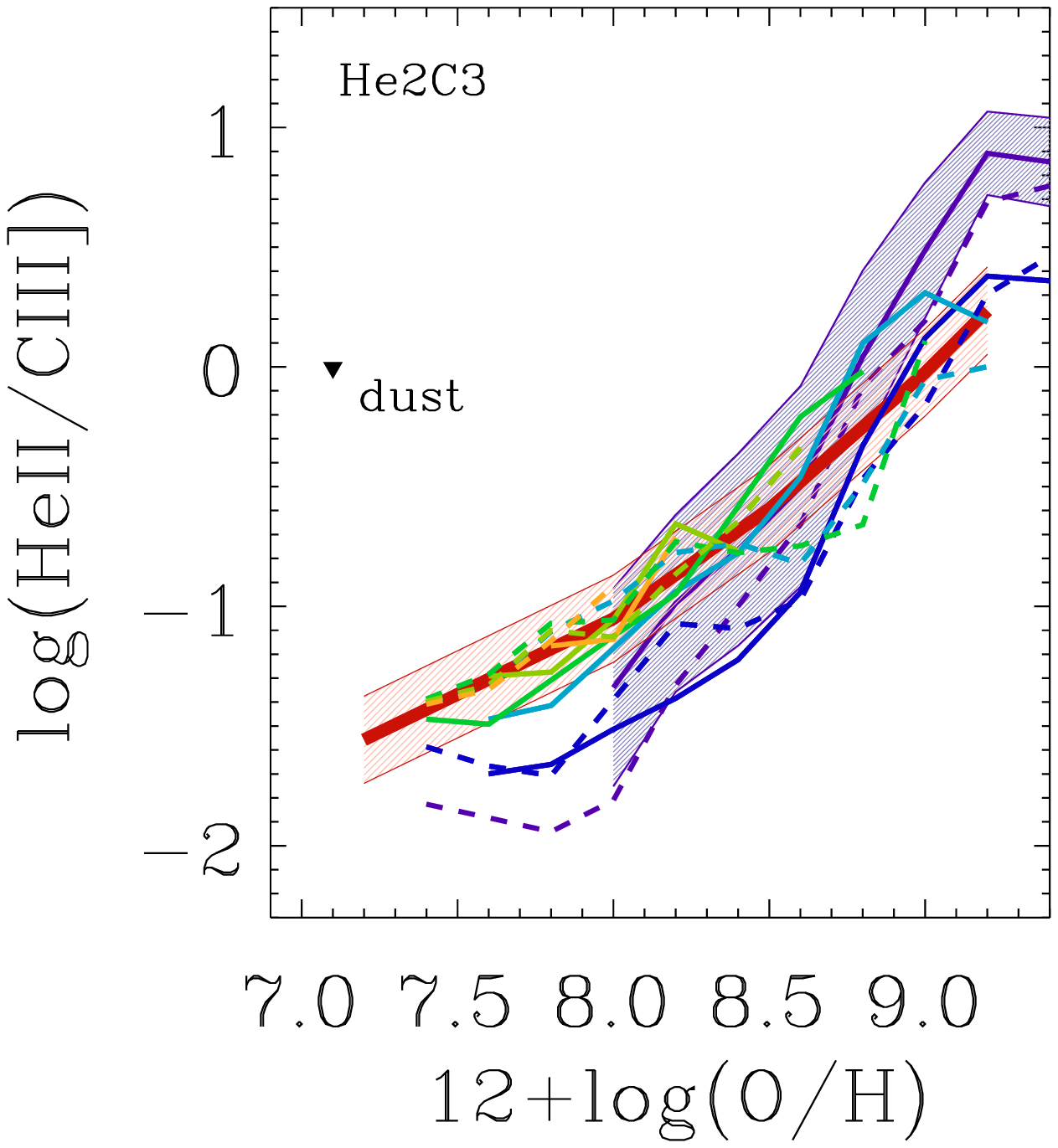,
  width=0.26\textwidth}\hspace{-.4cm}
\epsfig{file=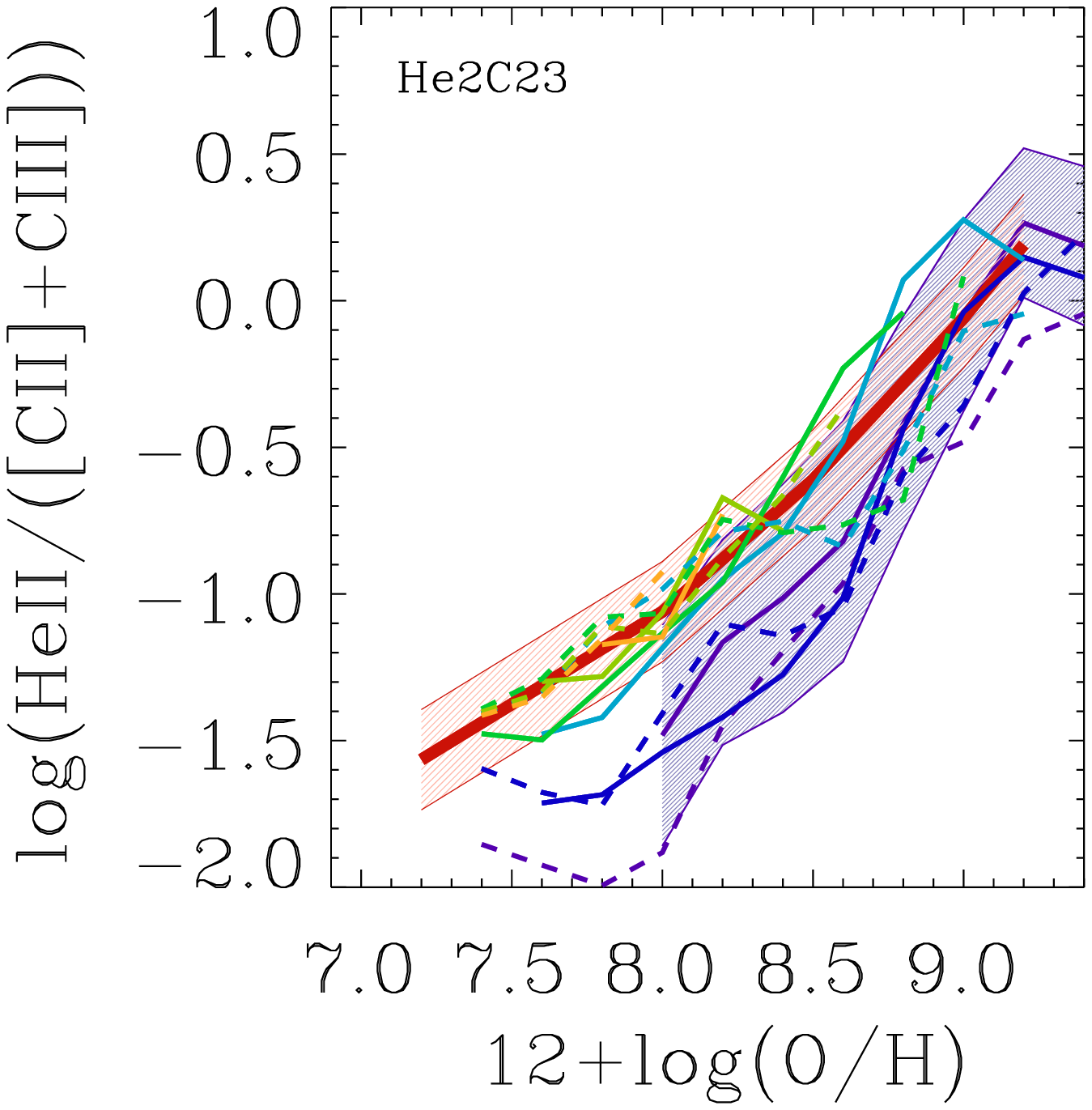,
  width=0.26\textwidth}\hspace{-.4cm}
\epsfig{file=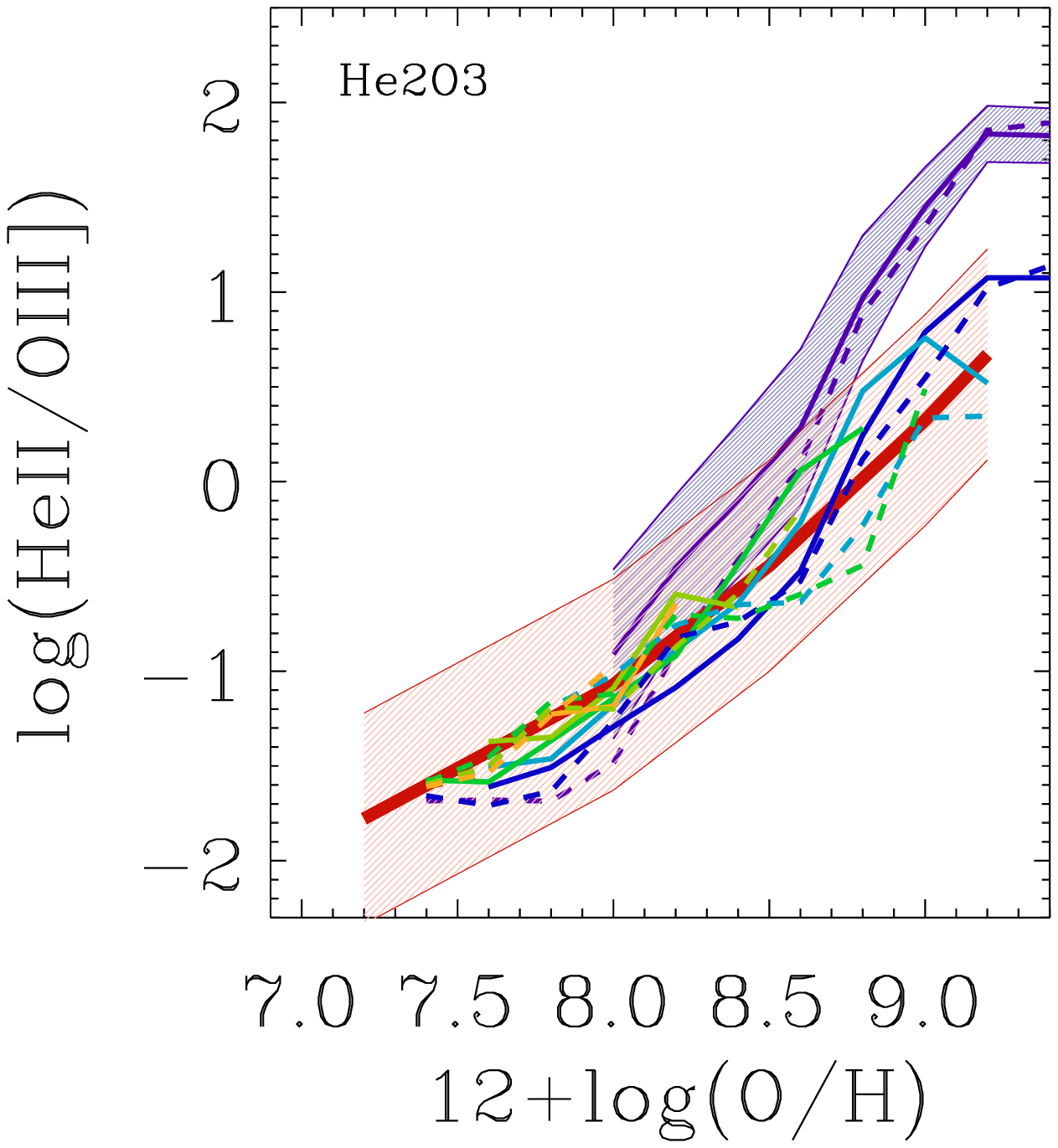,
  width=0.26\textwidth}\hspace{-.4cm}
\epsfig{file=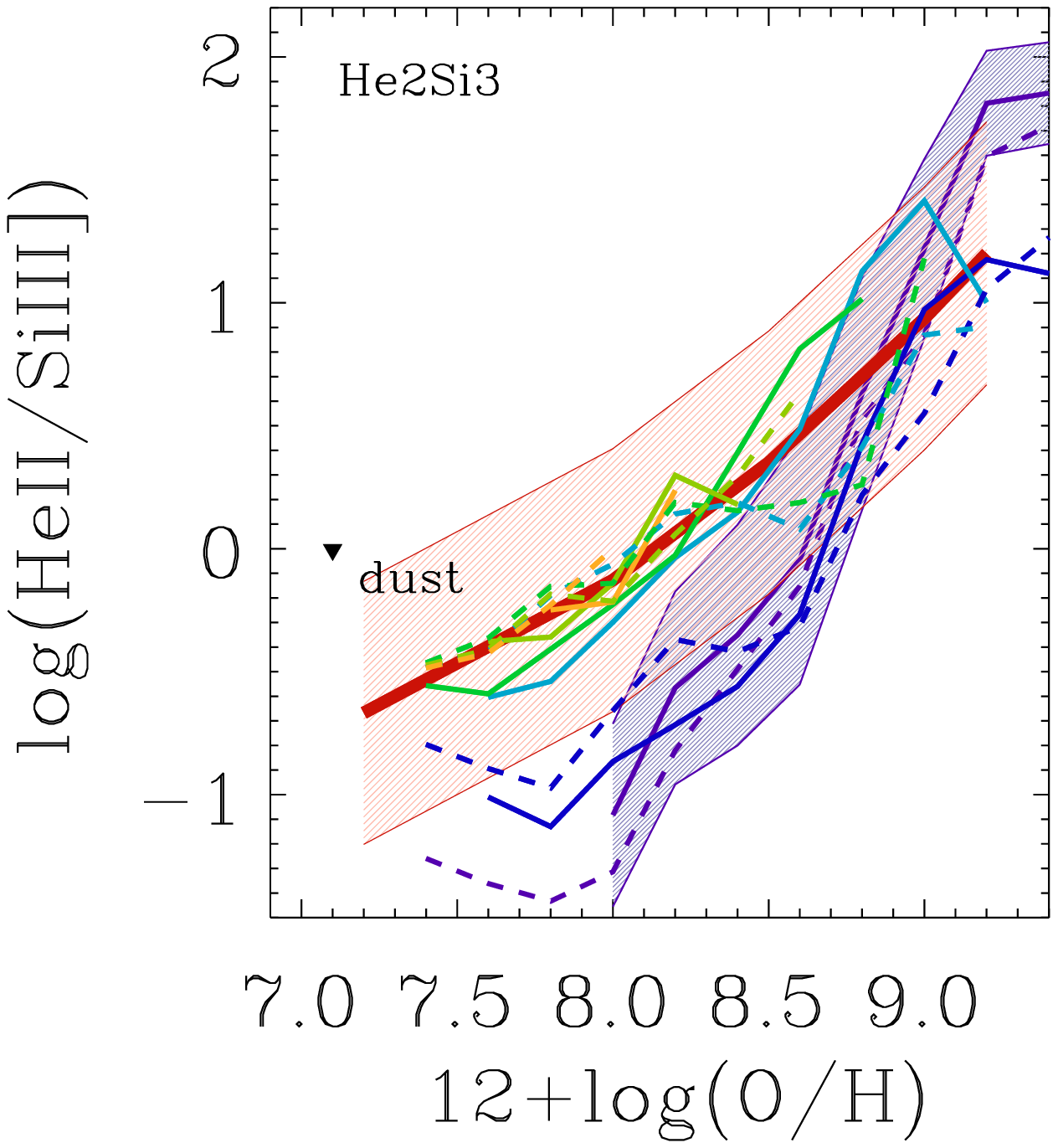,
  width=0.26\textwidth}\hspace{-.4cm}
\caption{Same as in Fig.~\ref{met_lineratio_evol}, but now for different
  UV ratios  C3O3, C4O3, C4C3-O3$^2$, N3O3, Si2C2, C3Si3, C2Si3,
  Si3O3, He2C3, He2C23, He2O3, He2Si3 (panels from top left to bottom
  right).}\label{met_uvlineratio_evol}      
\end{figure*}
\begin{table}
\centering
\begin{tabular}{ l | r | r | r | r}
Line ratio  &    $P_0$ 	&	$P_1$	&	$P_2$  	 &     $\chi^2$ \\
  \hline
  C3O3  &      1.632 &  $-0.716$ & 0.064 & 0.106 \\
  C4O3	&	32.852 & $-8.727$  & 0.575 &  0.324 \\
   C4C3-O3$^2$ &  34.485 & $-9.443$ & 0.639 & 0.534 \\
  N3O3 &	4.361 & $-2.305$ & 0.195 & 0.854  \\
  Si2C2 	&	4.770 & $-1.504$  & $-0.108$ & 0.089 \\
  C3Si2 & 0.369	 & 0.099  & $-0.004$ & 0.100 \\
  C2Si3	& 1.990	 & $-1.190$  & 0.105  & 0.252    \\
  Si3O3  &	1.248  & $-0.811$   & 0.067  &  0.096  \\
  He2C3	& 6.569  & $-2.713$  & 0.220  &  1.182 \\
  He2C23   & 5.764  & $-2.502$  & 0.206  & 1.171 \\
  He2O3	& 8.185 &  $-3.425$ & 0.284 &  1.557 \\
  He2Si3 & 6.938 & $-2.614$  & 0.216 &  1.764 \\

\end{tabular}
\caption{Parameters of the fits to UV-line ratios of llustrisTNG 
galaxies at $z=4$--8 used as metallicity indicators 
(thick red lines in Fig.~\ref{met_uvlineratio_evol}) with 
quadratic functions of the form $y = P_0 + P_1\,x +
P_2\,x^2$, where $x =  \logoh$ and $y$ is the line ratio. The
rightmost column quantifies the goodness of fit.}\label{table2} 
\end{table}

\subsection{UV-line ratios as tracers of interstellar metallicity 
  at different cosmic epochs}\label{UVlineratios} 

In Sections~\ref{opticalz0} and \ref{opticalhighz}, we saw how
optical-line ratios may be used to trace interstellar metallicity in
simulated galaxies at both low and high redshift. The predicted relations
between strong-line ratios and oxygen abundance are qualitatively 
consistent with published calibrations at $z=0$ and evolve strongly at
higher redshift. In this Section, we investigate which UV-line ratios 
may provide alternative metallicity tracers in high-redshift galaxies, particularly
when standard optical lines are redshifted outside the spectral window
accessible to near-IR spectrographs, such as \JWST/NIRSpec. 

For this analysis, we consider the following UV-line ratios, motivated 
both by recent review articles \citep{Kewley19, Maiolino19} and
predictions of photoionization models \citep[][]{Gutkin16, Byler18}:\\ 
 -- \ciii/\oiiiuv\ (hereafter simply C3O3)\\
 -- \civ/\oiiiuv\ (C4O3)\\
 -- C4O3 $\times$ C3O3 (C4C3-O3$^2$)\\
 -- \niii/\oiiiuv\ (N3O3)\\
 -- \silii/\cii\ (Si2C2)\\
 -- \siliii/\ciii\ (Si3C3)\\
 -- \cii/\siliii\ (C2Si3)\\
 -- \siliiid/\oiiiuv\ (Si3O3)\\
 -- \heii/\ciii\ (He2C3)\\
 -- \heii/(\ciii+\cii) (He2C23)\\
 -- \heii/\oiiiuv\ (He2O3)\\
 -- \heii/\siliiid\ (He2Si3)\\

By analogy with Fig.~\ref{met_lineratio_evol}, we show in 
Fig.~\ref{met_uvlineratio_evol} the evolution with redshift of the
relations between these UV-line ratios and oxygen abundance,
as predicted for the TNG50 and TNG100 galaxy populations. 
At $z=0$, all relations show an increase in ratio strength with
increasing metallicity, except for Si2C2 and Si3O3. As in the case 
of optical metallicity estimators, the relation for each UV-line ratio 
results from an interplay between relative enrichment in the involved
elements, ionization energies required and sensitivity of the involved
transitions to the drop in \Te\ as more coolants are added to the ISM.
Again, the change in the relations as $z$ increases can be traced 
back to the larger ionization parameter of simulated high-redshift  
galaxies relative to present-day ones (Section~\ref{opticalhighz}).
From $z=0$ to $z=4$, C4O3, C4C3-O3$^2$, N3O3,
He2C3, HeC23 and  He2O3 hardly change at fixed
oxygen abundance, because of the similar ionization energies of
the species involved in each ratio \citep[see fig.~1 in][]{Feltre16}. 
Instead, C3O3, Si2C2, C2Si3 and Si3O3 decrease, and Si2C2,
C3Si3 and He2Si3 increase over the same redshift interval, as the 
different ionization energies of the involved species are more sensitive 
to changes in \logU.

Beyond $z \sim 4$, the relations in Fig.~\ref{met_uvlineratio_evol}
all have positive slopes and evolve very little out to earlier epochs.
As in the case of optical metallicity estimators, we exploit this 
behaviour to provide fitting functions of the relations between UV-line 
ratios and oxygen abundance at $z \geq 4$ of the form of  $y = P_0 +
P_1\,x + P_2\,x^2$, where $x = \logoh$ and $y$ is the UV-line ratio 
under consideration (thick red line in each panel of Fig.~\ref{met_uvlineratio_evol}). 
The fit parameters $P_0$, $P_1$ and $P_2$ are listed in Table~\ref{table2}. 

The above calibrations of UV metallicity estimators of high-redshift 
galaxies are only mildly (if at all) affected by dust attenuation: applying 
a \citet{Calzetti00} attenuation curve with A$_V$ = 0.5~mag as in 
Section \ref{opticalhighz} implies changes of C4C3-O3$^2$, N3O3, 
C3Si3, Si3O3, He2C23 and He2O3 by less than 0.01~dex, while C2Si3 and 
C3O3 can increase by up to 0.08~dex and C4O3, He2C3, Si2C2 and 
He2Si3 decrease by up to 0.1~dex (as illustrated by the black arrow in 
the corresponding panels of Fig.~\ref{met_uvlineratio_evol}). 

The results of Fig.~\ref{met_uvlineratio_evol} suggest that the 
most promising UV metallicity estimators are N3O3, He2C3, 
He2C23 and He2O3, down to oxygen abundances as low as
$\logoh\sim7$. These UV-line ratios exhibit the least scatter combined 
with a strong dependence on metallicity over nearly two orders of 
magnitude, $7\la\logoh\la9$. A limiting factor may be the observability
of \heii, which tends to be weak in SF-dominated galaxies 
\citep[see fig.~15 of][]{Hirschmann22}.

\begin{figure*}
  \begin{center}
    {\bf Classical metallicity tracer}
  \end{center}\vspace{-0.4cm}
  \epsfig{file=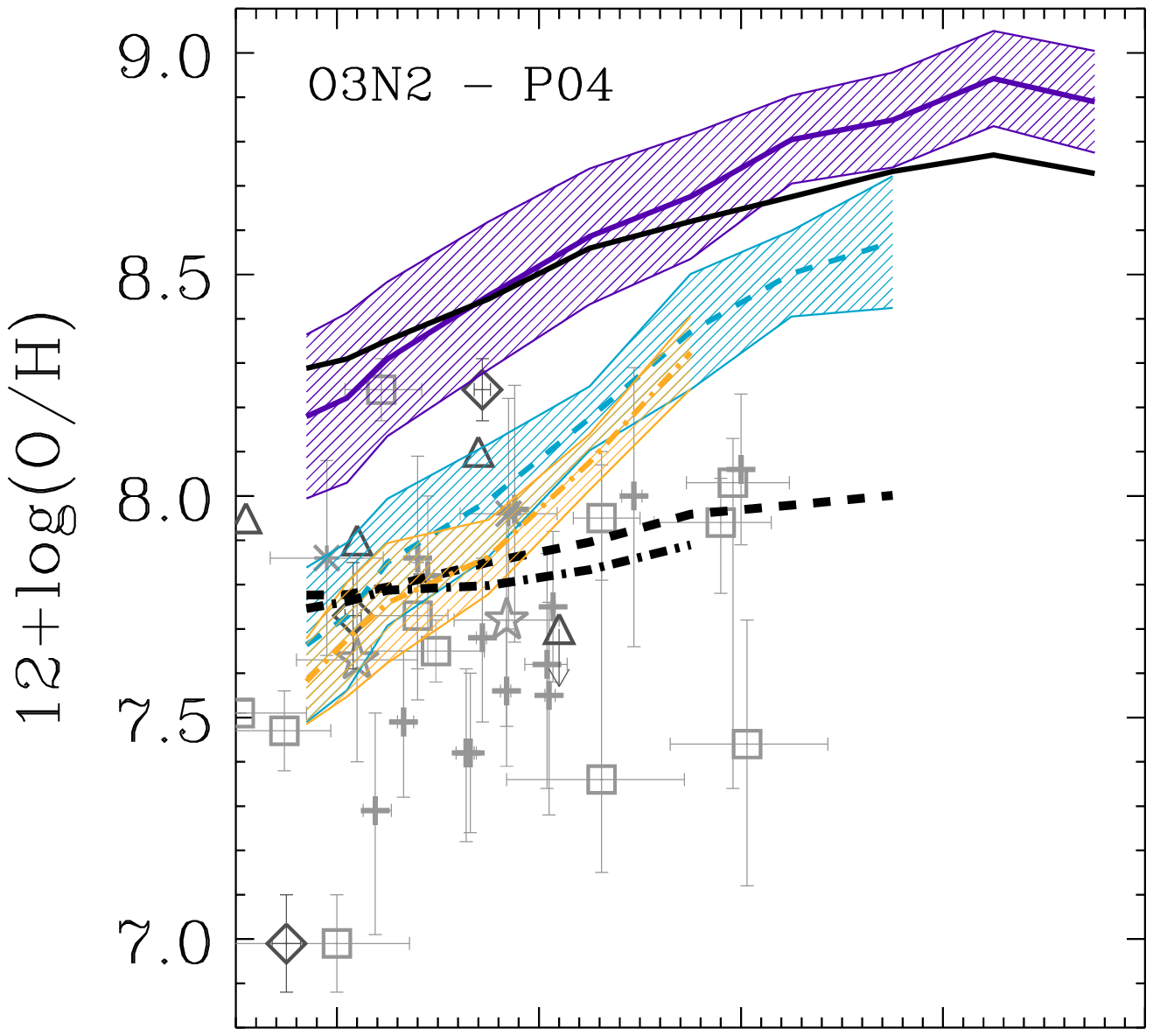,
    width=0.4\textwidth}\hspace{-2.cm}
  \epsfig{file=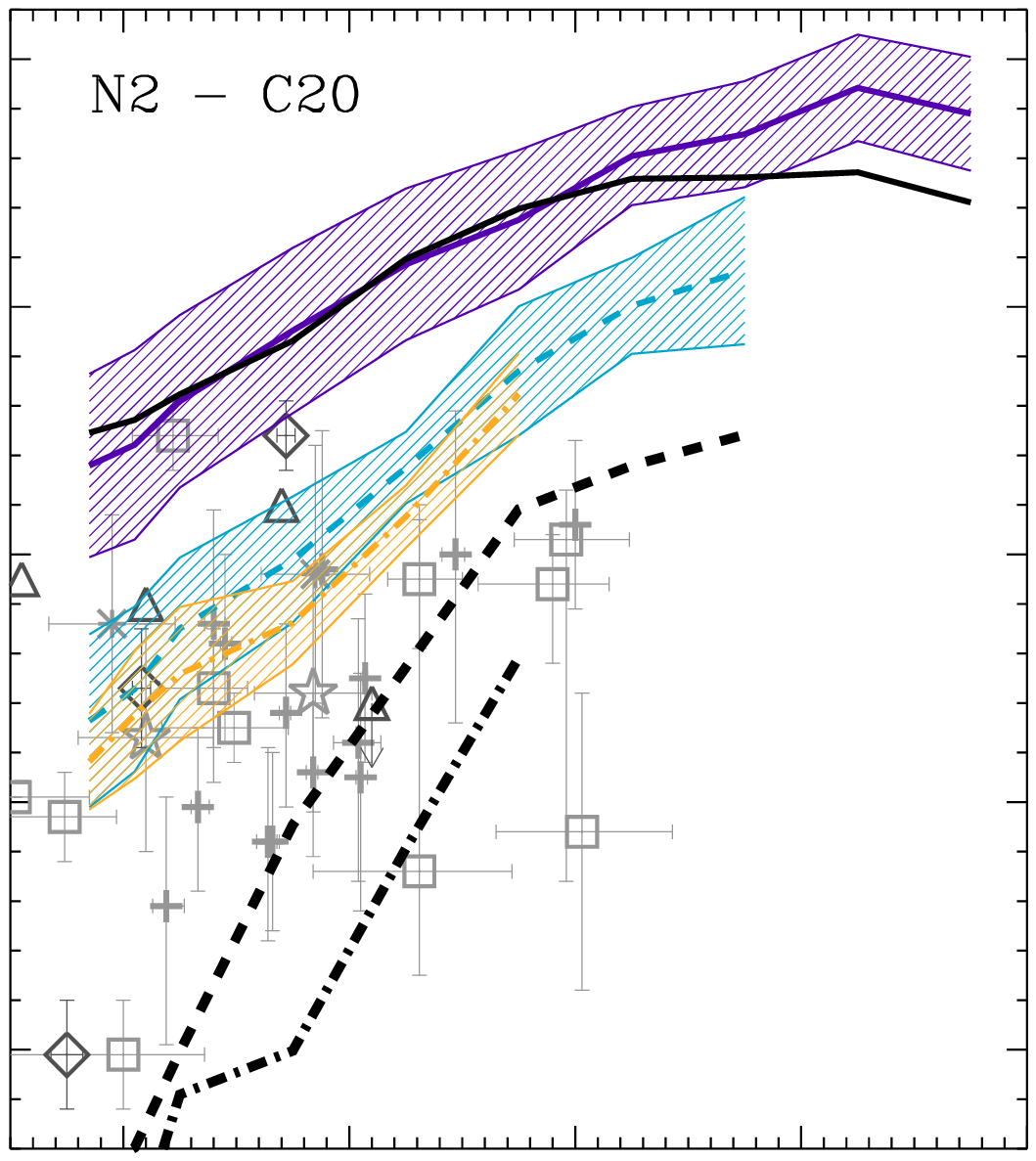,
    width=0.4\textwidth}\hspace{-2.cm}
  \epsfig{file=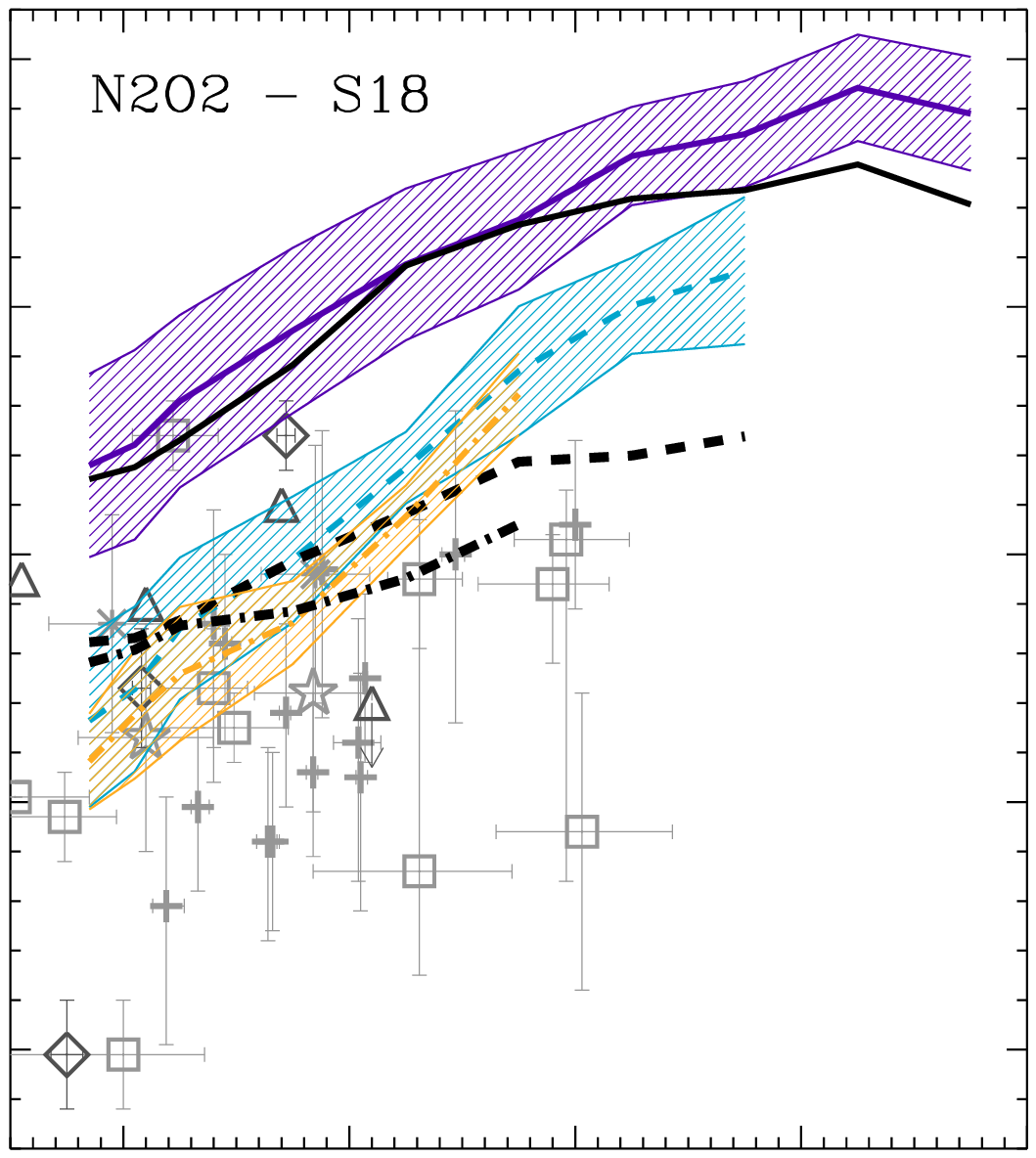,
    width=0.4\textwidth}\vspace{-1.2cm}
  \begin{center}
    {\bf \color{red}{Metallicity tracer of this work}}
  \end{center}\vspace{-0.4cm}
  \epsfig{file=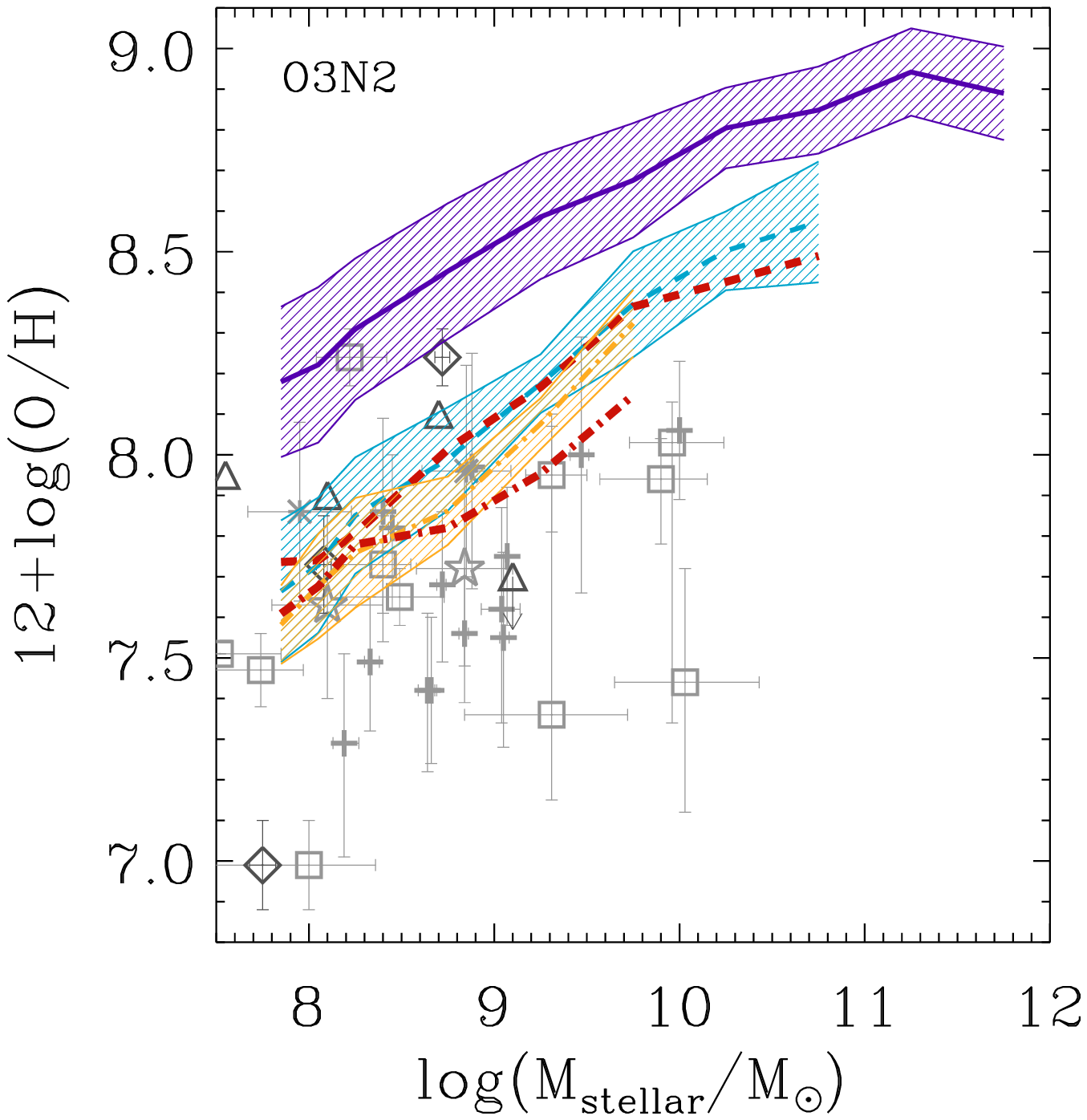,
    width=0.4\textwidth}\hspace{-2.cm}
  \epsfig{file=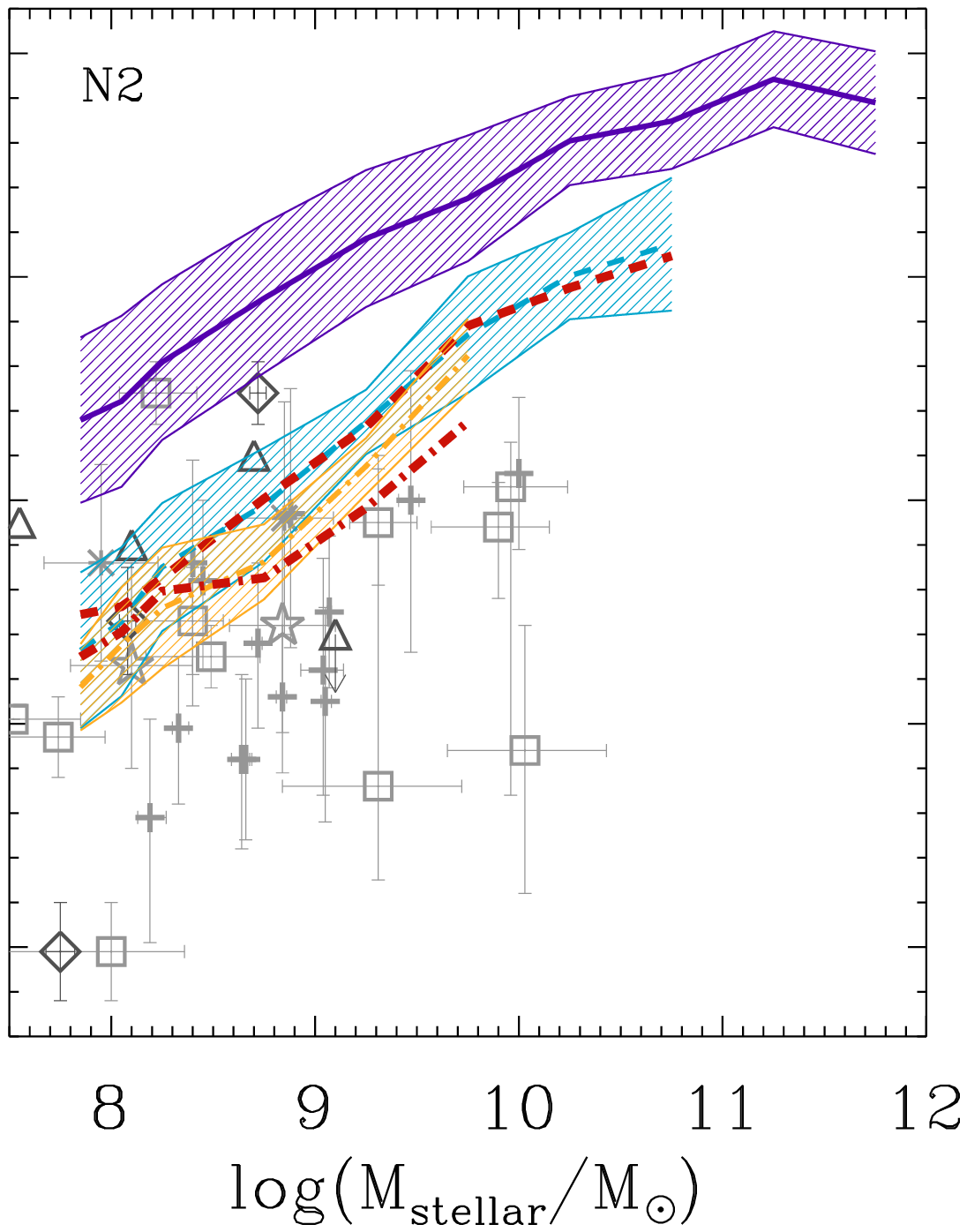,
    width=0.4\textwidth}\hspace{-2.cm}
  \epsfig{file=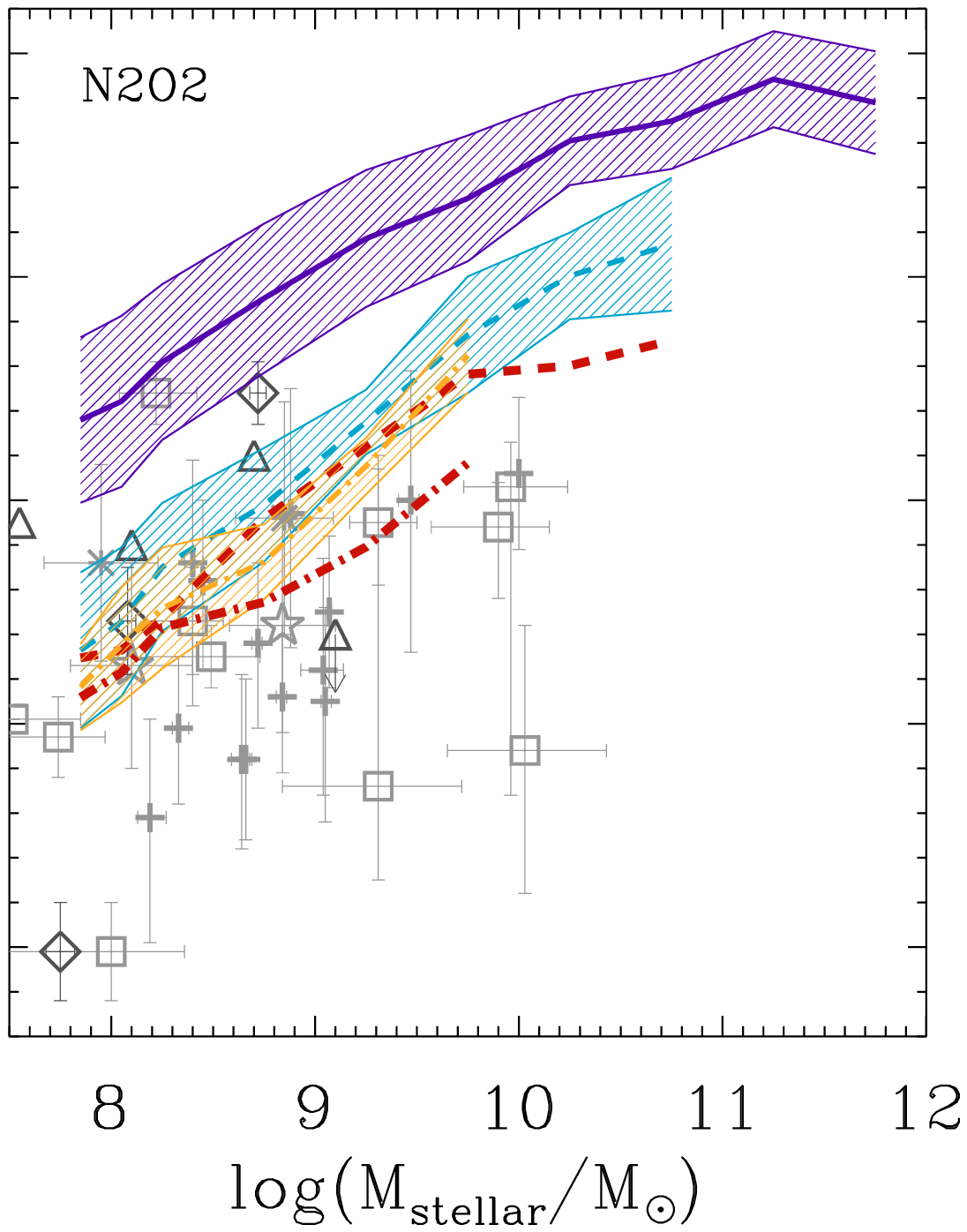,
    width=0.4\textwidth}\vspace{-0.2cm}\hspace{1cm}
 \epsfig{file=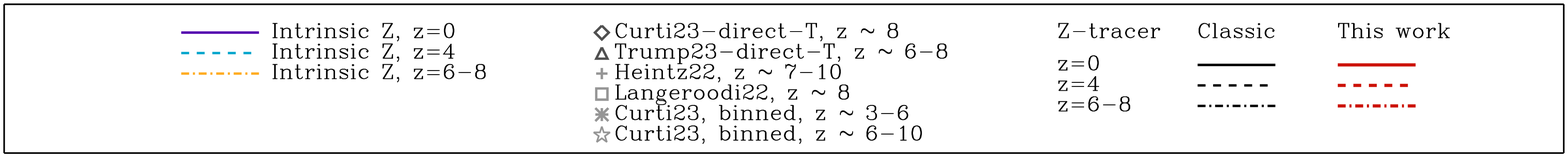,
    width=0.98\textwidth}\vspace{-1.cm}
\caption{Relation between \logoh\ and stellar mass for IllustrisTNG
  galaxies at three redshifts: $z = 0$ (solid lila line and shaded
  1$\sigma$-scatter in all panels the same), $ z = 4$  (dashed
  light-blue line and scatter)   and $z = 6$--8 (dashed-dotted orange
  line and scatter). Also shown in all panels are data of the z $\sim
  6$--8 galaxies observed with JWST, with \Te-based metallicity
  estimates from \citet[][diamonds]{Curti23} and
  \citet[][triangles]{Trump23} and with metallicity
  estimates (derived from strong line-ratios) from
  \citet[][squares]{Langeroodi22, Heintz22, Curti23}. Additionally
  overplotted are the relations  between \logoh\
  and stellar mass derived  from the simulated
  optical-line ratios of simulated galaxies, (i) using published $z=0$
  calibrations of the metallicity estimators O3N2
  \citep[][P04]{Pettini04}, N2 \citep[][C20]{Curti20}, N2O2
  \citep[][S18; black solid, dashed and dot-dashed lines for $z=0$, 4
  and 6--8, respectively, in top panels]{Sanders18}; and (ii) using
  the new, redshift-dependent calibrations proposed in
  Section~\ref{opticalhighz} (Table~\ref{table1}; red solid, dashed
  and dot-dashed lines for $z=0$, 4 and 6--8, respectively, in the
  bottom panels).}\label{MassMet1}         
\end{figure*}

\begin{figure*}
\begin{center}
    {\bf Classical Z-tracer}
  \end{center}\vspace{-0.4cm}
\epsfig{file=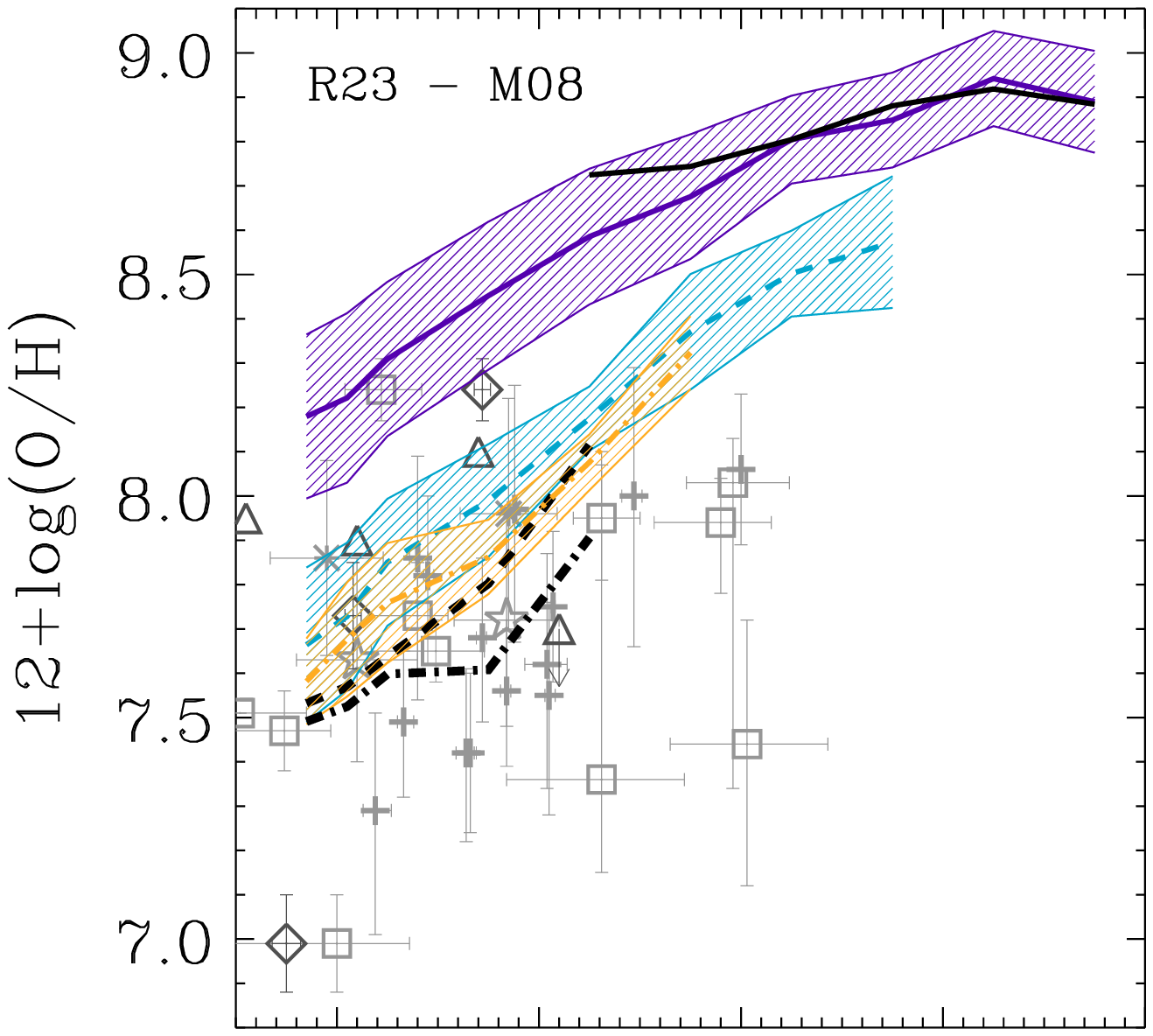,
  width=0.4\textwidth}\hspace{-2.cm}
\epsfig{file=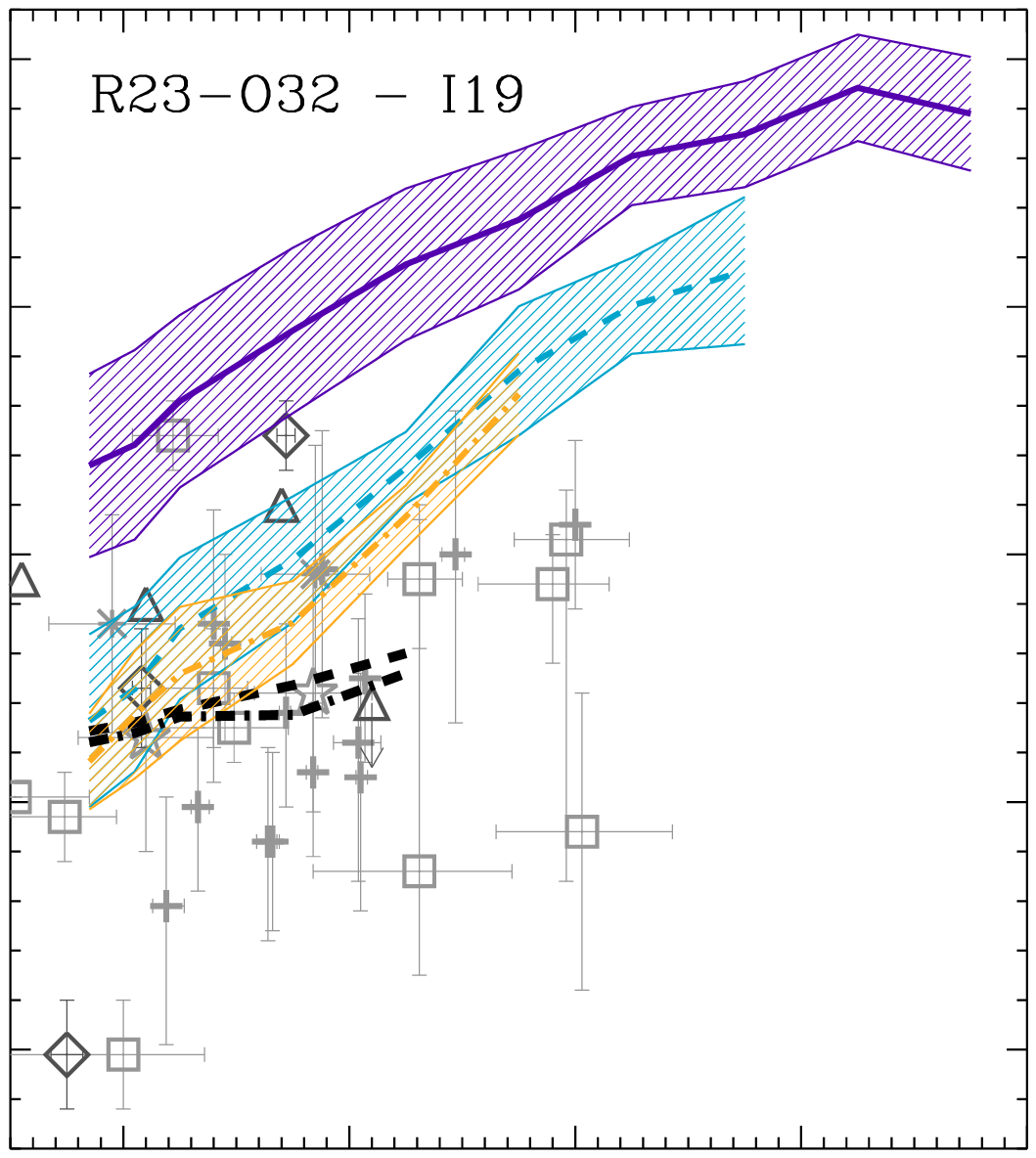,
  width=0.4\textwidth}\hspace{-2.cm}
\epsfig{file=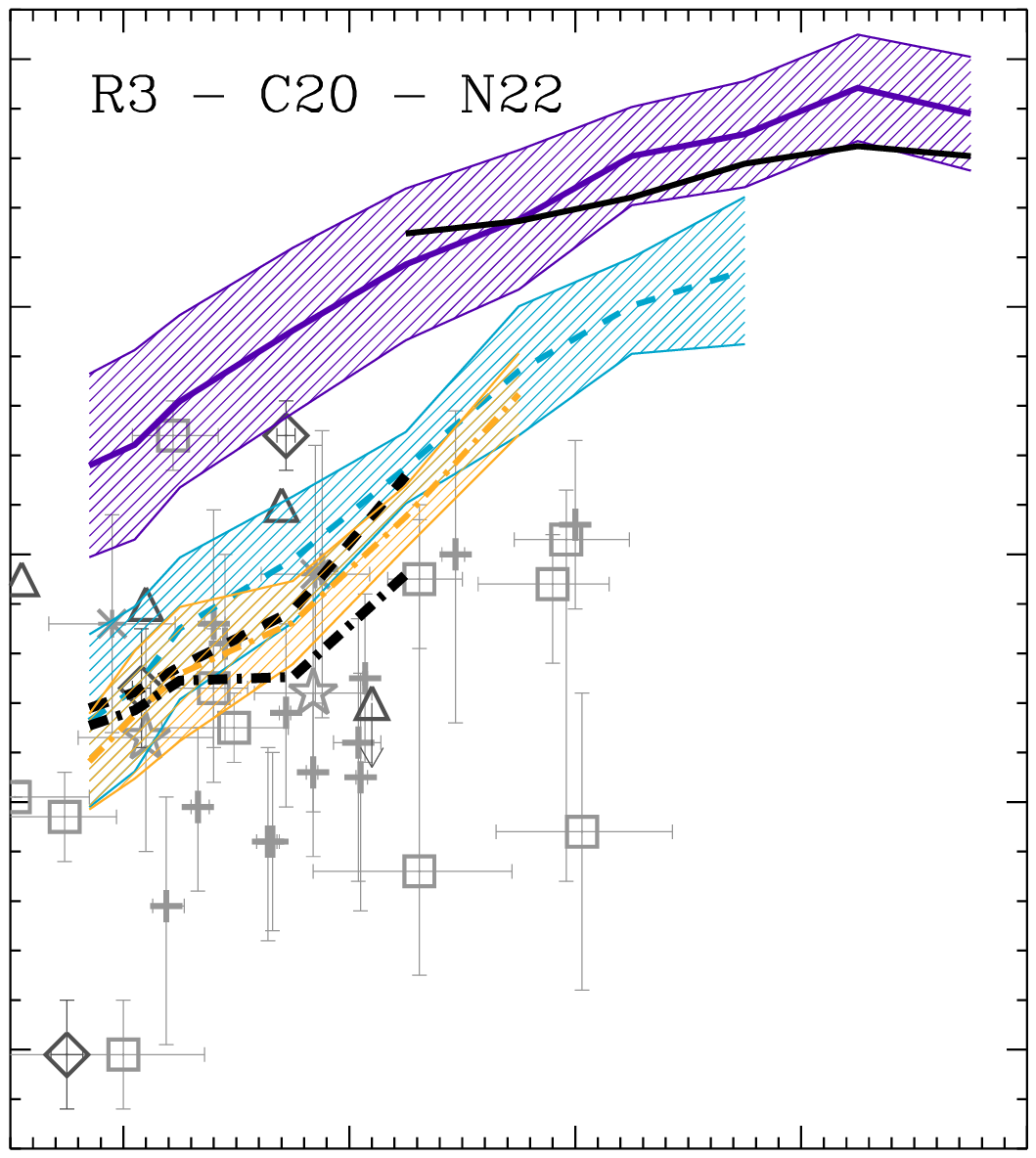,
  width=0.4\textwidth}\vspace{-1.2cm}
\begin{center}
    {\bf \color{red}{Metallicity tracer of this work}}
  \end{center}\vspace{-0.4cm}
  \epsfig{file=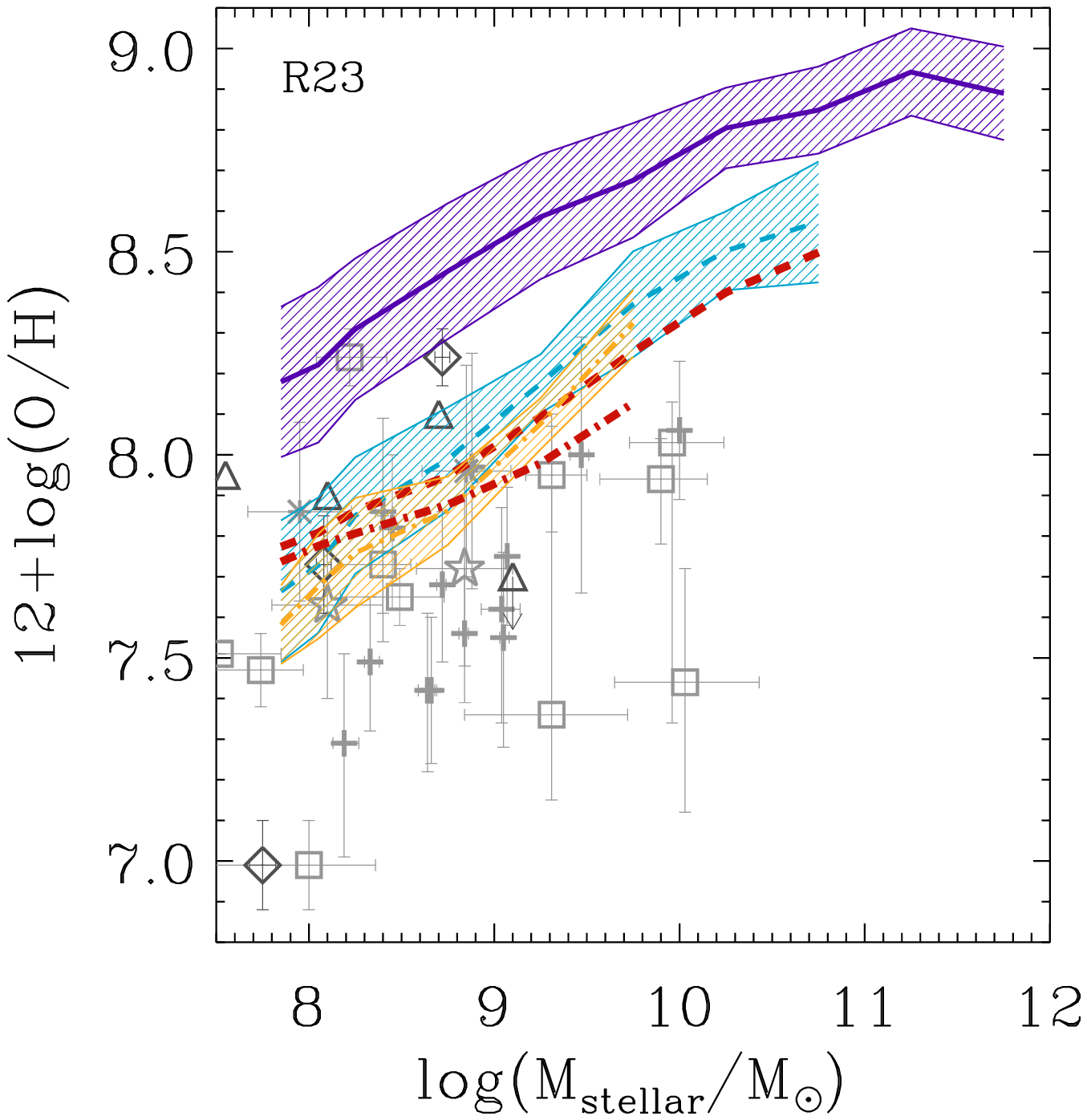,
  width=0.4\textwidth}\hspace{-2.cm}
\epsfig{file=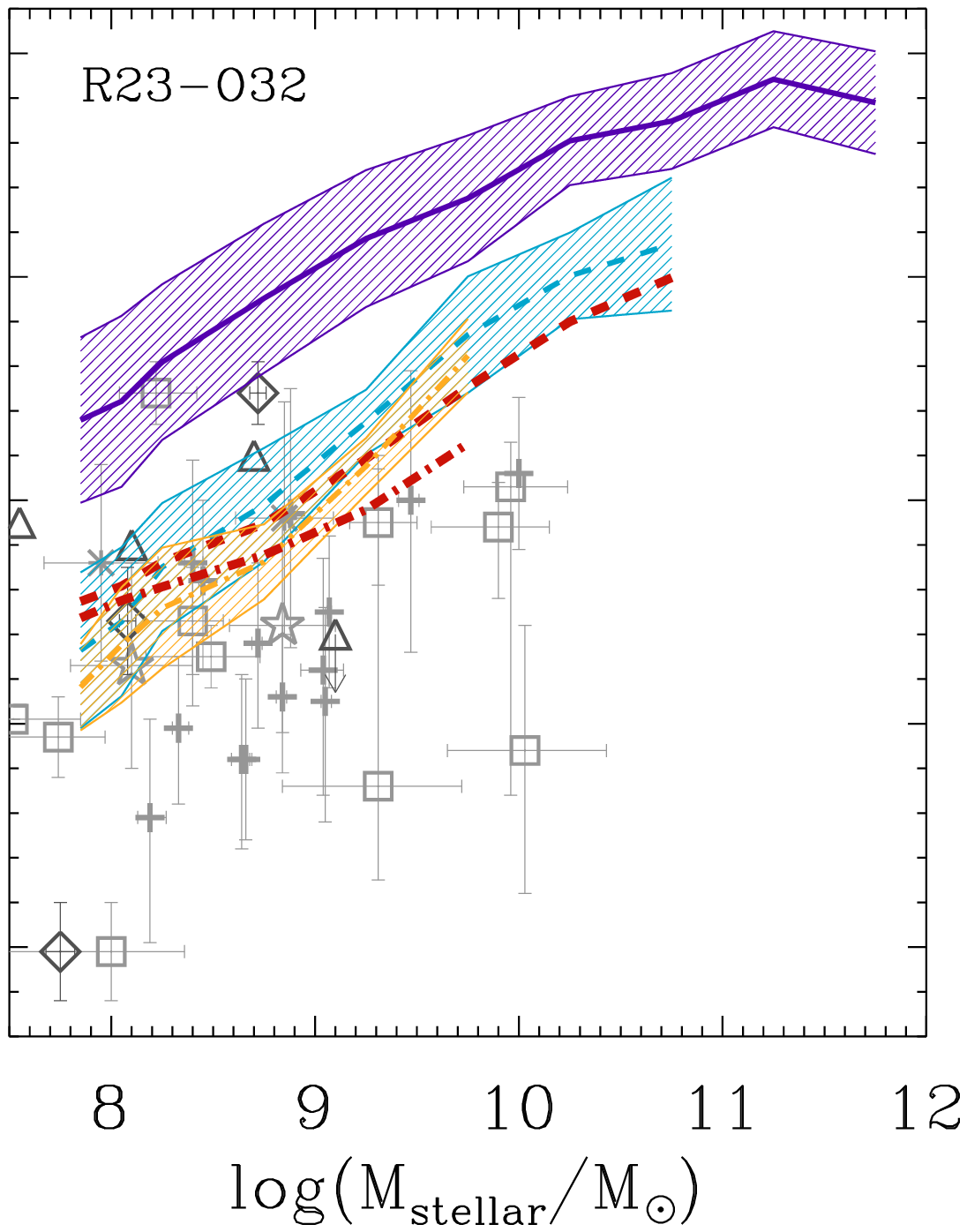,
  width=0.4\textwidth}\hspace{-2.cm}
\epsfig{file=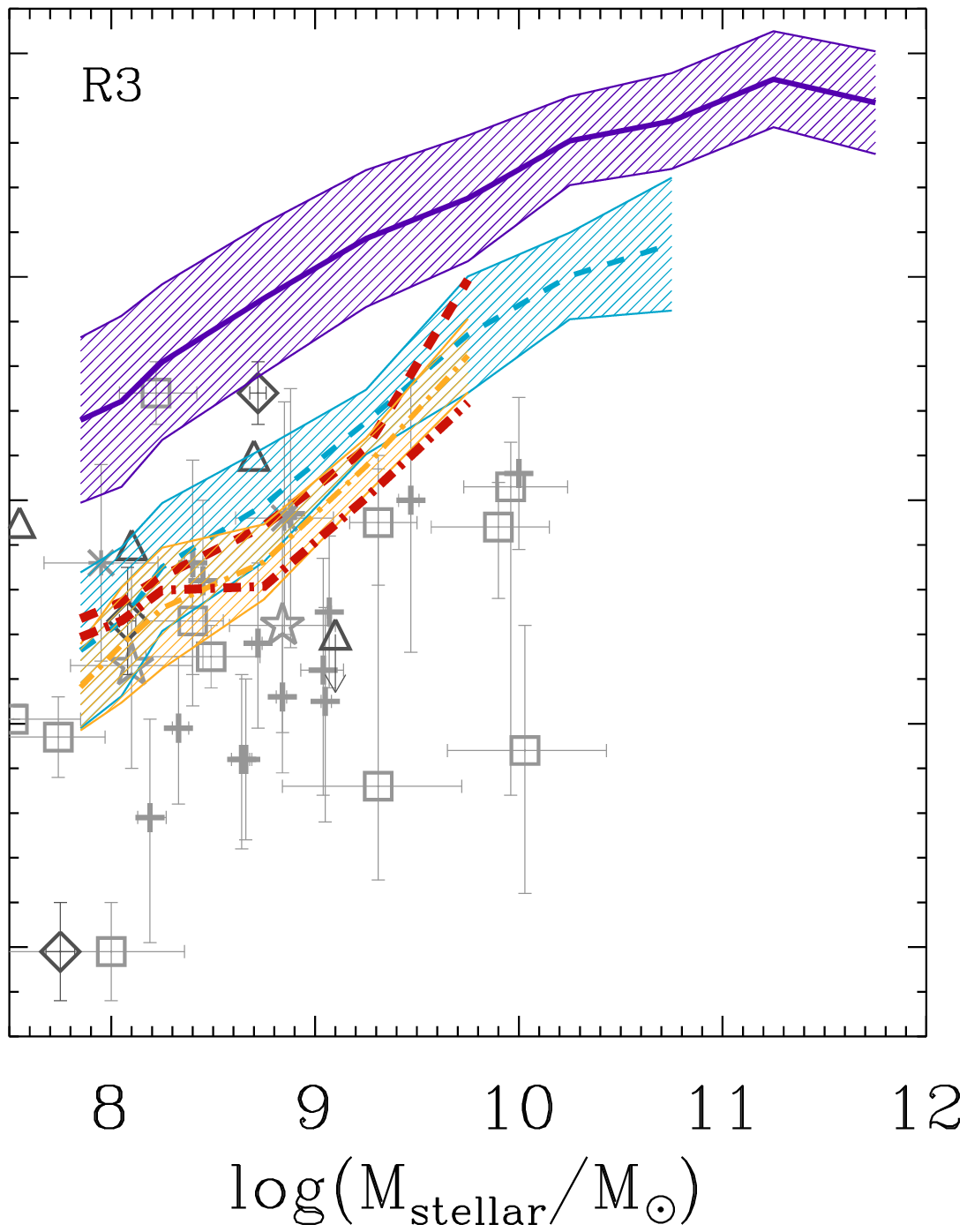,
  width=0.4\textwidth}\vspace{-0.2cm}\hspace{1cm}
 \epsfig{file=Legend_Fig5.eps,
    width=0.98\textwidth}\vspace{-1.cm}
\caption{Shown is the same as in Fig. \ref{MassMet1}, but for R23,
  R23-O32 and R3 as line-ratio tracers. Here, the black lines in the
  top panels were inferred from published $z=0$ calibrations of the
  R23 \citep[][M08]{Maiolino08},  R23-O32   \citep[][I19]{Izotov19}
  and R3 metallicity tracers (\citealt{Curti20}, C20, for $z=0$; 
  \citealt{Nakajima22}, N22, for $z\geq4$).}\label{MassMet2}          
\end{figure*}

It is worth noting that the trends followed by C3O3, C4O3, N3O3,
Si2C2, C2Si3, C3Si3 and He2C3 in Fig.~\ref{met_uvlineratio_evol}, 
obtained by combining the models of \citet{Gutkin16} and \citet{Feltre16} 
with IllustrisTNG simulations, are largely consistent with those shown by 
\citet{Byler18}  and \citet{Kewley19}, with perhaps two notable 
exceptions: (i) \citet{Byler18} find a strong dependence of C3Si3 on
metallicity with a small scatter, whereas our models show that at high 
redshift, the relation flattens dramatically; (ii) while \citet{Byler18} 
suggest that  Si2C2 is a strong metallicity tracer, our models predict
only a weak dependence on metallicity, with a change by less than 
0.4~dex over the whole range $7\la\logoh\la9$. As noted by \citet{Kewley19}, 
metallicity tracers including a silicon line should be used with
caution, because this element can be heavily depleted onto dust 
grains (for example, \citealt{Gutkin16} assume 90 per cent depletion; 
see their table~1). Since the amount of depletion may vary from galaxy 
to galaxy at different redshifts (and Si can be returned to the gas phase
via dust destruction), the use of Si diagnostics requires depletion to be 
known and the appropriate value to be used in the photoionization modelling. 
Thus, differences in the dust treatment are likely to be at least in part 
responsible for the discrepancy between our results and those of
\citet{Byler18} for Si-based metallicity estimators.

\subsection{Evolution of the mass-metallicity relation using
  different metallicity calibrations}\label{massmet}

{An important outcome of Section~\ref{opticalhighz} is that
  classical calibrations of optical-line ratios as metallicity
  estimators are predicted to evolve with redshift. In this 
  section, we discuss the biases introduced by the adoption of $z=0$ 
  calibrations at higher redshifts and the implications 
  for the derived cosmic evolution of the mass-metallicity relation.} 

The evolution of the mass-metallicity relation predicted by 
the IllustrisTNG simulations has been presented, discussed and 
shown to be in broad general agreement with $z<2$ observations
in \citet{Torrey19}.\footnote{{We find that the mass-metallicity 
relations derived from the TNG50 and TNG100 simulations
are largely consistent with one another and with \citet{Torrey19}.
\citet{Heintz22}, who make different assumptions on computing
metallicities of TNG galaxies, obtain different O abundances.}} In 
Figs~\ref{MassMet1} and \ref{MassMet2}, we show the corresponding
dependence of \logoh\ on stellar mass predicted for TNG
galaxies\footnote{{Given the similarity of the mass-metallicity 
relations of TNG50 and TNG100 galaxies, both samples were merged
to define a unique mass-metallicity relation of TNG galaxies and
improve the clarity of Figs.~\ref{MassMet1} and
  \ref{MassMet2}.}} at three redshifts: $z = 0$ (solid lavender line and  
shaded 1$\sigma$-scatter), $z=4$ (dashed light-blue line and 
scatter) and $z = 6$--8 (dashed-dotted orange line and scatter). 
This `intrinsic' mass-metallicity relation of simulated galaxies
(repeated  in all panels) exhibits little evolution at $z>4$.
The
different panels in both figures correspond to a selection of
different optical-line ratios used in the literature to estimate
metallicity: O3N2, N2, N2O2 in Fig.~\ref{MassMet1},  and R23, R23-O32
\citep[a new diagnostic defined as
$\mathrm{R23}-0.08\times\mathrm{O32}$ by][]{Izotov19} and R3 \citep[with
a specific calibration for local analogues of $z\ge4$ metal-poor galaxies
from][]{Nakajima22} in Fig.~\ref{MassMet2}. 
Also shown in all panels of both Figs~\ref{MassMet1} and \ref{MassMet2} are data
for the same  $z\approx6$--8 galaxies observed with \JWST\ as in
Fig.~\ref{met_lineratio_evol}, with \Te-based metallicity estimates
from  \citet[][diamonds]{Curti23} and \citet[][triangles]{Trump23}. 
These are scattered around the predicted $z=6$--8
relation.

We also report the \JWST\ results of
  \citet{Langeroodi22}, \citet{Heintz22} and \citet{Curti23}, who used
  different line ratios to estimate O abundances of $z=3$--10 galaxies
  (R23-O32 in \citealp{Langeroodi22} and mainly R3
  in \citealp{Heintz22} and \citealp{Curti23}). These empirical
  measurements exhibit a large scatter in metallicity at fixed stellar mass,
  probably because of the large uncertainties affecting estimates of both
  quantities. The derived O abundances are typically a few $\times0.1$~dex 
  below those predicted for TNG galaxies at $z=4$--8.   


In addition to these observations and the intrinsic mass-metallicity 
relation of TNG galaxies at different redshifts, we show in each 
panel the relation between \logoh\ and stellar mass derived in two
ways from the simulated optical-line ratios of  TNG galaxies. Firstly,
in the top panels of Figs.~\ref{MassMet1} and \ref{MassMet2}, we
use published $z=0$ calibrations of the metallicity estimators O3N2
\citep{Pettini04}, N2 \citep{Curti20} and  N2O2 \citep{Sanders18}, as well as
  R23 \citep{Maiolino08}, R23-O32 \citep{Izotov19} and R3 (\citealt{Curti20}
  for  $z=0$ and \citealt{Nakajima22} for $z\ge 4$), with results shown as black 
  solid, dashed and dot-dashed lines for $z=0$, 4 and 6--8.\footnote{We do not
  consider here the empirical calibrations  of \citet{Curti20} for 
  O3N2 and R23, which were derived over restricted  ranges of line
  ratios not overlapping with those of high-redshift TNG galaxies.}
Secondly, in the bottom panels of Figs.~\ref{MassMet1} and
  \ref{MassMet2}, we adopt the new, redshift-dependent calibrations
  proposed  in Section~\ref{opticalhighz} (Table~\ref{table1}), with
  results shown as red solid, dashed and dot-dashed lines for $z=0$, 4
  and 6--8, respectively.  

Figs.~\ref{MassMet1} and \ref{MassMet2} show that using classical
$z=0$ calibrations to  estimate O abundances from optical-line ratios
broadly retrieves the  intrinsic mass-metallicity relation at $z=0$
for stellar masses below  $\sim3\times 10^{10}\Msun$. Instead, the
abundances of more massive galaxies derived using O3N2, N2, N2O2 and 
R3 tend to be underestimated  by up to $\sim0.2$~dex. At redshifts $z
\geq 4$, the O abundances derived  using the $z=0$ calibrations of all
estimators except N2O2, and to a lesser extent also R3 (employing 
  the novel calibration of \citealp{Nakajima22} for metal-poor
  galaxies), are biased  downward by up to $\sim1$~dex, implying a
much stronger evolution of the  mass-metallicity relation than the
actual intrinsic one. Such a bias results from the sensitivity of
these line ratios (but not N2O2) to the cosmic evolution of the
ionization parameter, as seen in Section~\ref{opticalhighz}. 

Instead, the mass-metallicity relations derived when adopting the 
new, redshift-dependent calibrations of optical estimators proposed
in Section~\ref{opticalhighz} are in excellent agreement with the intrinsic 
relations in Figs.~\ref{MassMet1} and \ref{MassMet2}, suggesting that
these novel  calibrations represent a promising tool to interpret
spectroscopic surveys  of high-redshift galaxies.  

It is important to note that the results of Figs.~\ref{MassMet1} and
\ref{MassMet2} show only examples (O3N2, N2, N2O2, R23 and
R3) of the potential of our new  calibrations of classical metallicity
estimators to interpret high-redshift  galaxy spectra. We checked that
the predictions for the other line ratios  investigated in
Section~\ref{opticalhighz} (N2S2, S2, R2, RS23, O32 and O3S2) yield
qualitatively similar results in avoiding biases  introduced by the
use of standard $z=0$ calibrations. It is also worth  mentioning that
attempts have been made in the literature to account  for the
secondary dependence of optical-line ratios on ionization  parameter
\citep[see, e.g.,][and references therein]{Poetrodjojo18,
  Kewley19}. While those refined methodologies may improve the
accuracy of metallicities estimates in high-redshift galaxies, a more
thorough exploration goes beyond the scope of the present paper.

\section{Discussion}\label{discussion} 

In Section \ref{opticalz0}, we showed that the dependence of optical-line 
ratios on interstellar metallicity derived from the IllustrisTNG simulations of 
galaxy populations at $z=0$ agree qualitatively with published calibrations. 
This led us, in Sections \ref{opticalhighz} and \ref{UVlineratios}, to make
predictions about the dependence of optical- and UV-line ratios on oxygen
abundance in distant galaxies, out to $z \sim 8$, and propose associated 
diagnostics of interstellar metallicity to guide interpretations of new
spectroscopic surveys at high redshift. In Section \ref{massmet},
  we illustrated the biases introduced by the adoption of $z=0$ calibrations
  at higher redshifts and the implications for the derived cosmic
  evolution of the mass-metallicity relation. 
These results represent an important extension of our earlier work
focused on the use of IllustrisTNG  simulations to identify
diagnostics of the dominant ionizing sources in  galaxies and the
census of emission-line galaxies over cosmic time \citep{Hirschmann22}. 

An important outcome of the present work is the derivation of new 
calibration of metallicity estimators for high-redshift galaxies, aimed at 
providing a refined alternative to classical methods. In this context, in 
Section~\ref{calibrations} below, we briefly summarize known caveats of 
classical (empirical and  theoretical) metallicity estimators and, in 
Section~\ref{caveats},  the benefits and limitations our approach, to put into
perspective some quantitative differences between our predictions and
published work.

\subsection{Caveats of empirical and theoretical metallicity
  calibrations}\label{calibrations} 

The different methods used to derive metallicities from galaxy spectra,
whether `empirical' (i.e., based on the direct-\Te\ method) or `theoretical'
(based on photoionization models), each suffer from their own known 
limitations, which are likely the origin of discrepancies between the 
different calibrations of metallicity estimators (Fig.~\ref{opticalz0}). We
briefly review these here.

{\bf Empirical metallicity calibrations and direct-\Te\ method:} the
direct-\Te\ method combines an estimate of the electron temperature
from ratios of auroral to nebular forbidden-line intensities (e.g.
[OIII]$\lambda$4363/$\lambda$5007) with an estimate of the electron
density \nel\ from ratios of nebular forbidden-line intensities (e.g.
[OII]$\lambda$3729/$\lambda$3726). In general, the ions used to 
estimate \Te\ and \nel\ trace different ionized zones (e.g. O$^{2+}$ 
versus O$^+$), which forces the appeal to photoionization models to
compute the contributions of all ionic species (O$^0$, O$^+$, 
O$^{2+}$, etc.) to the total abundance of an element (O), based on the
observed line luminosities (e.g., \oi, \oii, \oiii, etc.). Such models are
typically calibrated on local \hii\ regions, whose properties may differ 
from those of chemically young galaxies at high redshift 
\citep[see section~5.1 of][]{Gutkin16}. 

Even in the Milky Way and nearby extragalactic \hii\ regions, the direct-\Te\ 
method appears to yield consistently lower abundances than those 
derived from metal-recombination lines, often considered as the `gold 
standard'  \citep{Peimbert17} of metallicity estimators \citep[e.g.][]{Peimbert93, 
Mathis99, Tsamis03}. This mismatch of typically $\sim0.2$~dex (which 
can reach $\sim0.6$~dex in some cases) is commonly referred to as the 
`abundance discrepancy factor' \citep[ADF; e.g.,][]{Garcia-Rojas05, 
Esteban09, Tsamis08, Mesa-Delgado10}. While a number of factors may
potentially cause this discrepancy (e.g., temperature fluctuations, departures 
from thermal equilibrium), its actual origin remains unknown (see 
\citealp{Kewley19, Maiolino19} for further details). 

Another complication of the direct-\Te\ method (perhaps less relevant 
in the context of young high-redshift galaxies) is that auroral lines are 
rarely seen in galaxies with O abundances greater than $\logoh\approx8.7$ 
and are sensitive to temperature gradients, which could lead to systematic 
underestimates of metallicity (by up to $\sim0.9$~dex) in more metal-rich 
environments \citep[see e.g.][]{Stasinska05, Kewley19, Katz23}. 

These various limitations can potentially bias metallicities 
estimated using strong-line ratios calibrated via the direct-\Te\ 
method, especially for metal-rich galaxies. In fact, while modern
spectrographs have enabled the assembly of \Te-based metallicity estimates 
for large galaxy samples \citep[e.g.,][]{Marino13, Curti20}, 
different calibrations based on fits to such metallicities still exhibit 
large scatter (Fig.~\ref{met_lineratio_z0} above and figure~8 
of \citealp{Kewley19}). 

{\bf Theoretical metallicity calibrations:} originally, purely theoretical 
calibrations of metallicity estimators were developed using
photoionization models to overcome discrepancies between different 
empirical calibrations. The main drawbacks of theoretical calibrations 
are the dependence on model assumptions (e.g., simplified 1D geometry, 
population synthesis modelling, ionization- versus density-bounded
\hii\ regions, dependence on current atomic data, simplified density 
structure) and the incorporation of additional components
(e.g., interstellar absorption, contribution by diffuse ionized gas). 
As a result, theoretical metallicity calibrations exhibit significant 
dispersion, just like empirical calibrations. The many adjustable 
parameters involved, which have competing effects on line ratios, 
hamper precise theoretical metallicity calibrations tailored to galaxies 
with unknown physical parameters at different cosmic epochs.

These difficulties can be partially overcome by using a Bayesian 
approach, in combination with large model libraries sampling the 
full space of adjustable parameters, to constrain metallicity from 
the simultaneous fitting of multiple emission lines (e.g., BEAGLE: 
\citealp{Chevallard16, Vidal22}; NEBULARBAYES: 
\citealp{Thomas18}; and BAGPIPES: \citealp{Carnall19}). While this 
approach provides statistical constraints on metallicity, these may 
end up being quite broad when the number of emission lines 
detected with high signal-to-noise ratio is limited.

\subsection{Caveats of our modelling approach}\label{caveats}

Our approach in this paper to identify metallicity diagnostics for 
high-redshift galaxy populations using IllustrisTNG simulations
alleviates some of the drawbacks of the direct-\Te\ method and 
the large photoionization-model grids described in Section~\ref{calibrations}.
Yet, this approach has its own caveats and limitations.

Firstly, although it allows
a drastic reduction of the parameter space of photoionization models
by using the constraints from cosmological evolution, it still suffers from the 
intrinsic limitations of photoionization models (Section~\ref{calibrations}).

Secondly, large-scale cosmological simulations (such as IllustrisTNG)
cannot directly resolve the multi-phase ISM nor many of the relevant
processes such as stellar and AGN feedback.
This forces the adoption of simplified and
ad-hoc sub-resolution models to describe baryonic processes, which
  are typically different for different simulations. Even if
the IllustrisTNG simulation suite has been extensively tested and
validated for low-redshift studies (Section~\ref{TNG}, and references
therein),  predictions for galaxies at redshifts greater than $z\sim3$
remain uncertain and can vary for different simulations and models
  (in particular the mass-metallicity relation; see,
  e.g., \citealp{Somerville15, Hirschmann16,
  Heintz22}). Different prescriptions of baryonic processes or in the 
  coupling of simulations with photoionization models might lead 
  to differences in the predicted ionization parameter, which strongly
  influences emission-line ratios.

{In fact, limitations arise} from the approximations entailed in 
the coupling with photoionization models. For example, the lack
of resolution of the ionized-gas component forces the adoption of 
approximate, galaxy-wide values of the hydrogen density, metallicity 
and dust-to-metal mass ratio (Section~\ref{ELmodels}). Moreover, in 
our approach, the ionization parameter is controlled by the star formation 
rate, consistently with empirical relations linking \logU, metallicity and 
specific SFR \citep[see section~2.3.1 of][and also the recent correlation 
between \logU\ and SFR surface density found by
\citealt{Reddy23}]{Hirschmann17}.  A weaker correlation between \logU\
and SFR would soften the predicted increase in \logU\ from low to high
redshift for IllustrisTNG galaxies.  

Despite these limitations, as shown in Section~\ref{opticalz0} above and
in \citet{Hirschmann22}, we have carefully and successfully validated our
methodology against numerous observational emission-line properties of 
galaxies at redshifts $z\approx0$--2, putting the current analysis and the
predicted metallicity diagnostics for high-redshift galaxies on firm
grounds.

\section{Summary}\label{summary} 

{Taking advantage of the multi-component emission-line catalogues of
IllustrisTNG galaxy populations presented in \citet{Hirschmann22},
we have investigated different optical and UV emission-line
diagnostics to estimate O abundances from observed spectra for both
present-day and high-redshift galaxies.  The emission-line catalogues
have been constructed using the IllustrisTNG cosmological 
simulations and self-consistently connecting them to modern,
state-of-the-art photoionization models \citep{Gutkin16, Feltre16,
  Hirschmann17,  Alarie19} based on the methodology of
\citet{Hirschmann17,   Hirschmann19, Hirschmann22}. This allows us to
compute the line emission from multiple components: young
star clusters,  AGN NLR, PAGB stellar populations and fast radiative
shocks. We can summarize our main results as follows: }
\begin{itemize}
\item For present-day IllustrisTNG galaxies, the
optical-line ratios \niioii\ (N2O2),
  \niisii\ (N2S2), \niiha\ (N2), \siiha\ (S2), \oiiihb/(\niiha) (O3N2),
  \oiiioii\ (O32), \oiiihb/(\siiha) (O3S2), \oiihb\ (R2),  \oiiihb\
  (R3), \oiiioiihb\ (with \hbox{[O\,{\sc iii}]}=\oiii+\oiiifour, R23) and \oiiihb +\siiha\ (RS23)
  are closely related to metallicity, in good agreement with empirical  
  and theoretical metallicity calibrations of SF galaxies published in the
  literature. Instead, for AGN-dominated, composite, shock- and 
  PAGB-dominated galaxies, the relations between optical-line ratios and
  metallicity can deviate from that for SF-dominated
  galaxies, and thus, should not be used as metallicity tracers for
  galaxy types other than SF. 

\item From $z=0$ to $z=4$, most considered optical-line ratios
  are predicted to strongly increase or decrease (except for N2O2 and N2S2) at
  fixed oxygen abundance. This mainly results from an evolution 
  of the ionization parameter, which increases for higher-redshift 
  galaxies due to their higher sSFR  and gas density.
  
   \item Also, several UV-line ratios, such as \heii/\ciii\ (He2C3),
  \heii/\oiiiuv\ (He2O3) and \niii/\oiii\ (N3O3), appear to strongly correlate
  with the O abundance of simulated galaxies at
  different cosmic epochs, suggesting that these provide powerful metallicity
  diagnostics for distant metal-poor galaxies. 

  \item At $z \geq 4$, the calibrations of both optical and 
    UV metallicity estimators exhibit almost no evolution with
    redshift. We find fairly good agreement between
    the predicted R2, R3, R23 and O32 calibrations 
    and those observationally estimated using the direct-\Te\ method for
    the $z=4$--8 galaxies presented in \citet{Curti23} and \citet{Sanders23}, in
    particular when accounting for observational-selection effects.
    We interpret the level of agreement between models and observations 
    as an encouraging illustration of the success of our
    approach. This motivated us to propose novel diagnostics for the
    interstellar metallicity in distant galaxies out to $z\sim8$, to
    guide interpretations of new spectroscopic surveys at high
    redshift. 
 
\item The last points entail an important consequence for the
  evolution of the mass-metallicity relation at high redshift: when
  metallicities of observed, high-redshift galaxies are estimated from
  $z=0$ calibrations, they can be biased downward by up to $\sim$1~dex 
  at a given galaxy stellar mass. This can lead to a much stronger evolution
  of the observed mass-metallicity relation than the actual one, hampering a
  meaningful comparison between models and observations.

\end{itemize}

{Overall, the multi-component optical and UV emission-line galaxy
 catalogues provide useful  insights into different optical and UV
 metallicity diagnostics for high-redshift galaxies. Specifically,
 the provided metallicity calibrations for high-redshift galaxies may
 guide the  interpretation of different near- and far-future emission-line
 surveys not only with  \JWST/NIRSpec, but also with for example,
 VLT/MOONs and ELT/Mosaic. Our results may contribute to obtain robust
 insights into the chemical enrichment of galaxies out 
 to cosmic dawn and, thus, provide accurate constraints to
 validate/invalidate uncertain models (e.g., stellar feedback) adopted
 in state-of-the-art cosmological simulations.}  

\section*{Data Availability}
 The data underlying this article are partly available in the article,
 will be also shared on reasonable request to the corresponding
 author. 

\section*{Acknowledgements}

The authors would like to thank Anna Feltre for providing us with the
AGN photo-ionisaiton models, and the BEAGLE team for fruitful
discussions.
MH acknowledges funding from the Swiss National Science Foundation
(SNF) via a PRIMA Grant PR00P2 193577 ``From cosmic dawn to high noon:
the role of black holes for young galaxies''. RSS is supported by the
Simons Foundation. 

\bibliographystyle{mnras}
\bibliography{Literaturdatenbank}

\label{lastpage}

\end{document}